\documentclass[twocolumn,prx,aps,superscriptaddress,longbibliography]{revtex4-2}
\usepackage{amsmath}
\usepackage{amssymb}
\usepackage{amsfonts}
\usepackage[dvips]{graphicx}
\usepackage{subfigure}
\usepackage{dcolumn}
\usepackage{txfonts}
\usepackage{bm}
\usepackage{makeidx}
\usepackage{color}
\usepackage{mathtools}
\usepackage{threeparttable}
\usepackage[colorlinks,linkcolor=blue,anchorcolor=blue,citecolor=blue,urlcolor=blue]{hyperref}

\begin{document}


\title{Quantum theory of photonic vortices and quantum statistics of twisted photons}
\author{Li-Ping Yang}

\affiliation{Center for Quantum Sciences and School of Physics, Northeast Normal University, Changchun 130024, China}

\author{Dazhi Xu}
\affiliation{Center for Quantum Technology Research and Key Laboratory of Advanced Optoelectronic Quantum Architecture and Measurement (MOE) and School of Physics, Beijing Institute of Technology, Beijing 100081, China}

\begin{abstract}
The topological charge of a photonic vortex is an essential quantity in singular optics and the critical parameter to characterize the vorticity of twisted light. However, the definition of the photonic topological charge remains elusive. Here we put forth a theoretical formalism to provide a comprehensive treatment of photonic vortices. We introduce quantum operators for the photon current density and helicity current density based on the continuity equations from the paraxial Helmholtz equation. Our formalism allows us to introduce flow velocity and circulation for photonic currents in parallel to their counterparts in superfluids. The quantized circulation of the photonic currents is conserved during propagation and it gives an explicit definition of the photonic topological charge as the winding number of a photonic vortex. In particular, we predict helicity current generated pure helicity vortices, in which the photon current vanishes. Finally, we show an interesting effect that the quantum statistics of twisted photon pairs are essentially determined by their spin states.
\end{abstract}

\maketitle

\section{Introduction}
In parallel to the outstanding advances in the orbital angular momentum (OAM) of light~\cite{allen1992orbital,allen1999iv,bliokh2015transverse}, photonic vortices in structured light with helical wavefront have also been studied intensively in both theory and experiments~\cite{molina2007twisted,Gotte2008Light,yu2011light,Naidoo2016controlled,devlin2017arbitrary,Forbes2021Structured}. Optical vortex beams have been routinely generated in lab~\cite{hayenga2019direct,zhang2020tunable,Sroor2020high,Cai2012Integrated,Wang2018Recent}. Single-photon source for vortex pulses~\cite{Chen2021Bright} and optical vortex lattice~\cite{Zhu2021optical,Du2017Chip} have also been achieved. Recently, the concept of the photonic vortex has also been generalized to spatiotemporal pulses~\cite{Jhajj2016spatiotemporal,Hancock2019free,Chong2020Generation}, in which the energy density varies spatially and temporally. Vortex beams and pulses have been widely applied in optical trapping~\cite{Gahagan1996Optical}, quantum communication and quantum information~\cite{Malik2016multiphoton,Ding2015Quantum,Malik2014Direct}, quantum computation~\cite{Babazadeh2017high}, bio-sensing~\cite{Zhuang2004Unraveling}, strong-field photoelectron ionization~\cite{Fang2021photo}, etc. However, the theoretical definition of the photonic topological charge, which is an essential quantity and critical parameter of photonic vortices, remains elusive.

Existing theories of photonic vortices have advanced over the past two decades to capture a plethora of phenomena related to phase singularities of light~\cite{COULLET1989optical,Soskin1997topological,Berry2001Polarization,Berry2004optical,Dennis2002polarization,freund2002polarization,Kotlyar2020Topological}. Based on the helical phase of the complex field-amplitude function, Berry defined the total vortex strength (photonic topological charge) as the signed sum of all the vortices threading a large loop including the propagating axis~\cite{Berry2004optical}. This seminal work has triggered extensive interest in exploring photonic vortices with fractional topological charges~\cite{Wen2019vortex,Leach2004observation,Tao2005Fractional,Guo2010Optical,Fang2017Fractional,Kotlyar2020Topological}. Akin to the quantum vortices in superfluids, the most basic quantity of a vortex is the corresponding current, which has not received the attention it deserves for photonic vortices. The photon current for a photonic polarization vortex has even been overlooked completely~\cite{Dennis2002polarization,freund2002polarization}. Without the associated current, the essential link between the photonic topological charge and the vorticity of twisted light is missing. On the other hand, the photon current defined as the gradient of the phase of the complex electric field has no clear physical meaning~\cite{Dennis2009Chapter}, since no continuity equation exists for this current.

\begin{figure*}
\centering
\includegraphics[width=12cm]{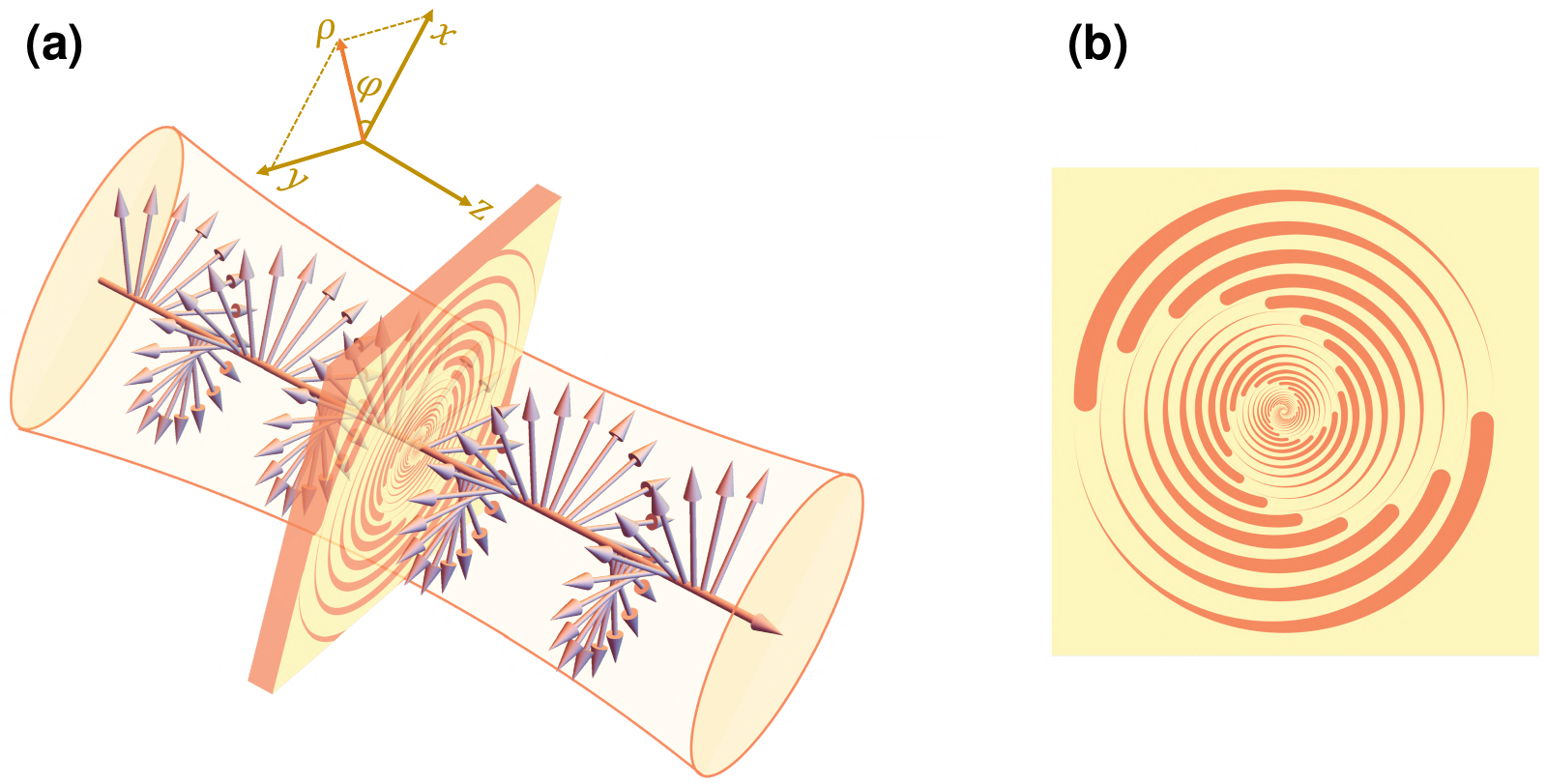}\caption{\label{fig:schematic} Schematic of (a) a twisted beam carrying orbital angular momentum and (b) the in-plane photonic current generated vortex in the transverse plane (b). The inset in (a) shows the cylindrical coordinate $\boldsymbol{r}=(\rho\cos\varphi,\rho\sin\varphi,z)$.}
\end{figure*}

In this paper, we put forth a theoretical formalism to provide a comprehensive treatment of photonic vortices (see the schematics in Fig.~\ref{fig:schematic}).  We introduce an effective photonic field operator, which allows us to handle photonic quantities (such as momentum, helicity, OAM, etc.) in real space within the standard framework of quantum mechanics. Specifically, building on previous important works in the optical Schr\"{o}dinger equation for paraxial light~\cite{Siviloglou2007observation,Wan2007dispersive,zhang2019particle}, we define quantum operators for the photon current density and helicity current density. In parallel to superfluids in condensed-matter physics, we introduce the flow velocities and the corresponding circulations for these two currents. The conserved and quantized circulation automatically connects the photonic topological charge to the winding number of the photonic vortex. We show a  particularly interesting result that a pure-helicity vortex with vanishing photon (particle) current can be obtained via the superposition of left and right circularly polarized laser beams. On the other hand, the quantum statistics of the twisted light remain poorly studied in both theory and experiments. Within the well-established theoretical framework for quantum optical coherence~\cite{Glauber1963coherence}, we show a particularly interesting effect that the quantum statistics of twisted photon pairs are strongly affected by their spin states. Two photons with the symmetric spin state tend to be bunched, and two photons with the anti-symmetric polarization state behave in a more anti-bunched way.

This article is structured as follows. In Sec. II, we begin by introducing a theoretical formalism for photonic quantities in real space. In Sec. III, we apply this formalism to investigate photonic vortices by defining photonic currents, flow velocities, and the corresponding circulations. In Sec. IV, we show how to apply our theory to study twisted laser beams commonly used in experiments. In Sec. V, we discuss the quantum coherence and quantum statistics of twisted light. We briefly summarize in Sec. VI.

\section{Photonic field operator and wave-packet function in real space}
Here, we introduce a quantum formalism to describe the photonic observables in striking parallel to their electronic counterparts. In particular, this rather general formalism leads us to define the photon current and helicity current via the continuity equation of paraxial beams (see Sec.~\ref{sec:vortex}). 
The quantized circulation of the current in a paraxial beam automatically gives the topological charge (winding number) of a photonic vortex~~\cite{Soskin1997topological,Soskin2001Singular,Berry2004optical,Dennis2009Chapter,Kotlyar2020Topological} (Sec.~\ref{sec:examples}). Moreover, this formalism enables us to investigate the quantum statistics of a twisted light (Sec.~\ref{sec:coherence}).

We start with the effective two-component field operator $\hat{\psi}(\boldsymbol{r})=[\hat{\psi}_{+}(\boldsymbol{r}),\hat{\psi}_{-}(\boldsymbol{r})]^T$ for photons in real space~\cite{yang2021quantum} with,
\begin{equation}
\hat{\psi}_{\lambda}(\boldsymbol{r})=\frac{1}{\sqrt{(2\pi)^{3}}}\int d^{3}k\hat{a}_{\boldsymbol{k},\lambda}e^{i\boldsymbol{k}\cdot\boldsymbol{r}},
\end{equation}
where $\hat{a}_{\boldsymbol{k},\lambda}$ is the annihilation operator for the plane-wave mode with wave vector $\boldsymbol{k}$ and the polarization index $\lambda$ ($\lambda=+1$ for left circular polarization and $\lambda=-1$ for right circular polarization). Using the basic equal-time commutation relations of the ladder operators $[\hat{a}_{\boldsymbol{k},\lambda},\hat{a}^{\dagger}_{\boldsymbol{k}',\lambda'}]=\delta_{\lambda\lambda'}\delta(\boldsymbol{k}-\boldsymbol{k}')$, we can verify that our introduced photonic field operator obeys the following bosonic commutation relations,
\begin{align}
[\hat{\psi}_{\lambda}(\boldsymbol{r}),\hat{\psi}_{\lambda'}^{\dagger}(\boldsymbol{r}')] & =\delta_{\lambda\lambda'}\delta(\boldsymbol{r}-\boldsymbol{r}'),\\
[\hat{\psi}_{\lambda}(\boldsymbol{r}),\hat{\psi}_{\lambda'}(\boldsymbol{r}')] & =[\hat{\psi}_{\lambda}^{\dagger}(\boldsymbol{r}),\hat{\psi}_{\lambda'}^{\dagger}(\boldsymbol{r}')]=0.
\end{align}

With this field operator, we can re-express and evaluate the photonic observables in real space, such as the photon number $\hat{N}=\int d^3r\hat{\psi}^{\dagger}(\boldsymbol{r})\hat{\psi}(\boldsymbol{r})$ and the linear momentum of light $\hat{\boldsymbol{P}}=\int d^3r \hat{\psi}^{\dagger}(\boldsymbol{r})\hat{\boldsymbol{p}}\hat{\psi}(\boldsymbol{r})$ ( $\hat{\boldsymbol{p}}=-i\hbar\boldsymbol{\nabla}$). We note that the field operator $\hat{\psi}(\boldsymbol{r})$ satisfies the wave equation, not a Schr{\"o}dinger-like equation and no probability continuity equation can be constructed~\cite{QED1982Landau}. Thus, in most cases, the integral kernel can not be interpreted as the corresponding density operator (see Appendix~\ref{sec:OAM}). However, as shown in Sec.~\ref{sec:paraxial}, the particle number density (PND) of a paraxial light can be well characterized by
\begin{equation}
\hat{n}(r)\equiv\hat{\psi}^{\dagger}(\boldsymbol{r})\hat{\psi}(\boldsymbol{r})=\sum_{\lambda=\pm}\hat{\psi}_{\lambda}^{\dagger}(\boldsymbol{r})\hat{\psi}_{\lambda}(\boldsymbol{r}).    
\end{equation} 
We note that not all physical quantities of light can have a simple and elegant form in this framework, such as the Hamiltonian $\hat{H}=\int d^3k\sum_{\lambda}\hbar\omega_{\boldsymbol{k}}\hat{a}^{\dagger}_{\boldsymbol{k},\lambda}\hat{a}_{\boldsymbol{k},\lambda}$ and the photonic spin operators, which can be handled more easily in $\boldsymbol{k}$ space~\cite{yang2021quantum}. However, our formalism offers significant convenience in dealing with quantities in real space, such as photonic helicity, OAM (see Appendix~\ref{sec:OAM}), vortices, and quantum coherence of light. In the circular-polarization representation, the helicity operator of photons is diagonal 
\begin{equation}
\hat{\Lambda} = \hbar\int d^{3}r\hat{\psi}^{\dagger}(\boldsymbol{r})\hat{\sigma}_z\hat{\psi}(\boldsymbol{r}),
\end{equation}
where $\hat{\sigma_z}$ is the Pauli matrix.   

The quantum state for a pulse or a laser beam (an extremely long pulse) can be constructed with the photon-wave-packet creation operator $
\hat{a}_{\xi}^{\dagger}=\int d^{3}k\sum_{\lambda}\xi_{\lambda}(\boldsymbol{k})\hat{a}_{\boldsymbol{k}\lambda}^{\dagger}$~\cite{yang2021quantum}. The pulse shape is determined by the normalized spectral  amplitude  function  (SAF) $\sum_{\lambda}\int d^3k|\xi_{\lambda}(\boldsymbol{k})|^2=1$. The quantum states for the most commonly encountered Fock-state and coherent-state pulses are given by $|n_{\xi}\rangle  =(\hat{a}_{\xi}^{\dagger})^{n}|0\rangle/\sqrt{n!}$ and $|\alpha_{\xi}\rangle =\hat{D}_{\xi}(\alpha)|0\rangle$, respectively, with
\begin{equation}
\hat{D}_{\xi}(\alpha)\equiv\exp\left(\alpha\hat{a}_{\xi}^{\dagger}-|\alpha|^{2}/2\right). \label{eq:displace}
\end{equation}
Utilizing the time evolution operator $\exp (-i\hat{H}t/\hbar)$, we can obtain a quantum state at time $t$ simply by replacing $\hat{a}_{\xi}^{\dagger}$ with
\begin{equation}
\hat{a}_{\xi}^{\dagger}(t) =\!\!\int\!\! d^{3}k\!\!\sum_{\lambda}\xi_{\lambda}(\boldsymbol{k})e^{-i\omega_{\boldsymbol{k}}t}\hat{a}_{\boldsymbol{k}\lambda}^{\dagger}=\!\!\int\!\! d^{3}r\!\!\sum_{\lambda}\tilde{\xi}_{\lambda}(\boldsymbol{r},t)\hat{\psi}_{\lambda}^{\dagger}(\boldsymbol{r}).   
\end{equation}
Here, the Fourier transformation of the SAF
\begin{equation}
\tilde{\xi}_{\lambda}(\boldsymbol{r},t)=\frac{1}{\sqrt{(2\pi)^{3}}}\int d^{3}k\xi_{\lambda}(\boldsymbol{k})e^{i(\boldsymbol{k}\cdot\boldsymbol{r}-\omega_{\boldsymbol{k}}t)}.
\end{equation}
also satisfies the wave equation. Within our introduced formalism, the mean value of a physical quantity can be obtained via the relations
\begin{align}
\hat{\psi}_{\lambda'}(\boldsymbol{r}')\left|n_{\xi}(t)\right\rangle  & =\sqrt{n}\tilde{\xi}_{\lambda'}(\boldsymbol{r}',t)\left|(n-1)_{\xi}(t)\right\rangle,\\
\hat{\psi}_{\lambda'}(\boldsymbol{r}')\left|\alpha_{\xi}(t)\right\rangle  & =\alpha\tilde{\xi}_{\lambda'}(\boldsymbol{r}',t)\left|\alpha_{\xi}(t)\right\rangle,
\end{align}
where we have used the identities
\begin{align*}
\left[\hat{\psi}_{\lambda'}(\boldsymbol{r}'),\left(\hat{a}_{\xi}^{\dagger}(t)\right)^{n}\right] & =n\tilde{\xi}_{\lambda'}(\boldsymbol{r}',t)\left(\hat{a}_{\xi}^{\dagger}(t)\right)^{n-1},\\
\left[\hat{\psi}_{\lambda'}^{\dagger}(\boldsymbol{r}'),\left(\hat{a}_{\xi}(t)\right)^{n}\right] & =-n\tilde{\xi}_{\lambda'}^{*}(\boldsymbol{r}',t)\left(\hat{a}_{\xi}(t)\right)^{n-1}.
\end{align*}

In experiments, the superposition of multiple laser beams has been routinely used to obtain various interesting structures. Here, we emphasize that the corresponding quantum state is not a simple superposition of the state of each beam. We take the superposition of two beams with strength $\alpha$ and $\alpha'$ and two-component SAFs $\xi$ and $\xi'$ as an example. The corresponding quantum state is given by
\begin{equation}
\left|\Psi\right\rangle =\frac{1}{\sqrt{\mathcal{N}}}\hat{D}_{\xi}(\alpha)\hat{D}_{\xi'}(\alpha')\left|0\right\rangle,
\end{equation}
where $\mathcal{N}$ is a normalization factor given in Eq.~(\ref{eq:NormFactor}). We can verify that $\hat{\psi}(\boldsymbol{r})\left|\Psi\right\rangle =\Psi(\boldsymbol{r},t)\left|\Psi\right\rangle$ (see Appendix~\ref{sec:MPwavefunction}). Here the two-component function $\Psi = [\Psi_+,\Psi_-]^T$, with $\Psi_{\pm}(\boldsymbol{r},t)\equiv\alpha\tilde{\xi}_{\pm}(\boldsymbol{r},t)+\alpha'\tilde{\xi}'_{\pm}(\boldsymbol{r},t)$, also obeys the wave equation. In the following, we only consider quantum pulses or beams. A randomly polarized light~\cite{Berry2001Polarization} or thermal light, which can not be described with a pure quantum state, will not be addressed in this work.

\section{Photonic vortices in paraxial beams \label{sec:vortex}}
Previously, the scalar electric field has been utilized to study the photonic vortices~\cite{Soskin1997topological,Soskin2001Singular,zhang2019particle}. In Ref.~\cite{Berry2004optical} Berry defined the total vortex strength of a paraxial scalar wave with complex amplitude $\psi(\boldsymbol{r})\propto\exp (im\varphi)$ as 
\begin{equation}
TC =\lim_{\rho\rightarrow\infty}\frac{1}{2\pi}\int_0^{2\pi}d\varphi\frac{\partial}{\partial\varphi}{\rm arg}\psi(\boldsymbol{r}),\label{eq:TC1}    
\end{equation}
which was rewritten with the scalar electric field later~\cite{Kotlyar2020Topological}, as
\begin{equation}
TC = \lim_{\rho\rightarrow\infty}\frac{1}{2\pi}{\rm Im}\int_0^{2\pi}d\varphi\frac{\partial E(\boldsymbol{r})/\partial\varphi}{E(\boldsymbol{r})}. \label{eq:TC2}   
\end{equation}
The strength of the phase singularity has also been called the topological charge of photonic vortices. We note that these two definitions are not exactly equivalent to each other. More importantly, the physical meaning of the defined topological charge in (\ref{eq:TC1}) and (\ref{eq:TC2}) is unclear.

There remain basic questions about photonic vortices having not been clarified. The obtained topological charge may not be conserved during propagation~\cite{Soskin1997topological,Gotte2008Light}. The existence of fractional topological charges in a uniformly polarized
light beam will cause confusion~\cite{Berry2004optical,Basistiy2004Synthesis,Kotlyar2020Topological}, because the electromagnetic field proportional to $\propto \exp (i\alpha \varphi)$ with non-integer $\alpha$ is multiple-valued. The strength of each vortex cannot be quantitatively analyzed by (\ref{eq:TC1}) and (\ref{eq:TC2}). The fundamental link between the photonic topological charges and winding numbers of photonic vortices is unclear. All these problems will be solved conclusively within our presented formalism. More importantly, we predict the pure-helicity vortex with vanishing photon flow, which can be measured with circular-polarization-sensitive devices.

\subsection{Photon current density and helicity current density~\label{sec:paraxial}}
The concept of vortex originates from fluid mechanics. In superfluids, the quantized circulation of the dissipationless superflow leads to quantum vortices with integer winding numbers~\cite{Annett2005super,pethick2008bose}. We note that the cornerstone of quantum vortices is the directly observable current density of the corresponding particle flow. Here, we introduce the photonic counterpart.

A paraxial laser beam propagating in the positive $z$ direction can be well described by a quasi-single-frequency two-component function $\Psi (\boldsymbol{r},t)=\Psi_{\rm PA}(\boldsymbol{r})\exp[i(k_0z-\omega_0 t)]$, where $\omega_0 = ck_0$ is its center frequency. The function $\Psi_{\rm PA}(\boldsymbol{r})$ slowly varying in $z$ satisfies the paraxial Helmholtz equation~\cite{Siegman1971An,Zhan2009cylindrical},
\begin{equation}
i\partial_{z}\Psi_{{\rm PA}}(\boldsymbol{r})=-\frac{1}{2k_0}\nabla_{T}^{2}\Psi_{{\rm PA}}(\boldsymbol{r})\label{eq:PHE}
\end{equation}
where $\boldsymbol{\nabla}_{T}=\boldsymbol{e}_{x}\partial_{x}+\boldsymbol{e}_{y}\partial_{y}$ is the differential operator in the $xy$ plane and $\boldsymbol{e}_j$ is the unit vector.
Because the parameter $t$ only contributes a phase factor to $\Psi (\boldsymbol{r},t)$, we can set the time at $t=0$ and take the coordinate $z$ as an effective "time" to study the dynamics of the propagating beam. This paraxial Helmholtz equation can be regarded as the effective Schr{\" o}dinger equation with effective mass $k_0$~\cite{Wan2007dispersive,zhang2019particle}. The two-component function $\Psi (\boldsymbol{r},t)$, which serves as the effective many-body wave function of paraxial light, is adequate to characterize a uniformly polarized or a vector vortex beam~\cite{Zhan2009cylindrical}. 

We now introduce an operator to characterize the photon current density in the $xy$ plane
\begin{equation}
\hat{\boldsymbol{j}}_N (\boldsymbol{r})=-\frac{i}{2k_{0}}\left\{ \hat{\psi}^{\dagger}(\boldsymbol{r})\boldsymbol{\nabla}_{T}\hat{\psi}(\boldsymbol{r})-\left[\boldsymbol{\nabla}_{T}\hat{\psi}^{\dagger}(\boldsymbol{r})\right]\hat{\psi}(\boldsymbol{r})\right\}.\label{eq:jN}
\end{equation}
With the help of the paraxial Helmholtz equation (\ref{eq:PHE}), we obtained the continuity equation (see Appendix~\ref{sec:hydrodynamic})
\begin{equation}
\frac{\partial}{\partial z}\langle\hat{n}\rangle+\boldsymbol{\nabla}_{T}\cdot\langle\hat{\boldsymbol{j}}_N\rangle=0,\label{eq:Continuity}
\end{equation}
which reveals the fact that the total particle number within a co-moving transverse slice does not change as the light propagates along the $z$ axis. We note that this conservation law is valid only under the paraxial approximation. Thus, for photonic vortices, multiple beams in superposition are required to be parallel to each other~\cite{Rozas1997experimental}. The photon current is a special dissipationless flow composed of non-interacting particles. Our formalism can also be generalized to vortices in non-linear media via adding an effective photon-photon interaction in the paraxial Helmholtz equation~\cite{Wan2007dispersive,zhang2019particle}
\begin{equation*}
i\partial_{z}\Psi_{{\rm PA}}=-\frac{1}{2k_0}\nabla_{T}^{2}\Psi_{{\rm PA}}-\kappa k_0|\Psi_{{\rm PA}}|^2\Psi_{{\rm PA}},
\end{equation*}
where the last term comes from the nonlinear refractive index change for a Kerr medium with a coefficient $\kappa$. In the following, we only focus on the free-space case.

To characterize the dynamics of the local polarization of a paraxial beam, we now introduce the helicity current density,
\begin{equation}
\hat{\boldsymbol{j}}_H(\boldsymbol{r})=-\frac{i}{2k_{0}}\left\{ \hat{\psi}^{\dagger}(\boldsymbol{r})\hat{\sigma}_z\boldsymbol{\nabla}_{T}\hat{\psi}(\boldsymbol{r})-\left[\boldsymbol{\nabla}_{T}\hat{\psi}^{\dagger}(\boldsymbol{r})\right]\hat{\sigma}_z\hat{\psi}(\boldsymbol{r})\right\}. \label{eq:Hcurrent}    
\end{equation}
From the paraxial Helmholtz equation, we obtain the corresponding continuity equation, 
\begin{equation}
\frac{\partial}{\partial z}\langle\hat{n}_H\rangle+\boldsymbol{\nabla}_{T}\cdot\langle\hat{\boldsymbol{j}}_H\rangle=0.\label{eq:HContinuity}
\end{equation}
Here, the helicity density in the circular polarization representation is given by $\hat{n}_H(\boldsymbol{r}) = \hat{\psi}^{\dagger}(\boldsymbol{r})\hat{\sigma}_z\hat{\psi}(\boldsymbol{r})$. Without loss of generality, both the helicity density and the corresponding current are divided by the constant $\hbar$. As shown in the following, pure helicity current and helicity vortices can be constructed via the superposition of two OAM beams.

Similar to the superfluid in a condensate~\cite{Annett2005super}, our defined PND and helicity density currents fundamentally stem from the spatially varying phase $\phi_{\pm} (\boldsymbol{r})$ of the paraxial many-photon wave-packet function $\Psi_{\pm}(\boldsymbol{r})=\left|\Psi_{\pm}(\boldsymbol{r})\right|\exp\left\{ i\left[\phi_{\pm}(\boldsymbol{r})+k_{0}z\right]\right\}$,
\begin{align}
\langle\hat{\boldsymbol{j}}_N(\boldsymbol{r})\rangle&=\frac{1}{k_{0}}\sum_{\lambda }\left|\Psi_{\lambda}(\boldsymbol{r})\right|^2\boldsymbol{\nabla}_{T}\phi_{\lambda}(\boldsymbol{r}),\\
\langle\hat{\boldsymbol{j}}_H(\boldsymbol{r})\rangle & =\frac{1}{k_{0}}\sum_{\lambda }\lambda\left|\Psi_{\lambda}(\boldsymbol{r})\right|^2\boldsymbol{\nabla}_{T}\phi_{\lambda}(\boldsymbol{r}).    
\end{align}

We emphasize that the continuity equation is essential to define a current. A similar photon current $\boldsymbol{j}=\rm{Im}\psi\boldsymbol{\nabla}\psi$ ($\psi$ is a complex scalar function) has been given in \cite{Dennis2009Chapter}. However, its physical meaning is unclear, because no continuity equation exists corresponding to this current. In principle, we can also introduce the currents for the photonic spin and OAM density~\cite{An2012universal,Sun2005definition}. However, the corresponding currents are rank-2 tensors, which are extremely difficult to measure in experiments. For a paraxial laser beam, the photon current (\ref{eq:jN}) and helicity current (\ref{eq:Hcurrent}) are enough to characterize its vorticity properties.

\subsection{Circulations of the density currents}
To characterize the vorticity of the two currents, we now introduce two flow velocities,
\begin{equation}
\boldsymbol{v}_{N}(\boldsymbol{r}) \equiv\frac{\langle\hat{\boldsymbol{j}}_N(\boldsymbol{r})\rangle}{\langle\hat{n}(\boldsymbol{r})\rangle},\ 
\boldsymbol{v}_{H}(\boldsymbol{r}) \equiv\frac{\langle\hat{\boldsymbol{j}}_H(\boldsymbol{r})\rangle}{\langle\hat{n}(\boldsymbol{r})\rangle}.
\end{equation}
and the corresponding circulation (vorticity flux)
\begin{equation}
\kappa_{N} \equiv \ointctrclockwise_{T}\boldsymbol{v}_{N}\cdot d\boldsymbol{r},\ \kappa_{H} \equiv \ointctrclockwise_{T}\boldsymbol{v}_{H}\cdot d\boldsymbol{r}, 
\end{equation}
where the integral is taken along a closed curve in the $xy$ plane in the counter-clockwise direction.
Usually, the flow velocity $\boldsymbol{v}_{N(H)}$ depends not only on the phase gradient $\boldsymbol{\nabla}_T\phi_{\lambda}(\boldsymbol{r})$, but also on the density distribution $|\Psi_{\lambda}(\boldsymbol{r})|^2$. Thus, the two circulations are path-dependent. For a focused beam, the beam waist increases and the density decreases when leaving the focal plane. Thus, the circulations $\kappa_N$ and $\kappa_H$ for a fixed closed-loop in the $xy$ plane are not conserved during propagation.

For a special case with $|\Psi_{+}(\boldsymbol{r})|/|\Psi_{-}(\boldsymbol{r})|=C$ ($C$ is a coordinate independent constant), we can obtain conserved path-independent circulations. In this case, the flow velocities reduce to a simplified form, 
\begin{align}
\boldsymbol{v}_N & = \frac{1}{k_0}\left[\frac{C^2}{1+C^2}\boldsymbol{\nabla}_T\phi_{+}(\boldsymbol{r})+\frac{1}{1+C^2}\boldsymbol{\nabla}_T\phi_{-}(\boldsymbol{r}) \right],\\
\boldsymbol{v}_H & = \frac{1}{k_0}\left[\frac{C^2}{1+C^2}\boldsymbol{\nabla}_T\phi_{+}(\boldsymbol{r})-\frac{1}{1+C^2}\boldsymbol{\nabla}_T\phi_{-}(\boldsymbol{r}) \right].
\end{align}
To show the conservation of the circulations $\kappa_N$ and $\kappa_H$, we introduce two velocities $\boldsymbol{v}_{\pm}=\boldsymbol{\nabla}_T\phi_{\pm}/k_0$ and the corresponding circulations $\kappa_{\pm}=\ointctrclockwise_{T}\boldsymbol{v}_{\pm}\cdot d\boldsymbol{r}$. From the hydrodynamic equations of $\boldsymbol{v}_{\pm}$ (see appendix~\ref{sec:hydrodynamic}), we have 
\begin{equation}
\frac{\partial}{\partial z}\kappa_{\pm}\!=\!\ointctrclockwise_{T}\boldsymbol{\nabla}_{T}\left[\frac{1}{2k^{2}|\Psi_{\pm}|}\nabla_{T}^{2}|\Psi_{\pm}|\!-\!\frac{1}{2}v_{\pm}^{2}\right]\cdot d\boldsymbol{r}\!=\!0,\!
\end{equation}
since it is the integral of a perfect differential of single-valued functions around a closed path. This guarantees that the circulations $\kappa_N$ and $\kappa_H$ are conserved during propagating. This is consistent with Kelvin’s theorem for an ideal classical fluid~\cite{Landau1987fluid}. We note that the paraxial approximation is essential to the conservation of circulation. Accidental phase singularity points can be generated by the superposition of two non-coaxial beams~\cite{Dennis2009Chapter}. These points are not stable and they will disappear on propagation.

We note that for another case with $\boldsymbol{\nabla}_{T}\phi_{+}(\boldsymbol{r})=\boldsymbol{\nabla}_{T}\phi_{-}(\boldsymbol{r})$, we have $\boldsymbol{v}_{N} =\boldsymbol{\nabla}_T\phi_{+}/k_0$ and conserved path-independent $\kappa_N$. However, the circulation of the helicity density current does not necessarily have these good properties. We do not present a detailed discussion about this case here. In the next subsection, we only focus on uniformly polarized paraxial beams.
We will show that there exists a direct link between the photonic topological charge and the winding number of a photonic vortex.

\subsection{Quantized circulation in uniformly polarized paraxial beams}
For a uniformly polarized beam, its many-photon wave-packet function $\Psi(\boldsymbol{r})=\alpha\tilde{\xi}(\boldsymbol{r})[c_{+},c_{-}]^T$ can be expressed by the product of a constant amplitude $\alpha$, a scalar wave-packet function $\tilde{\xi}(\boldsymbol{r})$, and a constant normalized two-component vector (i.e., $|c_{+}|^2+|c_{-}|^2=1$). The polarization degrees of freedom does not contribute to the photonic vortices. Thus, the positive-frequency part of a scalar electric field function $E(\boldsymbol{r})$ has been routinely used to study the corresponding photonic vortices~\cite{Soskin1997topological,Kotlyar2020Topological}. Here, we see that for a uniformly polarized beam, $\Psi(\boldsymbol{r})$ satisfies both conditions mentioned in preceding section, i.e., $|\Psi_{+}|/|\Psi_{-}|=|c_{+}|/|c_{-}|=C$ and $\boldsymbol{\nabla}_{T}\phi_{+}(\boldsymbol{r})=\boldsymbol{\nabla}_{T}\phi_{-}(\boldsymbol{r})$. We show that the circulations of the two currents are always quantized in this case and the associated photonic vortices can be characterized by an integer winding number.

With the re-expressed function $\tilde{\xi}(\boldsymbol{r})=|\tilde{\xi}(\boldsymbol{r})|\exp\left\{ i\left[\phi(\boldsymbol{r})+k_{0}z\right]\right\}$, we obtain the flow velocity for the photon current,
\begin{equation}
\boldsymbol{v}_{N}(\boldsymbol{r})=\frac{1}{k_{0}}\boldsymbol{\nabla}_{T}\phi(\boldsymbol{r}).
\end{equation}
The helicity density flow velocity is obtained by multiplying $\boldsymbol{v}_N$ with a constant $(C^2-1)/(C^2+1)$. The circulation of the photon current is given by
\begin{equation}
\kappa_N=\ointctrclockwise_{T}\boldsymbol{v}_{N}(\boldsymbol{r})\cdot d\boldsymbol{r}=\frac{1}{k_{0}}\delta\phi(\boldsymbol{r}),
\end{equation}
where $\delta\phi(\boldsymbol{r})$ is the change in the phase around this closed curve and the non-vanishing circulation stems from the multi-valuedness of the phase factor $\phi(\boldsymbol{r})$.  The wave-packet function of the light $\Psi (\boldsymbol{r})$ is to be determined uniquely; thus we must have $\delta\phi(\boldsymbol{r})=2\pi m$, where $m$ is an integer. Therefore, the
circulation $\kappa_N$ is quantized in
units of $2\pi/k_{0}=\lambda_0$, i.e.,
$\kappa_N=m\lambda_{0}$. This integer number corresponds to the winding number of the phase $\phi (\boldsymbol{r})$ around the closed loop. Similarly, the the
circulation $\kappa_H$ of the helicity current is also quantized in
units of $\lambda_0(C^2-1)/(C^2+1)$ with the same quantum number $m$.

Since the velocity is irrotational in the transverse plane, i.e., $\boldsymbol{\nabla}_{T}\times\boldsymbol{v}_{N}=0$ (but $\boldsymbol{\nabla}\times\boldsymbol{v}_{N}\neq0$), to obtain non-vanishing circulation, the fluid must contain vortices in the $xy$-plane, i.e., the multivalued phase factor induced diverging flow velocity. Thus, this integer number $m$ is also called the topological charge of photonic vortices~\cite{Soskin1997topological}. We also see that for a uniformly polarized beam, the obtained topological charge is always an integer. This is significantly different from previous results~\cite{Berry2004optical,Kotlyar2020Topological}, in which fractional topological charges can exist. As shown in Sec.~\ref{sec:BGbeam}, crossing a zero-amplitude point, $\psi (\boldsymbol{r})$ will experience an extra $\pm \pi$ phase jump, which will lead to singularities in the flow velocity. These singularities are essential to obtain quantized integer topological charges (see Sec.~\ref{sec:superposition}). We emphasize that the diverging flow velocity can not be detected in experiments. Only the well-behaved densities and the corresponding currents are physical observables. 

\begin{figure}
\centering
\includegraphics[width=4cm]{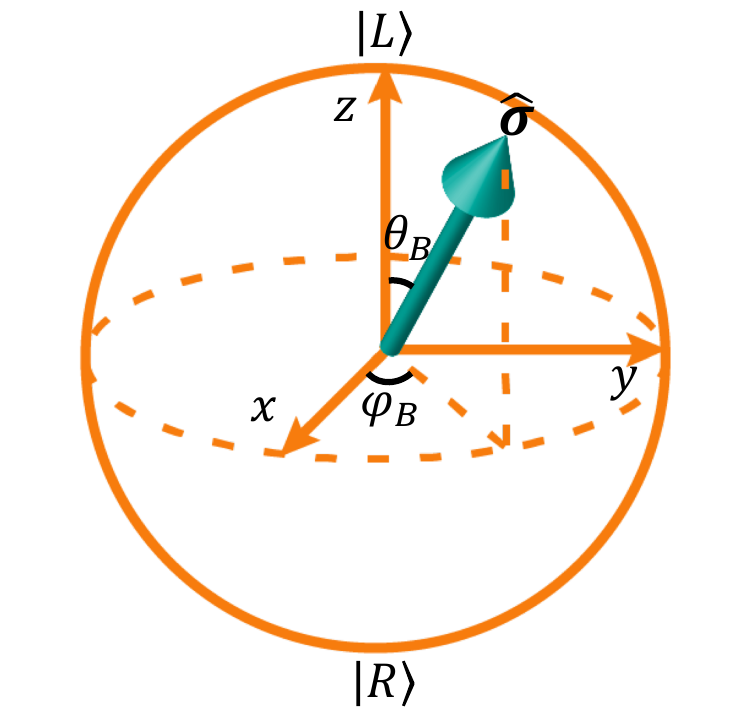}
\caption{\label{fig:Bloch} Bloch sphere for photonic polarizations. In a circular-polarization representation, the     north and south poles denote the left and right circular polarized states, respectively. The polarization of an arbitrary pure state can be characterized by the eigen state of the  operator $\hat{\boldsymbol{\sigma}}=\{\hat{\sigma}_x\sin\theta_B\cos\varphi_B,\hat{\sigma}_x\sin\theta_B\sin\varphi_B,\hat{\sigma_z}\cos\theta_B\}$.}
\end{figure}

\subsection{Pure helicity vortices}
Previously, the normalized Stokes parameters have been utilized to characterized polarization singularities in vector beams~\cite{freund2002polarization,Dennis2002polarization,Vyas2013polarization,Otte2016higher}. However, for the same beam, there exist three types of polarization vortices depending on the selected polarization basis~\cite{Freund2002stokes}. Our introduced helicity current and the associated helicity vortices are uniquely defined. More importantly, we show that there exist pure helicity-vortex beams, in which the photon current and the corresponding vortices vanish. 

We first consider a beam by superposition of two circularly polarized beams. The corresponding many-photon wave-packet function is given by $\Psi(\boldsymbol{r}) =\alpha \tilde{\xi}(\boldsymbol{r})[e^{i\phi(\boldsymbol{r})},e^{-i\phi(\boldsymbol{r})}]^T/\sqrt{2}$, where the scalar function $ \tilde{\xi}(\boldsymbol{r})$ does not contribute to the currents in the $xy$ plane, i.e., $\boldsymbol{\nabla}_T\tilde{\xi}(\boldsymbol{r})=0$. We can easily verify that $\langle\hat{\boldsymbol{j}}_N\rangle = 0$ and $\boldsymbol{v}_N = 0$. However, the helicity density current and the corresponding flow velocity do not vanish,
\begin{equation}
\langle\hat{\boldsymbol{j}}_H(\boldsymbol{r})\rangle = \frac{\langle\hat{n}(\boldsymbol{r})\rangle}{k_0}\boldsymbol{\nabla}_T\phi(\boldsymbol{r}),\ \boldsymbol{v}_H(\boldsymbol{r}) = \frac{1}{k_0} \boldsymbol{\nabla}_T\phi(\boldsymbol{r}),  
\end{equation}
where the PND is given by $\langle\hat{n}(\boldsymbol{r})\rangle = |\alpha\tilde{\xi}(\boldsymbol{r})|^2$. Thus, pure helicity vortices will be obtained from the quantized circulation of the helicity current. This is similar to the net spin current in condensed-matter physics. We also note that the helicity density vanishes at every point in this case, i.e., 
\begin{equation*}
 \langle\hat{\boldsymbol{n}}_H(\boldsymbol{r})\rangle =\frac{1}{2} |\alpha\tilde{\xi}(\boldsymbol{r})|^2 [e^{-i\phi(\boldsymbol{r})},e^{i\phi(\boldsymbol{r})}]\hat{\sigma}_z[e^{i\phi(\boldsymbol{r})},e^{-i\phi(\boldsymbol{r})}]^T=0.   
\end{equation*}

We can also construct a pure-helicity vortex with non-vanishing helicity density. With the help of the Bloch sphere, the polarization of a uniformly polarized beam can be characterized by a vector $\hat{\boldsymbol{\sigma}}=\{\hat{\sigma}_x\sin\theta_B\cos\varphi_B,\hat{\sigma}_x\sin\theta_B\sin\varphi_B,\hat{\sigma_z}\cos\theta_B\}$ as shown in Fig.~\ref{fig:Bloch}. Here, $\theta_B$ and $\varphi_B$ are the polar angle and azimuthal angle in the Bloch space, respectively, not in the real space. The two normalized eigen states of $\hat{\boldsymbol{\sigma}}$ are given by~\cite{Naidoo2016controlled}
\begin{align}
\left|\uparrow\right\rangle & = [\cos\frac{\theta_B}{2}e^{-i\varphi_B/2},\sin\frac{\theta_B}{2}e^{i\varphi_B/2}]^T,\\
\left|\downarrow\right\rangle & = [-\sin\frac{\theta_B}{2}e^{-i\varphi_B/2},\cos\frac{\theta_B}{2}e^{i\varphi_B/2}]^T.
\end{align}
We note that, all the operators are expressed in the circular-polarization representation. Thus, for $\theta_B=0\ {\rm or}\ \pi$, $\left|\uparrow\right\rangle$ and $\left|\downarrow\right\rangle$ are circularly polarized states. For $\theta_B=\pi/2$, $\left|\uparrow\right\rangle$ and $\left|\downarrow\right\rangle$ are linearly polarized states. Otherwise, $\left|\uparrow\right\rangle$ and $\left|\downarrow\right\rangle$ will be elliptically polarized.

Now we construct a beam by the superposition of two elliptically polarized light with the many-photon wave-packet function
\begin{equation}
\Psi(\boldsymbol{r}) = \alpha\tilde{\xi}(\boldsymbol{r})\left[c_{\uparrow}e^{i\phi(\boldsymbol{r})}\left|\uparrow\right\rangle + c_{\downarrow} e^{-i\phi(\boldsymbol{r})}\left|\downarrow\right\rangle\right].\label{eq:MPWF-helicityvortex}  
\end{equation}
For the case $|c_{\uparrow}|=|c_{\downarrow}|=1/\sqrt{2}$, we have zero photon current $\langle\hat{\boldsymbol{j}}_N(\boldsymbol{r})\rangle=0$ but non-vanishing helicity density current
\begin{equation}
 \langle\hat{\boldsymbol{j}}_H(\boldsymbol{r}) \rangle=  \frac{\langle\hat{n}(\boldsymbol{r})\rangle}{k_0}\cos\theta_B\boldsymbol{\nabla}_T\phi(\boldsymbol{r}). 
\end{equation}
The corresponding PND and helicity density are given by
\begin{equation}
\langle\hat{n}(\boldsymbol{r})\rangle \!=\! |\alpha\tilde{\xi}(\boldsymbol{r})|^2,\ \langle\hat{n}_H(\boldsymbol{r})\rangle\! =\! |\alpha\tilde{\xi}(\boldsymbol{r})|^2 \!\sin\theta_B\cos[ 2\phi(\boldsymbol{r})\!+\!\phi_0],  
\end{equation}
where the constant phase $\phi_0$ is determined by the relative phase between $c_{\uparrow}$ and $c_{\downarrow}$. We can verify that the circulation of the helicity current is quantized in units of $\lambda_0\cos\theta_B$.

The helicity density can be measured with a pixelated polarization filter array, which has been routinely used in polarization-sensitive imaging~\cite{Kulkarni2012polarization,Garcia2017polarization}. On the other hand, atoms can have asymmetric circularly polarized light-induced transitions. Thus, our predicted pure helicity current and pure helicity vortex can be detected via imaging the created atomic vortex via two-photon Raman processes~\cite{anderson2006quantized}.

\begin{figure*}
\centering
\includegraphics[width=13cm]{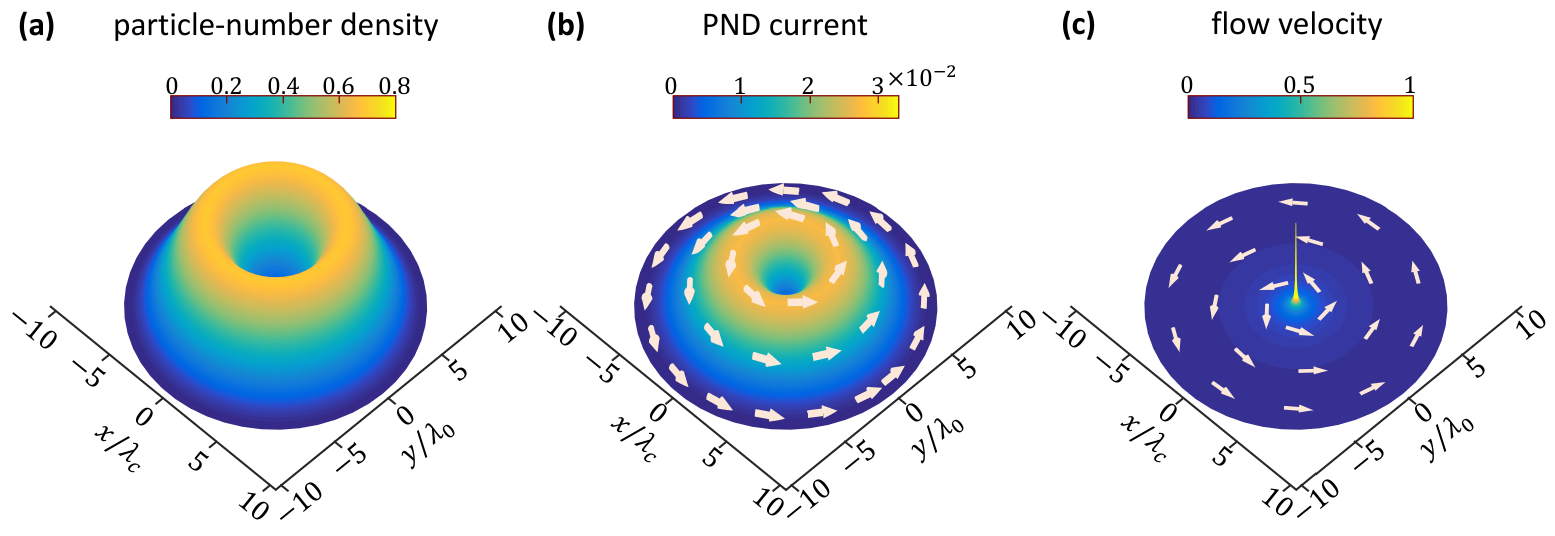}\caption{\label{fig:LGbeam} Photonic vortex in a paraxial Laguerre-Gaussian beam. (a) Rescaled particle number density $\langle\hat{n}(\rho,\varphi,0)\rangle/|\alpha \mathcal{N}_{pm}|^2$. (b) Rescaled photon current density $\langle\hat{\boldsymbol{j}}_N(\rho,\varphi,0)\rangle/|\alpha \mathcal{N}_{pm}|^2$. (c) Flow velocity $\boldsymbol{v}_N(\rho,\varphi,0)$. The diverging velocity at the center leads to the photonic vortex. Here, the parameters in the simulation are taken as $p=1$, $m=1$, and $w_0 = 10\lambda_0$. The length is in the unit of $\lambda_0$ (center wave length of the beam).}
\end{figure*}

\section{Photonic vortices in paraxial beams \label{sec:examples}}
Recently, the quantized topological charge of photonic vortices in paraxial beams or pulses has been attracting increasing interest~\cite{Zhan2009cylindrical,Rosales2018review,Shen2019optical}. We now apply the theory presented in the preceding section to investigate the properties of vortices in a paraxial Laguerre-Gaussian (LG) beam and a Bessel-Gaussian (BG) beam, which are widely used in experiments. We explicitly calculate the associated PND, current density, and specifically the flow velocity. We then analyze the flow-velocity singularities due to the ill-defined phase at the zeros of the many-photon wave-packet function and the contribution of these singularities to the winding number of a photonic vortex. In particular, we exemplify that no fractional topological charge exists in a uniformly polarized beam. Finally, we show the pure-helicity current and pure-helicity vortex in a superposition of two BG beams.  


\subsection{Laguerre-Gaussian beam}
We now apply the theory we presented in the previous section to study the properties of photonic vortices in a paraxial linearly polarized LG beam. The corresponding many-photon wave-packet function is given by $\Psi (\boldsymbol{r})= \alpha\tilde{\xi}_{{\rm LG},pm}(\rho,\varphi,z)[1,1]^T/\sqrt{2}$, where the complex constant $\alpha $ denotes the strength of the beam. The pulse profile is characterized by the wave-packet function~\cite{Enderlein2004unified}
\begin{align}
\tilde{\xi}_{{\rm LG},pm}(\rho,\varphi,z)= &\mathcal{N}_{pm}\left(\frac{1}{q(z)}\right)^{2p+|m|+1}\left|q(z)\right|^{2p}\left[\frac{\sqrt{2}\rho}{w_{0}}\right]^{|m|}\nonumber \\
&\!\!\!\!\!\!\!\!\!\!\!\!\!\!\!\!\!\!\!\!\times L_{p}^{|m|}\left(\frac{2\rho^{2}}{w_{0}^{2}\left|q(z)\right|^{2}}\right)\exp\left[ik_{0}z\!-\!\frac{\rho^{2}}{w_{0}^{2}q(z)}\!+\!im\varphi\right],    
\end{align}
where $L_p^m(x)$ is the associated Laguerre polynomials with a non-negative integer $p$
and an integer $m$, $q(z)=1+iz/z_{R}$, $z_{R}=kw_{0}^{2}/2=\pi w_{0}^{2}/\lambda_0$ is the Rayleigh length, and $w_{0}$ is the Gaussian beam waist radius. The constant factor $\mathcal{N}_{pm}$ is determined by the normalization condition~$\int_V d^{3}r|\tilde{\xi}_{{\rm LG},pm}(\rho,\varphi,z)|^2=1$ and $V$ is the effective volume of the laser beam. 

In the previous section, we have shown that both the current and flow velocity are in the transverse plane perpendicular to the propagating direction. Without loss of generality, we only consider the focal plane ($z=0$) in the following. The PND in the $z=0$ plane is given by
\begin{equation}
\langle\hat{n}(\rho,\varphi,0)\rangle = \left|\alpha\mathcal{N}_{pm}\left[\frac{\sqrt{2}\rho}{w_{0}}\right]^{|m|}L_{p}^{|m|}\left(\frac{2\rho^{2}}{w_{0}^{2}}\right)e^{-\rho^{2}/w_{0}^{2}}\right|^{2}.
\end{equation}
For $m\neq 0$, the mean PND vanishes on the $z$-axis with scaling~$\sim \rho^{2|m|}$.   
We show the rescaled PND $\langle\hat{n}(\rho,\varphi,0)\rangle/|\alpha \mathcal{N}_{pm}|^2$ in Fig.~\ref{fig:LGbeam}~(a). We see that there is a hole at the center.

The corresponding in-plane photon current density is given by
\begin{equation}
\langle\hat{\boldsymbol{j}}_N(\rho,\varphi,0)\rangle = \langle\hat{n}(\rho,\varphi,0)\rangle \times\frac{m\lambda_{0}}{2\pi\rho}\boldsymbol{e}_{\varphi},\label{eq:j_LG}
\end{equation}
which only has a tangent component. We see that this current density is proportional to the PND and the integer $m$. It is also modulated by the function $1/\rho$.  We note that the photon current is well-defined on the whole $xy$-plane and it vanishes on the $z$-axis due to the vanishing PND ($\lim_{\rho\rightarrow 0}\langle\hat{n}\rangle \propto \rho^{2|m|}$). The rescaled photon current $\langle\hat{\boldsymbol{j}}_T(\rho,\varphi,0)\rangle/|\alpha \mathcal{N}_{pm}|^2$ is shown in Fig.~\ref{fig:LGbeam}~(b). For $m>0$, it flows in a counter-clockwise direction. For a linearly polarized beam, both the helicity density and helicity current density are zero (not shown).

The flow velocity of the photon current is given by,
\begin{equation}
\boldsymbol{v}_{N}=\frac{m\lambda_{0}}{2\pi\rho}\boldsymbol{e}_{\varphi}+\sum_{j}\frac{\lambda_{0}}{2}\delta(\rho-\rho_{{\rm ZA},j})\boldsymbol{e}_{\rho}.
\end{equation}
The Laguerre polynomial $L_p^{|m|}(x)$ has $p$ zeros~\cite{GATTESCHI2002Asymptotics}, which lead to $p$ zero-PND circles with radius $\rho_{ZA,j}$ in the $xy$ plane (not shown in Fig.~\ref{fig:LGbeam}), i.e., $L_p^{|m|}(2\rho^2_{ZA,j}/w_0^2)=0$. We can see that there are two types of singularities in the flow velocity. The first type is a singularity point lying at the center of a vortex as shown in Fig.~\ref{fig:LGbeam} (c) and the second type (not shown) comes from the $\pi$-phase jump when the wave-packet function $\Psi$ cross a zero-value curve. A closed velocity singularity curve does not contribute to the winding number of a vortex, because any integral loop in the $xy$ plane will always cross the singularity curve an even number of times. The accumulated phases cancel each other. Thus, the topological winding number of the vortex in this LG beam is $m$. 

We note that the diverging flow velocity is unphysical and cannot be observed in experiments. These singularities are fully due to the ill-defined phase of the complex wave-packet function $\Psi(\boldsymbol{r})$ at the zero-value points. On the other hand, no singularity exists in the truly observable PND and photon current density.

\begin{figure*}
\centering
\includegraphics[width=13cm]{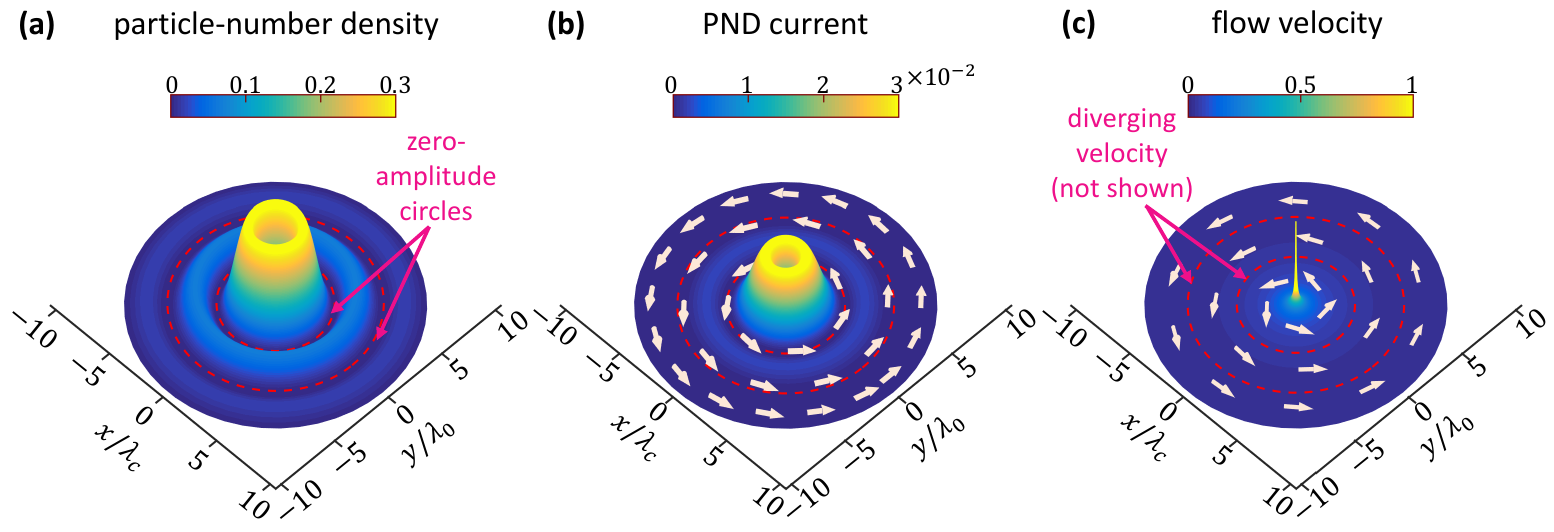}\caption{\label{fig:BGbeam} Photonic vortex in a Bessel-Gaussian beam. (a) Rescaled particle number density $\langle\hat{n}(\rho,\varphi,0)\rangle/|\alpha \mathcal{N}_{m}|^2$. Red dashed curves denote the zero-amplitude circles at $\rho_{{\rm ZA},j}$ resulting from $J_{m}(\beta\rho_{{\rm ZA},j})=0$. (b) Rescaled photon current density $\langle\hat{\boldsymbol{j}}_N(\rho,\varphi,0)\rangle/|\alpha \mathcal{N}_{m}|^2$. (c) Flow velocity $\boldsymbol{v}_N(\rho,\varphi,0)$. The diverging radial component of the flow velocity proportional to $\boldsymbol{e}_{\rho}\delta(\rho-\rho_{{\rm ZA},j})$ on the zero-amplitude circles is not shown. Here, the parameters in the simulation are taken as $m=1$, $w_0 = 10\lambda_0$, and $\theta_p = 0.05\pi$.}
\end{figure*}

\subsection{Bessel-Gaussian beam\label{sec:BGbeam}}
In this section we study the vortex in a paraxial linearly polarized BG beam. The corresponding many-photon wave-packet function is given by $\Psi (\boldsymbol{r})= \alpha\tilde{\xi}_{{\rm BG},m}(\rho,\varphi,z)[1,1]^T/\sqrt{2}$, where the complex constant $\alpha $ denotes the strength of the beam. The pulse profile is characterized by the wave packet function~\cite{gori1987bessel}
\begin{align}
\tilde{\xi}_{{\rm BG},m}(\rho,\varphi,z)  =&\mathcal{N}_{m}\frac{1}{q(z)}J_{m}\left(\frac{\beta\rho}{q(z)}\right)\nonumber\\
&\!\!\!\!\!\!\!\!\!\!\times \exp\left\{ ik_{0}z\left[1\!-\!\frac{\sin^{2}\theta}{2q(z)}\right]\!-\!\frac{\rho^{2}}{w_{0}^{2}q(z)}\!+\!im\varphi\right\}\!, \label{eq:WPFBG}
\end{align}
where $J_{m}(x)$ is the Bessel function of the first kind with an integer $m$, $\beta=k_0\sin\theta$ is the scaling factor of the Bessel function, and $\theta$ is the half-angle of a conical wave that forms the Bessel beam. The constant factor $\mathcal{N}_{m}$ is determined by the normalization condition~$\int_V d^{3}r|\tilde{\xi}_{{\rm BG},m}(\rho,\varphi,z)|^2=1$. 

The PND of a BG beam in $z=0$ plane is given by
\begin{equation}
\langle\hat{n}(\rho,\varphi,0)\rangle = \left|\alpha\mathcal{N}_{m}J_{m}\left(\beta\rho\right)e^{-\rho^{2}/w_{0}^{2}}\right|^{2}.\label{eq:PNDBG}
\end{equation}
We show the rescaled PND $\langle\hat{n}(\rho,\varphi,0)\rangle/|\alpha \mathcal{N}_{p}|^2$ in Fig.~\ref{fig:BGbeam}~(a). For $m\neq 0$, there is a hole with vanishing PND on the $z$-axis. On the other hand, a Bessel function $J_m(x)$ has an infinite number of real zeros, and thus a Bessel beam will have an infinite number of zero-amplitude circles in the transverse plane (see the red dashed curves). Here $\rho_{{\rm ZA},j}$ denotes the radius of the $j$th circle.

The corresponding photon current density is given by
\begin{equation}
\langle\hat{\boldsymbol{j}}_{N}(\rho,\varphi,0)\rangle = \langle\hat{n}(\rho,\varphi,0)\rangle \times\frac{m\lambda_{0}}{2\pi\rho}\boldsymbol{e}_{\varphi}.\label{eq:j_BG}
\end{equation}
The rescaled photon current $\langle\hat{\boldsymbol{j}}_T(\rho,\varphi,0)\rangle/|\alpha \mathcal{N}_{p}|^2$ is shown in Fig.~\ref{fig:BGbeam}~(b). We see that this current is modulated by the PND and it flows in a counter-clockwise direction for $m>0$. Similar to the LG beam, the flow velocity for a BG light beam also has a radial component
\begin{equation}
\boldsymbol{v}_{N}=\frac{m\lambda_{0}}{2\pi\rho}\boldsymbol{e}_{\varphi}+\sum_{j}\frac{\lambda_{0}}{2}\delta(\rho-\rho_{{\rm ZA},j})\boldsymbol{e}_{\rho}.\label{eq:v_BG}
\end{equation}
where the diverging radial velocity comes from the $\pi$-phase jump at zeros of the Bessel function and $\rho_{ZA,j}$ denotes the radius of the $j$th zero-amplitude circle. As explained in preceding section, only the first term in $\boldsymbol{v}_N$ will contribute to the vortex. Thus, the topological winding number of  the vortex in this BG beam is $m$.

\begin{figure*}
\centering
\includegraphics[width=13cm]{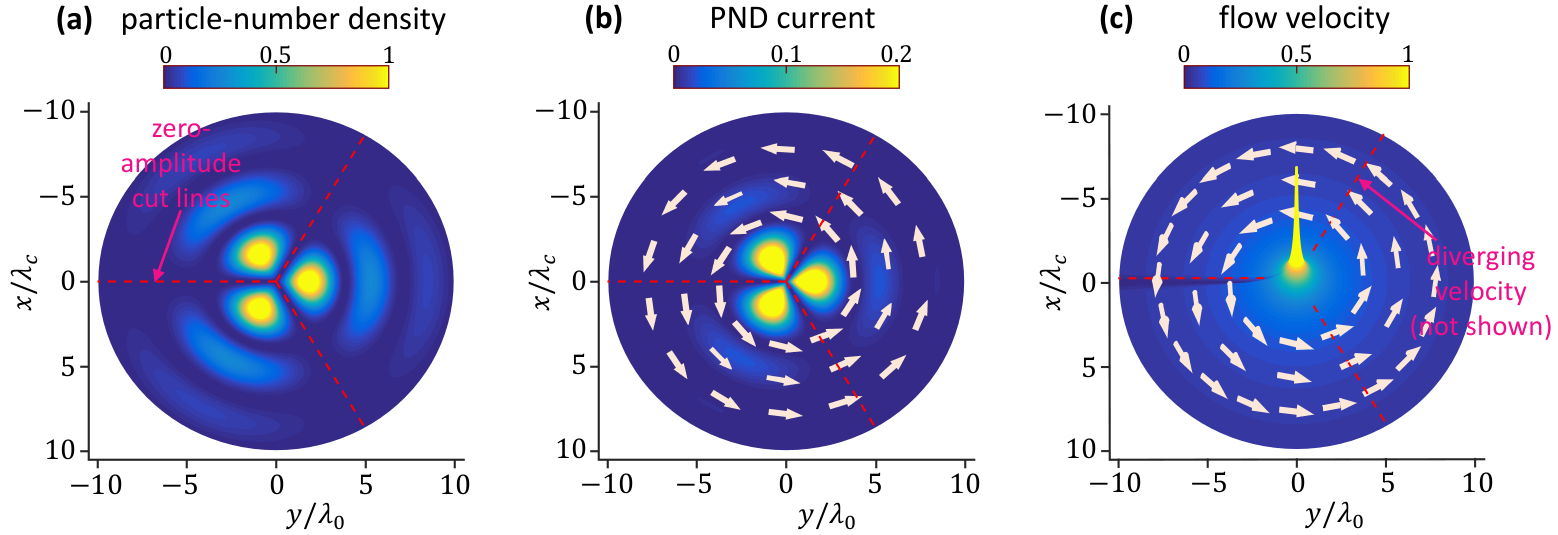}\caption{\label{fig:mixedbeam} Photonic vortex in a superposition of two Bessel-Gaussian beams. (a) Rescaled particle number density. Red-dashed lines denote the zero-amplitude cut lines at $\phi = \pi(2j-1)\pi/|m-n|$ ($j=1,2,...,|m-n|$) [see Eq.~(\ref{eq:v_mixed})]. (b) Rescaled photon current density. (c) Flow velocity $\boldsymbol{v}_N(\rho,\varphi,0)$. Here, the parameters in the simulation are taken as $p=1$, $m=1$, $n=4$, $w_0 = 10\lambda_0$, and $\theta_p = 0.05\pi$. }
\end{figure*}

\begin{figure*}
\centering
\includegraphics[width=13cm]{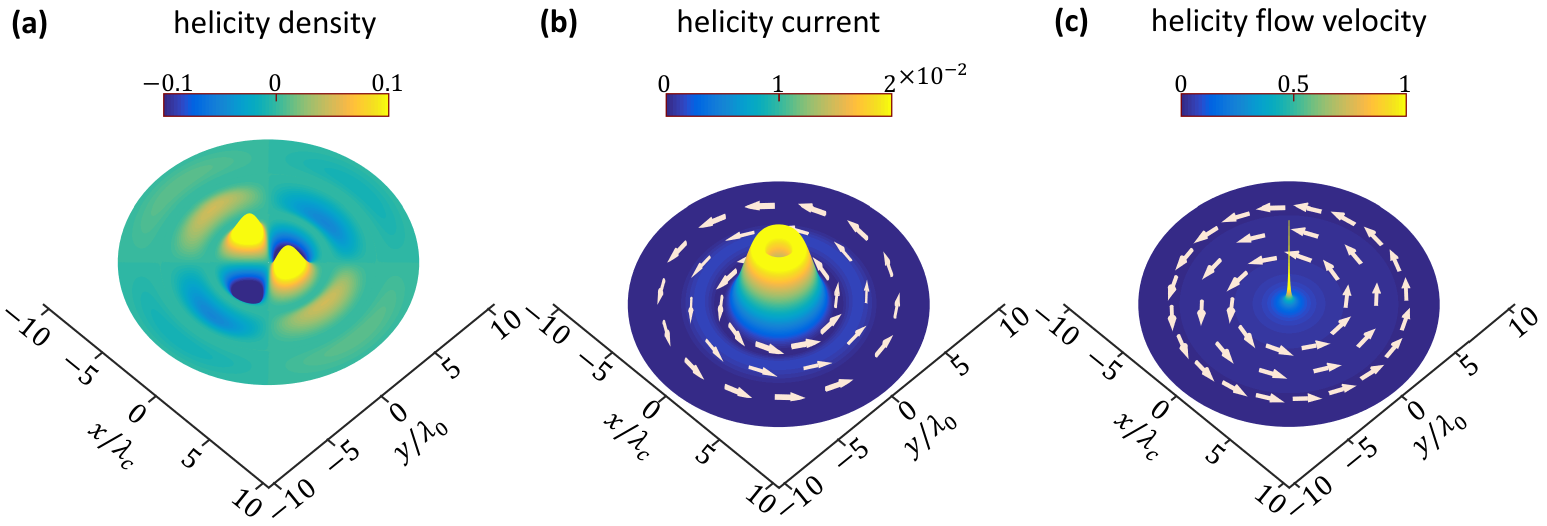}\caption{\label{fig:helicitybeam} Pure helicity vortex in a superposition of two Bessel-Gaussian beams. (a) Re-scaled helicity density $\langle\hat{n}_H(\rho,\varphi,0)\rangle/|\alpha \mathcal{N}_{m}|^2$. (b) Re-scaled helicity current density $\langle\hat{\boldsymbol{j}}_H(\rho,\varphi,0)\rangle/|\alpha \mathcal{N}_{m}|^2$. (c) Helicity flow velocity $\boldsymbol{v}_H(\rho,\varphi,0)$. The diverging terms due to crossing the zero-amplitude curves are not shown. Here, the parameters in the simulation are taken as $m=1$, $w_0 = 10\lambda_0$, and $\theta_p = 0.05\pi$.}
\end{figure*}

\subsection{Fractional topological charge controversy\label{sec:superposition}}
In previous studies, photonic vortices with fractional topological charge were found in both theory~\cite{Berry2004optical,Kotlyar2020Topological} and experiments~\cite{Leach2004observation,Wen2019vortex}. In Sec.~\ref{sec:vortex}, we showed that the circulation of a paraxial beam is not quantized and even not conserved in most cases. However, for a uniformly polarized beam, the circulation is conserved during propagation and the corresponding winding number must be an integer. Now we show that half-integer topological charges obtained in a uniformly polarized beam are due to the improper evaluation of the circulation.

To resolve the controversy, we look at a simple example of the superposition of two co-axial beams with the same polarization, \textcolor{red}{using an idealized common radial envelope to isolate the phase discontinuity produced by the cut lines}, but with different helical phase factors. \textcolor{red}{For physical paraxial modes with $m\neq n$, the radial beam profiles are generally different; therefore, the common-envelope form below should be understood as an idealized model for the cut-line topology rather than as a globally exact pair of paraxial modes.} The many-photon wave-packet function \textcolor{red}{in this idealized model} can be expressed in the form $\Psi(\boldsymbol{r})=[c_{+},c_{-}]^T\alpha\tilde{\xi}(\boldsymbol{r})$ with normalized constants $|c_{+}|^2+|c_{-}|^2=1$. In the $z=0$ plane, the wave-packet function is given by,
\begin{align}
\tilde{\xi}(\rho,\varphi,0) & =\eta(\rho)\left(e^{im\varphi}+e^{in\varphi}\right)\\
   & = 2\eta(\rho)\cos\frac{m-n}{2}\varphi e^{i(m+n)\varphi/2},
\end{align}
where $\eta(\rho)$ is \textcolor{red}{a common reference envelope in this idealized model}. In Fig.~\ref{fig:mixedbeam}~(a), we plot the rescaled PND for \textcolor{red}{this equal-envelope example} with $m=1$ and $n=4$. Here we see that there are $|m-n|=3$ zero-amplitude cut lines from the center to infinity (the red-dashed lines). The photon current density is given by [see Fig.~\ref{fig:mixedbeam}(b)]
\begin{equation}
\langle\hat{\boldsymbol{j}}_{N}(\rho,\varphi,0)\rangle = \langle\hat{n}(\rho,\varphi,0)\rangle \times\frac{(m+n)}{2}\times\frac{\lambda}{2\pi\rho} \boldsymbol{e}_{\varphi}.   
\end{equation}
Different from Eqs.~(\ref{eq:j_LG}), a half-integer enters the photon current here. 

We note that the winding number of the photonic vortex is still an integer in this case. The flow velocity now contains the contributions from the cut lines,
\begin{align}
\boldsymbol{v}_{N} = & \frac{m+n}{2}\frac{\lambda_{0}}{2\pi\rho}\boldsymbol{e}_{\varphi}-\sum_{j=1}^{|m-n|}(-1)^j\frac{\lambda_{0}}{2}\delta\left(\varphi\!-\!\frac{2j\!-\!1}{|m\!-\!n|}\pi\right)\boldsymbol{e}_{\varphi}. \label{eq:v_mixed}  
\end{align}
As explained in previous sections, the first term determines the position of the vortex core and the second term comes from the $\pi$-phase jump crossing the cut lines.

The half-integer topological charge has been obtained~\cite{Kotlyar2020Topological} in a linear combination of optical vortices with the definition in Eq.~(\ref{eq:TC2}). However, we note that this fractional topological charge is because the definition in Eq.~(\ref{eq:TC2}) has not taken the contribution from the \textcolor{red}{second} term in Eq.~(\ref{eq:v_mixed}), \textcolor{red}{namely, the $\pi$-phase jump when the integration contour crosses a zero-amplitude cut line}. Different from the closed zero-amplitude curve, the open zero-amplitude lines can contribute to the winding number of the vortex. If $m+n$ is an even integer, then $|m-n|$ must also be an even integer and the second term vanishes. The topological winding number of the vortex is $(m+n)/2$. If $m+n$ is an odd integer, then $|m-n|$ must also be an odd integer and the second term will contribute an extra $\pi$-phase. The topological winding number of the vortex is $(m+n+1)/2$. Similarly, a cut line connecting two vortices~\cite{kovalev2021propagation} will also contribute to the circulation of each vortex. \textcolor{red}{Thus, once the phase jumps across all cut lines are included, the winding number evaluated on a closed contour is integer.}

\textcolor{red}{For a physical paraxial superposition with $m\neq n$, the two components generally have different radial envelopes, which may be denoted by $\eta_m(\rho)$ and $\eta_n(\rho)$. The exact zero-amplitude set is then determined by $\eta_m(\rho)e^{im\varphi}+\eta_n(\rho)e^{in\varphi}=0$ and need not consist of the straight radial cut lines of the equal-envelope model above. Nevertheless, the central point remains unchanged: whenever the destructive-interference cut considered in the circulation is crossed, its associated $\pi$-phase jump must be included. The equal-envelope construction above therefore serves only as a transparent illustration of the contribution that was omitted in the half-integer calculation.}

\subsection{Helicity vortex}
In this section, we investigate the pure-helicity current and pure-helicity vortex in superposition of two BG beams. The many-photon wave-packet function in the focal plane is given by
\begin{equation}
\!\!\!\!\Psi(\rho,\varphi,0) \!=\!\alpha\mathcal{N}_{m}J_{m}\!\left(\beta\rho\right)e^{-\rho^2/w_0}\!\left(c_{\uparrow}e^{im\varphi}\left|\uparrow\right\rangle\!+\!c_{\downarrow}e^{-im\varphi}\left|\downarrow\right\rangle\right)\!,\label{eq:Psi_hilicity}
\end{equation}
with $|c_{\uparrow}|^2=|c_{\downarrow}|^2=1/2$. The PND in this plane is the same as a BG beam as given in Eq.~(\ref{eq:PNDBG}) [see Fig.~\ref{fig:BGbeam} (a)]. Different from the PND, the helicity in this plane has also been modulated by the azimuth angle
\begin{equation}
\langle\hat{n}_H(\rho,\varphi,0)\rangle=\langle\hat{n}(\rho,\varphi,0)\rangle\sin\theta_B\cos(2m\varphi+\phi_0).\label{eq:n_Helicity}   
\end{equation} 
We plot the helicity density in the focal plane in Fig.~\ref{fig:helicitybeam} (a) with $\theta_B=\pi/4$ and $\phi_0=0$. The zero-amplitude circles due to the zeros of the Bessel function also exist (not shown).

The photon current vanishes in a pure-helicity vortex beam described by (\ref{eq:Psi_hilicity}). However, the helicity current is given by
\begin{equation}
\langle\hat{\boldsymbol{j}}_{H}(\rho,\varphi,0)\rangle = \langle\hat{n}(\rho,\varphi,0)\rangle\cos\theta_B \times\frac{m\lambda_{0}}{2\pi\rho}\boldsymbol{e}_{\varphi}.\label{eq:j_helicity}
\end{equation}
The helicity flow velocity $\boldsymbol{v}_H$ is obtained by multiplying the velocity $\boldsymbol{v}_N$ in Eq.~(\ref{eq:v_BG}) by a factor $\cos\theta_B$ [see 
Fig.~\ref{fig:helicitybeam} (c)]. Similar to the vortex of the photon current in a BG beam, the zero-amplitude circles do not contribute to the winding number of the vortex. The topological charge of this pure helicity vortex is also $m$.

From Eqs. (\ref{eq:n_Helicity}) and (\ref{eq:j_helicity}) we see that the superposition of two circularly polarized beams with $\theta_B=0\ {\rm or}\ \pi$ gives the largest helicity current but vanishing helicity density and the superposition of two linearly polarized beams with $\theta_B=\pi/2$ gives the largest helicity density but vanishing helicity current. The superposition of two elliptically polarized beams leads to non-vanishing helicity density and current.


\section{Quantum coherence of twisted light \label{sec:coherence}}
In his seminal work~\cite{Glauber1963coherence}, Glauber introduced a succession of quantum correlation functions and quantum coherence functions for optical light, laying the foundation for modern quantum optics. We now re-express the quantum coherence functions of light with our introduced field operator $\hat{\psi}(\boldsymbol{r})$. We also propose a quantum correlation function for the photonic helicity density, which can be directly measured in experiments. When applying to twisted photon pairs, we find an interesting phenomenon that two photons with a symmetrical spin state tend to be bunched, and two photons with an anti-symmetric polarization state behave in a more anti-bunched manner. The quantum statistics of a twisted photon pair are strongly dependent on its spin (polarization) state.

The quantum statistics of light are essentially characterized by its high-order correlations, which can be measured via high-order quantum interference experiments~\cite{Zanthier2021highorder}. Without loss of generality,
we only take the most commonly used second-order correlation in coincidence
measurements as an example
\begin{align}
G^{(2)}(\boldsymbol{r},\boldsymbol{r}')
 & =\sum_{\lambda\lambda'}\langle\hat{\psi}_{\lambda}^{\dagger}(\boldsymbol{r})\hat{\psi}_{\lambda'}^{\dagger}(\boldsymbol{r}')\hat{\psi}_{\lambda'}(\boldsymbol{r}')\hat{\psi}_{\lambda}(\boldsymbol{r})\rangle.
\end{align}
The corresponding second-order coherence function in real space is given by
\begin{equation}
g^{(2)}(\boldsymbol{r},\boldsymbol{r}')=\frac{G^{(2)}(\boldsymbol{r},\boldsymbol{r}')}{\langle\hat{n}(\boldsymbol{r})\rangle\langle\hat{ n}(\boldsymbol{r}')\rangle}.
\end{equation}
We emphasize that a paraxial approximation is also required here; otherwise the physical meaning of $\hat{n}(\boldsymbol{r})$ is unclear. In Glauber's original work, the electric-field operator was used to define the quantum coherence function. It is actually based on the energy-density correlation function instead of photon-number-density correlation addressed here. On the other hand, we only consider equal-time correlation and coherence functions in the following. The time dependence can be easily recovered with the field operator $\hat{\psi}_{\lambda}(\boldsymbol{r},t)$ in the Heisenberg picture or with time-varying quantum states in the Schr{\"o}dinger picture.

We now propose a new correlation function corresponding to the photonic helicity density,
\begin{equation}
G_{H}^{(2)}(\boldsymbol{r},\boldsymbol{r}')=\sum_{\lambda\lambda'}\lambda\lambda'\langle\hat{\psi}_{\lambda}^{\dagger}(\boldsymbol{r})\hat{\psi}_{\lambda'}^{\dagger}(\boldsymbol{r}')\hat{\psi}_{\lambda'}(\boldsymbol{r}')\hat{\psi}_{\lambda}(\boldsymbol{r})\rangle,
\end{equation}
which can be exploited to reveal the correlation information about the polarization degrees of freedom of light. Here, we do not consider the correlation functions for the photonic
spin and OAM densities. A $3\times3$ correlation matrix is needed to fully characterize the corresponding quantum correlations~\cite{yang2021quantum}. On the other hand, it is currently extremely challenging to measure the spin or OAM density correlation at the few-photon level.

We first check the quantum coherence of the pure-helicity vortex laser beam with the many-photon wave-packet function (\ref{eq:MPWF-helicityvortex}). We can verify that a coherent-state light (even with highly sophisticated structure) has a constant quantum coherence function, i.e., $g^{(2)}(\boldsymbol{r},\boldsymbol{r}')=1$. Its helicity correlation function is simply the product of the helicity density at each point, i.e., $G_{H}^{(2)}(\boldsymbol{r},\boldsymbol{r}') = \langle\hat{n}_H(\boldsymbol{r})\rangle\langle\hat{n}_H(\boldsymbol{r}')\rangle$. Nontrivial quantum coherence only exists in truly quantum light, such as Fock-state photon pulses and squeezed light pulses. In the following, we focus on entangled photon pairs~\cite{Torres2003quantum,Saleh2000photonpair,Liu2019photonpair}, which have been extensively explored for testing of Bell’s inequality~\cite{Shalm2015strong}, quantum
key distribution~\cite{Antonio2007keydistribution}, ghost imaging~\cite{cai2005ghost,Chan2009twocolor}, etc.

The quantum state of a photon pair can be written as~\cite{loudon2000quantum},
\begin{equation}
\left|P_{\xi}\right\rangle \!=\!\frac{1}{\sqrt{2}}\!\sum_{\lambda_{1}\lambda_{2}}\!\int\!\! d^{3}k_{1}\!\!\!\int \!\!d^{3}k_{2}\xi_{\lambda_{1},\lambda_{2}}(\boldsymbol{k}_{1},\boldsymbol{k}_{2})\hat{a}_{\boldsymbol{k}_{2},\lambda_{2}}^{\dagger}\hat{a}_{\boldsymbol{k}_{1},\lambda_{1}}^{\dagger}\!\left|0\right\rangle.\!
\end{equation}
The normalization requirement of the state $|P_{\xi}\rangle$ requires $\sum_{\lambda_{1}\lambda_{2}}\int d^{3}k_{1}\int d^{3}k_{2}\left|\xi_{\lambda_{1},\lambda_{2}}(\boldsymbol{k}_{1},\boldsymbol{k}_{2})\right|^{2}=1$. On the other hand, photons are bosons; thus the two-photon SAF should also be symmetric, i.e., $\xi_{\lambda_{1},\lambda_{2}}(\boldsymbol{k}_{1},\boldsymbol{k}_{2})=\xi_{\lambda_{2},\lambda_{1}}(\boldsymbol{k}_{2},\boldsymbol{k}_{1})$. Via the Fourier transformation, we can also reexpress $|P_{\xi}\rangle$ with the field operators,\begin{equation}
\left|P_{\xi}\right\rangle \!=\!\!\frac{1}{\sqrt{2}}\!\sum_{\lambda_{1}\lambda_{2}}\!\int\!\!\! d^{3}r_{1}\!\!\!\!\int\!\!\! d^{3}r_{2}\tilde{\xi}_{\lambda_{1},\lambda_{2}}(\boldsymbol{r}_{1},\boldsymbol{r}_{2})\hat{\psi}_{\lambda_{2}}^{\dagger}\!(\boldsymbol{r}_{2})\hat{\psi}_{\lambda_{1}}^{\dagger}\!(\boldsymbol{r}_{1})\!\left|0\right\rangle,\!
\end{equation}
where the two-photon wave packet in real space
\begin{equation}
\tilde{\xi}_{\lambda_{1},\lambda_{2}}(\boldsymbol{r}_{1},\boldsymbol{r}_{2})\!=\!\frac{1}{(2\pi)^{3}}\!\!\int\!\!\!d^3k_1\!\!\!\int\!\!\!d^3k_2\xi_{\lambda_{1},\lambda_{2}}(\boldsymbol{k}_{1},\boldsymbol{k}_{2})e^{i(\boldsymbol{k}_{1}\cdot\boldsymbol{r}_{1}+\boldsymbol{k}_{2}\cdot\boldsymbol{r}_{2})},
\end{equation}
is also normalized $\sum_{\lambda_{1}\lambda_{2}}\int d^{3}r_{1}\int d^{3}r_{2}\left|\tilde{\xi}_{\lambda_{1},\lambda_{2}}(\boldsymbol{r}_{1},\boldsymbol{r}_{2})\right|^{2}=1$ and symmetric.

We now give the quantum correlation and coherence functions for an arbitrary two-photon pulse. The PND and helicity density correlations are given by
\begin{equation}
G^{(2)}(\boldsymbol{r},\boldsymbol{r}') =  2\sum_{\lambda\lambda'}\left|\tilde{\xi}_{\lambda\lambda'}(\boldsymbol{r},\boldsymbol{r}')\right|^{2}
\end{equation}
and
\begin{equation}
G^{(2)}_H(\boldsymbol{r},\boldsymbol{r}') =  2\sum_{\lambda\lambda'}\lambda\lambda'\left|\tilde{\xi}_{\lambda\lambda'}(\boldsymbol{r},\boldsymbol{r}')\right|^{2},  
\end{equation}
respectively. Here the factor $2$ comes from the fact that there are two photons in the pulse. The quantum coherence function can be easily obtained with the help of the PND at each point $\left\langle \hat{n}(\boldsymbol{r})\right\rangle = 2\sum_{\lambda\lambda'}\int dr'\left|\tilde{\xi}_{\lambda\lambda'}(\boldsymbol{r},\boldsymbol{r}')\right|^{2}$. The helicity density is  given by $\left\langle \hat{n}_H(\boldsymbol{r})\right\rangle = 2\sum_{\lambda\lambda'}\lambda\int dr'\left|\tilde{\xi}_{\lambda\lambda'}(\boldsymbol{r},\boldsymbol{r}')\right|^{2}$.

\begin{figure}
\centering
\includegraphics[width=8.5cm]{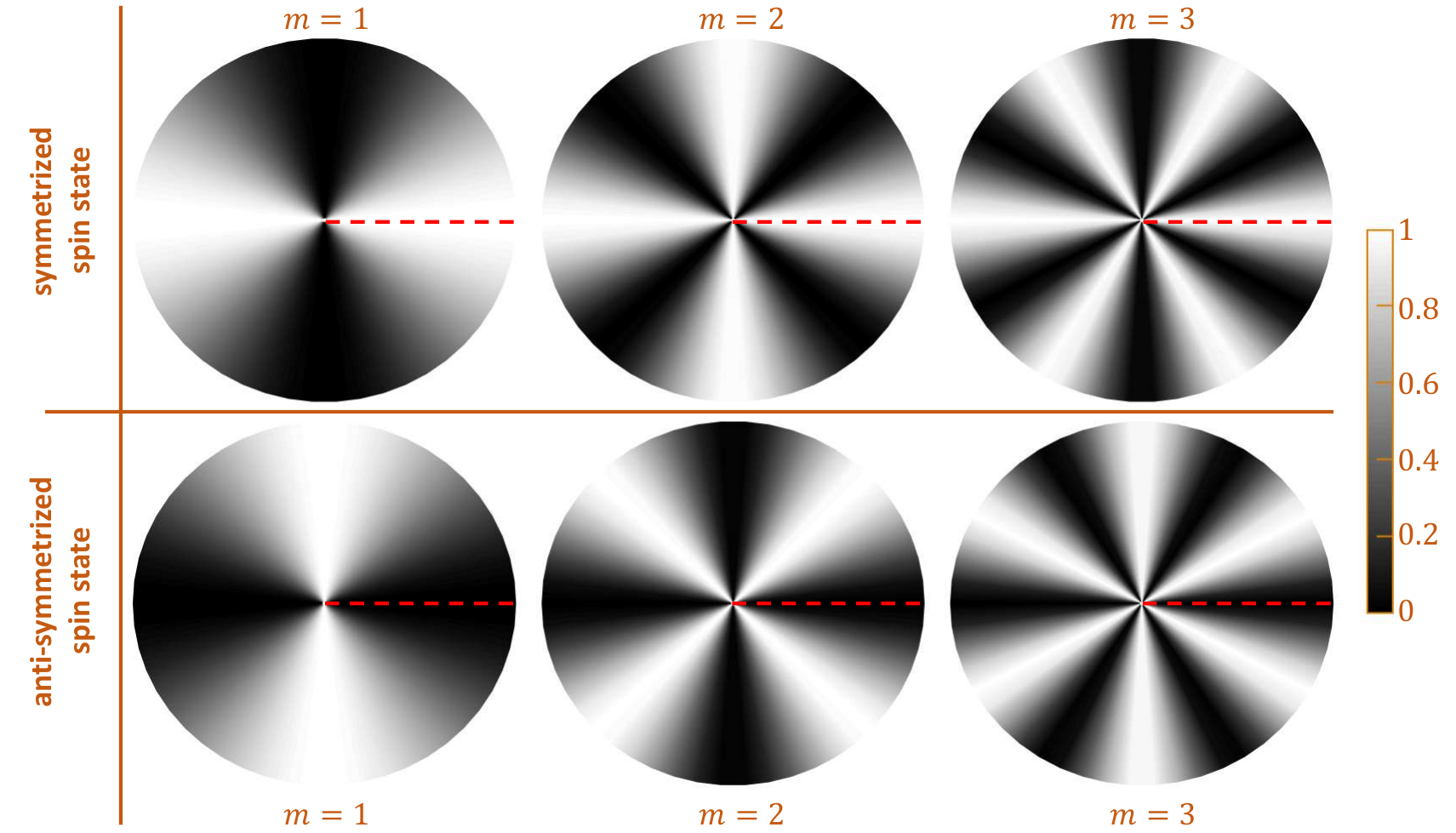}
\caption{\label{fig:coherence} Quantum coherence of twisted photon pairs. Each disk shows the $g^{(2)}$ as a function of azimuthal angle difference of the two photons~$\varphi-\varphi'\in [0,2\pi)$ and the red dashed line denotes the case $\varphi-\varphi'=0$. The top and bottom rows denote the $g^{(2)}$ function of twisted photon pairs with symmetric and antisymmetric spin states, respectively.
} 
\end{figure}

\subsection*{Spin-state dependent quantum statistics}
It is well known that statistics of identical particles are strongly dependent on their spin property. For fermions with the same spin state, the two-particle correlation vanishes when $|\boldsymbol{r}-\boldsymbol{r}'|\rightarrow 0$, i.e., $g^{(2)}(\boldsymbol{r},\boldsymbol{r})=0$. For photons in a thermal light, we have $g^{(2)}(\boldsymbol{r},\boldsymbol{r})=2$. Here, we show an exotic phenomenon for entangled twisted photon pairs. The spatial distribution of the quantum coherence function $g^2(\boldsymbol{r},\boldsymbol{r}')$ is controlled by the photonic spin state. The $g^{(2)}$-function is reversed when the photonic spin state changes from symmetric to anti-symmetric.  

We now apply our theory to investigate the quantum correlations of elliptically polarized photon pairs with a symmetric
spin state~$\sim \left|\uparrow\right\rangle\otimes\left|\downarrow\right\rangle+\left|\downarrow\right\rangle\otimes\left|\uparrow\right\rangle$.
For simplicity, we consider a photon pair with the two-photon SAF,
\begin{equation}
\xi_{\lambda_{1},\lambda_{2}}(\boldsymbol{k}_{1},\boldsymbol{k}_{2})=\mathcal{N}\Theta_{\lambda_{1}\lambda_{2}}\eta(\boldsymbol{k}_{1})\eta(\boldsymbol{k}_{2})\left[e^{im(\varphi_{k_{1}}-\varphi_{k_{2}})}+{\rm c.c}\right],
\end{equation}
Here the function $\eta(\boldsymbol{k})$, which determines the pulse shape and transverse-plane distribution, is independent of $\varphi_k$~\cite{yang2021quantum}. The polarization of the photon pair is described by a symmetric
$2\times2$ matrix 
\begin{equation}
\Theta=\left[\begin{array}{cc}
\Theta_{++} & \Theta_{+-}\\
\Theta_{-+} & \Theta_{--}
\end{array}\right]=\left[\begin{array}{cc}
-\sin\theta_{B}e^{-i\varphi_{B}} & \cos\theta_{B}\\
\cos\theta_{B} & \sin\theta_{B}e^{i\varphi_{B}}
\end{array}\right].
\end{equation}
The normalization of state $|P_{\xi}\rangle $ requires
$\int d^{3}k\left|\eta(\boldsymbol{k})\right|^{2}=1$ and $\mathcal{N}=\left[4(1+\delta_{m,0})\right]^{-1/2}$. This type of photon pair can be generated by a two-atom light source~\cite{Fearn1989twophoton}. Deterministic photon pairs can be generated via bundle-emission processes~\cite{chang2016determinstic,Bin2020Nphoton}.

The two-photon wave-packet in real space can be expressed in the form of (see Appendix~\ref{sec:twophoton})
\begin{equation}
\tilde{\xi}_{\lambda\lambda'}(\boldsymbol{r},\boldsymbol{r}')=\mathcal{N}\Theta_{\lambda\lambda'}\tilde{\eta}_m(\boldsymbol{r})\tilde{\eta}_{-m}(\boldsymbol{r}')\left[e^{im(\varphi-\varphi')}+{\rm c.c}\right],
\end{equation}
where we have used the property of the Bessel function $J_m(x)=(-1)^mJ_{-m}(x)$.
The PND and helicity density are given by $\left\langle \hat{n}(\boldsymbol{r})\right\rangle =2\left|\tilde{\eta}_m(\boldsymbol{r})\right|^{2}=2\left|\tilde{\eta}_{-m}(\boldsymbol{r})\right|^{2}$ and $\left\langle \hat{n}_{H}(\boldsymbol{r})\right\rangle =0$, respectively. This vanishing net helicity density results from the fact that we have required the two photons in the pulse to have opposite helicity. The corresponding correlation functions are given by,
\begin{align}
G^{(2)}(\boldsymbol{r},\boldsymbol{r}') & =8\mathcal{N}^{2}\left|\tilde{\eta}_m(\boldsymbol{r})\right|^{2}\left|\tilde{\eta}_m(\boldsymbol{r}')\right|^{2}\left\{ 1+\cos[2m(\varphi-\varphi')]\right\} ,\\
G_{H}^{(2)}(\boldsymbol{r},\boldsymbol{r}') & =-G^{(2)}(\boldsymbol{r},\boldsymbol{r}') \cos2\theta_{B}.
\end{align}
We obtain a simple relation between $G^{(2)}(\boldsymbol{r},\boldsymbol{r}')$ and $G_H^{(2)}(\boldsymbol{r},\boldsymbol{r}')$ for noninteracting photons in free space. A more sophisticated and interesting photon helicity correlation structure can be obtained in a nonlinear medium, such as the Rydberg-atom array induced photon-photon interaction~\cite{chang2021quantum,zhang2021photon}. In experiments, the helicity correlation can be measured via nano-scale quantum sensors~\cite{Kalhor2020probe}.

The striking property of the twisted photon pair is that its quantum coherence function is now modulated by the azimuthal angle difference of the two photons,
\begin{equation}
g^{(2)}(\boldsymbol{r},\boldsymbol{r}')=\frac{1}{2(1+\delta_{m,0})}\left\{ 1+\cos[2m(\varphi-\varphi')]\right\}. \label{eq:g2symmetry}
\end{equation}
For the $m=0$ case, the quantum coherence function reduces to a constant $g^{(2)}=1/2$ for an untwisted two-photon Fock state. For the $m\neq 0$ case, the $g^{(2)}$ function reaches it maximum $1$ when $\varphi-\varphi'=0$. Thus, the two photons in a twisted photon pair tend to be bunched compared to a regular untwisted pair. For a twisted photon pair with symmetric spin state~$\left|\uparrow\right\rangle\otimes \left|\uparrow\right\rangle$ or $\left|\downarrow\right\rangle\otimes \left|\downarrow\right\rangle$, the quantum coherence function is the same as Eq.~(\ref{eq:g2symmetry}). However, the helicity density is not zero for these two cases, $\langle \hat{n}_{H}(\boldsymbol{r})\rangle =\pm2\left|\tilde{\eta}_m(\boldsymbol{r})\right|^{2}\cos\theta_{B}$.

Another interesting phenomenon is that the quantum statistics of twisted photon pairs are reversed if their spin state is anti-symmetric~$\sim  \left|\uparrow\right\rangle\otimes\left|\downarrow\right\rangle-\left|\downarrow\right\rangle\otimes\left|\uparrow\right\rangle$. The corresponding two-photon SAF is given by
\begin{equation}
\xi_{\lambda_{1},\lambda_{2}}(\boldsymbol{k}_{1},\boldsymbol{k}_{2})=\mathcal{N}\Theta_{\lambda_{1}\lambda_{2}}\eta(\boldsymbol{k}_{1})\eta(\boldsymbol{k}_{2})\left[e^{im(\varphi_{k_{1}}-\varphi_{k_{2}})}-{\rm c.c.}\right],
\end{equation}
where $m\neq0$. The polarization of the photon pair is now described
by an anti-symmetric constant matrix 
\begin{equation}
\Theta=\left[\begin{array}{cc}
\Theta_{++} & \Theta_{+-}\\
\Theta_{-+} & \Theta_{--}
\end{array}\right]=\left[\begin{array}{cc}
0 & 1\\
-1 & 0
\end{array}\right].
\end{equation}
The quantum coherence of photons for this type of twisted photon pairs is given by
\begin{equation}
g^{(2)}(\boldsymbol{r},\boldsymbol{r}')=\frac{1}{2}\left\{ 1-\cos[2m(\varphi-\varphi')]\right\}. \label{eq:g2antisymmetry}
\end{equation}
In contrast to Eq.~(\ref{eq:g2symmetry}), the $g^{(2)}$ function now reaches it minimum $0$ when $\varphi-\varphi'=0$. Thus, the two photons in a twisted photon pair tend to be antibunched compared to a regular untwisted pair.


In Fig.~\ref{fig:coherence}, we contrast the $g^{(2)}$ function for twisted photon pairs with a symmetric spin state (the top row) and with an anti-symmetric spin state (the bottom row). We find that the two rows are precisely complementary to each other. Thus, the quantum statistics of twisted photon pairs are reversed when their spin state changes from symmetric to anti-symmetric. Here, we use two simple examples to show the influence of the photon spin on the quantum statistics of light. The SAF of entangled photon pairs generated via an spontaneous parametric down-conversion process will be more complicated~\cite{Arnaut2000photonpair,Keller1997theory}. However, we have revealed the essential feature of spin-dependent statistics in twisted photon pairs.

In the upper two examples, the spin and spatial degrees freedom of the photon pairs are highly entangled. Thus, the photon-number-density and helicity-density correlations exhibit almost the same correlation behavior. However, the properties of these two correlations could be significantly different in more general cases, such as randomly polarized Fock-state photon pairs. Quantum correlation can still exist in the photon-number density, but the helicity density will be completely uncorrelated.

\section{Conclusion}
We have put forth a quantum framework for structured quantum light in real space by introducing the effective field operator of photons. We exploited our presented theoretical formalism to study the photonic vortices. We showed the analogy between the photonic vortex and its counterpart in superfluids by defining quantum operators for the photonic currents. We also give an unambiguous definition of the topological charge of a photonic vortex---the winding number of the photonic currents. We predicted the pure helicity vortex, which can be measured in experiments. 

We also studied the quantum statistics of twisted light with our proposed theoretical formalism. We showed an interesting effect: The statistical behaviors are essentially different for twisted photon pairs with symmetric and antisymmetric spin states. Entangled twisted photons, which possess a spatially varying quantum coherence function in the transverse plane, can be utilized to enhance resolution in quantum imaging~\cite{maurer2011spatial} and detect the texture of a target in the quantum-illumination-based radar system~\cite{Lloyd2008Enhanced,Tan2008quantum}.

\section*{ACKNOWLEDGMENTS}
L.P.Y was funded by National Key R\&D Program of China (Grant No. 2021YFE0193500). D.X. was supported by NSFC Grant No.12075025.

\appendix 
\section{Photonic OAM operators in real space \label{sec:OAM}}
Historically, the discovery of the optical OAM has boosted the development of optical phase singularity research~\cite{Dennis2009Chapter}. Here, we show how to evaluate the OAM of light in real space within our proposed theoretical framework in this work. The directly observable part of the photonic OAM is given by $\hat{\boldsymbol{L}}^{{\rm obs}} =\varepsilon_{0}\int d^{3}r\hat{E}_{\perp}^{j}(\boldsymbol{r},t)(\boldsymbol{r}\times\boldsymbol{\nabla})\hat{A}_{\perp}^{j}(\boldsymbol{r},t)$~\cite{yang2020quantum}, {\color{red}which, in the two-component helicity representation, can be written in the fully real-space nonlocal form
\begin{equation}
\begin{aligned}
\hat{\boldsymbol{L}}^{{\rm obs}}={}&\int d^{3}r\,\hat{\psi}^{\dagger}(\boldsymbol{r})(\boldsymbol{r}\times\hat{\boldsymbol{p}})\hat{\psi}(\boldsymbol{r})\\
&-\int d^{3}r\int d^{3}r'\,\hat{\psi}^{\dagger}(\boldsymbol{r})\boldsymbol{K}(\boldsymbol{r}-\boldsymbol{r}')\hat{\psi}(\boldsymbol{r}'),
\end{aligned}
\label{eq:OAM}
\end{equation}
where $\hat{\boldsymbol p}=-i\hbar\boldsymbol\nabla$, $\boldsymbol K(\boldsymbol R)=\mathrm{diag}[\boldsymbol K_{+}(\boldsymbol R),\boldsymbol K_{-}(\boldsymbol R)]$, and
\begin{equation}
\boldsymbol K_{\lambda}(\boldsymbol R)=\frac{\hbar}{(2\pi)^3}\int d^3k\,e^{i\boldsymbol k\cdot\boldsymbol R}
\left[\boldsymbol k\times\boldsymbol{\mathcal A}_{\lambda}(\boldsymbol k)\right],
\end{equation}
with the Berry connection $\boldsymbol{\mathcal A}_{\lambda}(\boldsymbol k)=i\boldsymbol e_{\lambda}^{*}(\boldsymbol k)\cdot\boldsymbol\nabla_{\boldsymbol k}\boldsymbol e_{\lambda}(\boldsymbol k)$.}

\textcolor{red}{In a Cartesian coordinate, the three components of the differential part $\hat{\boldsymbol{\mathfrak{l}}}\equiv\boldsymbol r\times\hat{\boldsymbol p}$ are given by}
\begin{align}
\hat{\mathfrak{l}}_{x} & =-i\hbar\left(y\frac{\partial}{\partial z}-z\frac{\partial}{\partial y}\right),\\
\hat{\mathfrak{l}}_{y} & =-i\hbar\left(z\frac{\partial}{\partial x}-x\frac{\partial}{\partial z}\right),\\
\hat{\mathfrak{l}}_{z} & =-i\hbar\left(x\frac{\partial}{\partial y}-y\frac{\partial}{\partial x}\right).
\end{align}
It is more convenient to evaluate the OAM of a paraxial pulse or beam in a cylindrical coordinate as shown in Fig.~{\ref{fig:schematic}} (a). \textcolor{red}{Then these three differential components are given by}
\begin{align}
\hat{\mathfrak{l}}_{x} &  =-i\hbar\left(\rho\sin\varphi\frac{\partial}{\partial z}-z\sin\varphi\frac{\partial}{\partial\rho}-\frac{z}{\rho}\cos\varphi\frac{\partial}{\partial\varphi}\right), \label{eq:lx}\\
\hat{\mathfrak{l}}_{y} & =i\hbar\left(\rho\cos\varphi\frac{\partial}{\partial z}-z\cos\varphi\frac{\partial}{\partial\rho}+\frac{z}{\rho}\sin\varphi\frac{\partial}{\partial\varphi}\right),\label{eq:ly}\\
\hat{\mathfrak{l}}^{(0)}_{z} & = -i\hbar\frac{\partial}{\partial\varphi}.
\end{align}

For a laser pulse or a beam, the mean value of the OAM is obtained by simply replacing the field operators $\hat{\psi}(\boldsymbol{r})$ and $\hat{\psi}^{\dagger}(\boldsymbol{r})$ with the functions $\Psi (\boldsymbol{r})$ and $\Psi^{*} (\boldsymbol{r})$, respectively, \textcolor{red}{in the full nonlocal expression in Eq.~(\ref{eq:OAM})}. This is also the reason that the OAM of light has usually been handled classically~\cite{barnett1994orbital,berry1998paraxial,li2020spin}. We emphasize that no paraxial approximation~\cite{barnett1994orbital,berry1998paraxial} is required for the global quantities. Our theory provides a powerful and versatile tool to handle the OAM of all classes of quantum structured light pulses in real space, specifically for the one with spatiotemporal optical vortices~\cite{Jhajj2016spatiotemporal,Hancock2019free,Chong2020Generation}.

\textcolor{red}{We note that the nonlocal integrands in Eq.~(\ref{eq:OAM}) cannot in general be interpreted as a local photonic OAM density, which by definition is given by}
\begin{align}
\hat{\boldsymbol{l}}_{M}^{{\rm obs}}(\boldsymbol{r}) & =\varepsilon_{0}\hat{E}_{\perp}^{j}(\boldsymbol{r})(\boldsymbol{r}\times\boldsymbol{\nabla})\hat{A}_{\perp}^{j}(\boldsymbol{r}).
\end{align}
With the plane-wave expansion of $\hat{\boldsymbol{E}}_{\perp}$ and $\hat{\boldsymbol{A}}_{\perp}$~\cite{yang2021quantum}, we have
\begin{align}
\hat{\boldsymbol{l}}_{M}^{{\rm obs}}(\boldsymbol{r}) \approx &\frac{-i\hbar}{(2\pi)^{3}}\sum_{\lambda\lambda'}\int d^{3}k\int d^{3}k'\sqrt{\frac{\omega_{\boldsymbol{k}}}{\omega_{\boldsymbol{k}'}}}\boldsymbol{e}^{*}(\boldsymbol{k},\lambda)\cdot\boldsymbol{e}(\boldsymbol{k}',\lambda') \nonumber \\ 
&\times\hat{a}_{\boldsymbol{k},\lambda}^{\dagger}e^{-i\boldsymbol{k}\cdot\boldsymbol{r}}(\boldsymbol{r}\times\boldsymbol{\nabla})\hat{a}_{\boldsymbol{k}',\lambda'}e^{i\boldsymbol{k}'\cdot\boldsymbol{r}},
\end{align}
where we have neglected the counterrotating wave terms~\textcolor{red}{$\hat a_{\boldsymbol{k},\lambda}^{\dagger}\hat{a}_{\boldsymbol{k}',\lambda'}^{\dagger}$
and $\hat{a}_{\boldsymbol{k},\lambda}\hat{a}_{\boldsymbol{k}',\lambda'}$}. In the paraxial limit, we have $\boldsymbol{e}^{*}(\boldsymbol{k},\lambda)\cdot\boldsymbol{e}(\boldsymbol{k}',\lambda')\approx\delta_{\lambda\lambda'}\exp\left[i\lambda(\varphi_{k}-\varphi_{k'})\right]$,
where we have used the relations
\begin{equation}
\boldsymbol{e}(\boldsymbol{k},\lambda)\!=\!e^{-i\lambda\varphi_{k}}\!\cos^{2}\!\frac{\theta_{k}}{2}\boldsymbol{e}_{\lambda}-e^{i\lambda\varphi_{k}}\!\sin^{2}\!\frac{\theta_{k}}{2}\boldsymbol{e}_{-\lambda}-\frac{1}{\sqrt{2}}\sin\theta_{k}\boldsymbol{e}_{z},\!
\end{equation}
and $\boldsymbol{e}_{\lambda}=(\boldsymbol{e}_{x}+i\lambda\boldsymbol{e}_{y})/\sqrt{2}$~\cite{yang2021quantum}. Thus, even for a quasi-single-frequency paraxial beam with $\sqrt{\omega_{\boldsymbol{k}}/\omega_{\boldsymbol{k}'}}\approx1$, the OAM density only reduces to
\begin{equation}
\hat{\boldsymbol{l}}^{\rm obs}(\boldsymbol{r},t)\approx\hat{\tilde{\psi}}^{\dagger}(\boldsymbol{r})(\boldsymbol{r}\times\hat{\boldsymbol{p}})\hat{\tilde{\psi}}(\boldsymbol{r}),
\end{equation}
where
\begin{equation}
\hat{\tilde{\psi}}_{\lambda}(\boldsymbol{r})\equiv \frac{1}{\sqrt{(2\pi)^{3}}}\int d^{3}k\hat{a}_{\boldsymbol{k},\lambda}e^{i(\boldsymbol{k}\cdot\boldsymbol{r}-\lambda\varphi_k)}  
\end{equation}
\textcolor{red}{The last two equations are a paraxial, quasi-single-frequency reduction in the specified polarization frame and should not be interpreted as a general exact replacement of the full nonlocal operator in Eq.~(\ref{eq:OAM}).}
It is more convenient to evaluate both the photonic spin and OAM densities in $\boldsymbol{k}$-space. The angular momentum density of light will be of increasing interest for experiments in the near future.

\section{Quantum state of the superposition of multiple laser beams \label{sec:MPwavefunction}}
A laser beam can be regarded as a quantum light pulse with an extremely long pulse length and enormous number of photons. The corresponding quantum state is given by
\begin{equation}
\left|\alpha_{\xi}\right\rangle =\hat{D}_{\xi}(\alpha)|0\rangle =e^{-|\alpha|^2/2}\sum_{n=0}^{\infty}\frac{\alpha^{n}}{n!}\left(\hat{a}_{\xi}^{\dagger}\right)^{n}\left|0\right\rangle.
\end{equation}
where the operator $\hat{D}_{\xi}$ is given in Eq.~(\ref{eq:displace}) and $\hat{a}^{\dagger}_{\xi}$ is the photon-wave-packet creation operator. It can be easily verified that two operators $\hat{D}_{\xi_1}(\alpha_1)$  and $\hat{D}_{\xi_2}(\alpha_2)$ commute with each other, i.e., $[\hat{D}_{\xi_1}(\alpha_1),\hat{D}_{\xi_2}(\alpha_2)]=0$, because only creation operators $\hat{a}^{\dagger}_{\boldsymbol{k}}$ are involved in $\hat{D}_{\xi}$.

The superposition of multiple laser beams can be described by the quantum state
\begin{align}
|\Psi\rangle = & \frac{1}{\sqrt{\mathcal{N}}}\prod_j \hat{D}_{\xi_j}(\alpha_j)|0\rangle \\ = & \frac{1}{\sqrt{\mathcal{N}}}\exp \left[\sum_j \left(\alpha_j\hat{a}_{\xi_j}^{\dagger}-\frac{1}{2}|\alpha_j|^{2}\right)\right]|0\rangle.    
\end{align}
Using the relation,
\begin{equation}
e^{\sum_j\alpha^*_j\hat{a}_{\xi_j}}e^{\sum_j\alpha_j\hat{a}_{\xi_j}^{\dagger}}=e^{\sum_j\alpha_j\hat{a}_{\xi_j}^{\dagger}}e^{\sum_j\alpha^*_j\hat{a}_{\xi_j}}e^{\sum_{j,j'}\alpha^*_j\alpha_{j'}\left[\hat{a}_{\xi_j},\hat{a}_{\xi_{j'}}^{\dagger}\right]},
\end{equation}
we obtain the normalization factor
\begin{equation}
\mathcal{N} = \exp\left(\sum_{j\neq j'}\alpha^*_j\alpha_{j'}[\hat{a}_{\xi_j},\hat{a}_{\xi_{j'}}^{\dagger}]\right), \label{eq:NormFactor}
\end{equation}
with the overlap of the SAFs of the two pulses
\begin{align}
[\hat{a}_{\xi_j},\hat{a}_{\xi_{j'}}^{\dagger}] & = \sum_{\lambda} \int d^3k\xi^*_{j,\lambda}(\boldsymbol{k})\xi_{j',\lambda}(\boldsymbol{k})\\
& =\sum_{\lambda} \int d^3r\tilde{\xi}^*_{j,\lambda}(\boldsymbol{r},t)\tilde{\xi}_{j',\lambda}(\boldsymbol{r},t).
\end{align}

Utilizing the relation
\begin{equation*}
\left[a_{\boldsymbol{k}},\hat{D}_{\xi}(\alpha)\right] = \alpha\xi(\boldsymbol{k})\hat{D}_{\xi}(\alpha),   
\end{equation*}
we show that $\hat{\psi}(\boldsymbol{r})\left|\Psi\right\rangle =\Psi(\boldsymbol{r},t)\left|\Psi\right\rangle$ with
$\Psi(\boldsymbol{r},t) =\sum_j \alpha_j \tilde{\xi}_j(\boldsymbol{r},t)$
satisfying the wave equation
$\left(\nabla^{2}-\frac{1}{c^{2}}\frac{\partial^{2}}{\partial t^{2}}\right)\Psi(\boldsymbol{r},t)=0$.
We note that our presented formalism is significantly different from the previous one based on a six-component photon wave function~\cite{Bia1994On,Sipe1995photon}. Our introduced two-component field operator $\hat{\psi}(\boldsymbol{r})$ satisfies the wave equation~\cite{yang2021quantum} and is compatible with light-matter interaction. However, the six-component wave function satisfies a Schr{\" o}dinger-like equation and is in-compatible with the light-matter interaction.

In the circular-polarization representation, a linearly polarized laser pulse can be described by a state
\begin{align}
\left|\alpha_{\xi,\lambda=1}\right\rangle & = \exp \left[\frac{1}{\sqrt{2}}\alpha(\hat{a}^\dagger_{\xi,+}+\hat{a}^\dagger_{\xi,-})-\frac{|\alpha|^2}{2}\right]|0\rangle, \\
\left|\alpha_{\xi,\lambda=2}\right\rangle & = \exp \left[\frac{i}{\sqrt{2}}\alpha(\hat{a}^\dagger_{\xi,+}-\hat{a}^\dagger_{\xi,-})-\frac{|\alpha|^2}{2}\right]|0\rangle,  
\end{align}
with
\begin{align}
\hat{\psi}(\boldsymbol{r})\left|\alpha_{\xi,\lambda=1}\right\rangle &= \frac{1}{\sqrt{2}} [\alpha\xi(\boldsymbol{r},t),\alpha\xi(\boldsymbol{r},t)]^T \left|\alpha_{\xi,\lambda=1}\right\rangle,\\
\hat{\psi}(\boldsymbol{r})\left|\alpha_{\xi,\lambda=2}\right\rangle &= \frac{i}{\sqrt{2}} [\alpha\xi(\boldsymbol{r},t),-\alpha\xi(\boldsymbol{r},t)]^T \left|\alpha_{\xi,\lambda=2}\right\rangle.
\end{align}
Elliptically polarized quantum pulses can be constructed in a similar way.

\section{Hydrodynamics of vortices in paraxial laser beams\label{sec:hydrodynamic}}
A laser beam can be regarded as a coherent-state pulse with an extremely
long pulse length, i.e., very narrow linewidth in the frequency domain. In this case, we obtain the Helmholtz equation for this quasi-single-frequency
beam
\begin{equation}
\left(\nabla^{2}+k_{0}^{2}\right)\Psi_{\pm}(\boldsymbol{r},t)=0.
\end{equation}
For a well-collimated beam, we can take the paraxial approximation.
To extract the primary propagating factor out of the wave-packet function, we let
\begin{equation}
\Psi_{\pm}(\boldsymbol{r},t)=\Psi_{{\rm PA},\pm}(\boldsymbol{r})e^{i(k_{0}z-\omega t)},
\end{equation}
where the function $\Psi_{{\rm PA},\pm}(\boldsymbol{r})$ is a slowly
varying function of $z$, i.e., $\partial_{z}\Psi_{{\rm PA},\pm}\ll\Psi_{{\rm PA},\pm}/\lambda_{0}\sim k\Psi_{{\rm PA},\pm}$.
Using the relations
\begin{equation}
\partial_{z}^{2}\Psi_{{\rm PA},\pm}\ll k\partial_{z}\Psi_{{\rm PA},\pm}\ll k_{0}^{2}\Psi_{{\rm PA},\pm},
\end{equation}
we obtain the paraxial Helmholtz equation
\begin{equation}
i\partial_{z}\Psi_{{\rm PA},\pm}(\boldsymbol{r})\approx-\frac{1}{2k_{0}}\nabla_{T}^{2}\Psi_{{\rm PA},\pm}(\boldsymbol{r}),
\end{equation}
where we have neglected the $\partial_{z}^{2}\Psi_{{\rm PA},\pm}$
term.

Similar to the probability current in quantum mechanics, we can define a photon current in the $xy$ plane for a laser beam
\begin{align}
\langle\hat{\boldsymbol{j}}_N(\boldsymbol{r})\rangle &\! =\!-\frac{i}{2k_{0}}\!\!\sum_{\lambda}\!\left[\!\Psi_{{\rm PA},\lambda}^{*}(\boldsymbol{r})\boldsymbol{\nabla}_{T}\Psi_{{\rm PA},\lambda}(\boldsymbol{r})\!-\!\Psi_{{\rm PA},\lambda}(\boldsymbol{r})\boldsymbol{\nabla}_{T}\Psi_{{\rm PA},\lambda}^{*}(\boldsymbol{r})\!\right] \nonumber\\
 & =-\frac{i}{2k_{0}}\sum_{\lambda}\left[\Psi^{*}_{\lambda}(\boldsymbol{r})\boldsymbol{\nabla}_{T}\Psi_{\lambda}(\boldsymbol{r})-\Psi_{\lambda}(\boldsymbol{r})\boldsymbol{\nabla}_{T}\Psi^{*}_{\lambda}(\boldsymbol{r})\right]\nonumber \\
 & =-\frac{i}{2k_{0}}\left\langle \hat{\psi}^{\dagger}(\boldsymbol{r})\boldsymbol{\nabla}_{T}\hat{\psi}(\boldsymbol{r})-\left[\boldsymbol{\nabla}_{T}\hat{\psi}^{\dagger}(\boldsymbol{r})\right]\hat{\psi}(\boldsymbol{r})\right\rangle.
\end{align}
The paraxial Helmholtz equation gives the continuity equation (\ref{eq:Continuity}) with $\langle\hat{n}\rangle = \sum_{\lambda}|\Psi _{\lambda}(\boldsymbol{r})|^2$.

For a laser beam, the wave-packet function is a continuous function of $\boldsymbol{r}$. We can always rewrite it as $\Psi_{\pm}(\boldsymbol{r})=|\Psi_{\pm}(\boldsymbol{r})|\exp[i\phi_{\pm}(\boldsymbol{r})]$. Here, we emphasize that the norm of $\Psi_{\pm}$ is still continuous, but the phase $\phi_{\pm}$ can have singularity points and discontinuous steps. Splitting the paraxial Helmholtz equation into real and imaginary parts~\cite{pethick2008bose,Rozas1997propagation}, we obtain two equations
\begin{equation}
\frac{\partial}{\partial z}\phi_{\pm}=\frac{1}{2k_0}\left[\frac{1}{|\Psi_{\pm}|}\nabla_{T}^{2}|\Psi_{\pm}|-(\boldsymbol{\nabla}_{T}\phi_{\pm})^{2}\right],
\end{equation}
and
\begin{equation}
\frac{\partial}{\partial z}|\Psi_{\pm}|=-\frac{1}{2k_0}\left[|\Psi_{\pm}|\nabla_{T}^{2}\phi_{\pm}-2(\boldsymbol{\nabla}_{T}|\Psi_{\pm}|)\cdot(\boldsymbol{\nabla}_{T}\phi_{\pm})\right].
\end{equation}
The second one can be used to derive the continuity equations (\ref{eq:Continuity}) and (\ref{eq:HContinuity}).

The divergence of the first one gives the hydrodynamic equations for $\boldsymbol{v}_{\pm}\equiv\boldsymbol{\nabla}_T\phi_{\pm}/k_0$,
\begin{equation}
\frac{\partial}{\partial z}\boldsymbol{v}_{\pm}=\boldsymbol{\nabla}_{T}\left[\frac{1}{2k_0^{2}|\Psi_{\pm}|}\nabla_{T}^{2}|\Psi_{\pm}|-\frac{1}{2}v_{\pm}^{2}\right], \label{eq:HE1}
\end{equation}
which can be rewritten as an analog of the Euler equation
\begin{equation}
\frac{\partial}{\partial z}\boldsymbol{v}_{\pm}+\boldsymbol{v}_{\pm}\cdot\boldsymbol{\nabla}_{T}\boldsymbol{v}_{\pm}=\boldsymbol{\nabla}_{T}\left[\frac{1}{2k_0^{2}|\Psi_{\pm}|}\nabla_{T}^{2}|\Psi_{\pm}|\right], \label{eq:HE2}
\end{equation}
without an external force term. The term on the right hand side has been
referred to as the quantum pressure term, which describes forces due
to spatial variations in the magnitude of the wave-packet function~\cite{pethick2008bose}.

\section{Wave packet in real space \label{sec:twophoton}}
It can be easily verified that the Fourier transformation of the SAF~$\xi(\boldsymbol{k})=\eta(\boldsymbol{k})\exp(im\varphi_k)$
can be written in the form
\begin{align}
\tilde{\xi}(\boldsymbol{r}) & =\int d^{3}k\xi(\boldsymbol{k})e^{i\boldsymbol{k}\cdot\boldsymbol{r}}=\int d^{3}k\eta(\boldsymbol{k})e^{im\varphi_k}e^{i\boldsymbol{k}\cdot\boldsymbol{r}}\\
 & =\int_{-\infty}^{\infty}\!\!\!\!dk_{z}\!\!\int_{0}^{\infty}\!\!\!\!\rho_{k}d\rho_{k}\!\!\int_{0}^{2\pi}\!\!\!\!\!d\varphi_{k}\frac{\eta(\boldsymbol{k})}{\sqrt{(2\pi)^{3}}}e^{i\left[k_{z}z+\rho\rho_{k}\cos(\varphi-\varphi_{k})+m\varphi_{k}\right]} \nonumber\\
 & \equiv\tilde{\eta}_m(\boldsymbol{r})e^{im\varphi},
\end{align}
where  
\begin{equation}
\tilde{\eta}_m(\boldsymbol{r})=\frac{i^{m}}{\sqrt{(2\pi)}}\int_{-\infty}^{\infty}dk_{z}\int_{0}^{\infty}\rho_{k}d\rho_{k}\eta(\boldsymbol{k})e^{ik_{z}z}J_{m}(\rho\rho_{k}),
\end{equation}
is independent of $\varphi$ and $J_{n}(x)$ is the $n$th Bessel function of the first kind. We can verify that $\psi(\boldsymbol{r})$
is also normalized, i.e., 
\begin{align}
\int d^{3}r\left|\eta_m(\boldsymbol{r})\right|^{2} & =\int d^{3}r\left|\tilde{\xi}(\boldsymbol{r})\right|^{2} =\int d^{3}k\left|\eta(\boldsymbol{k})\right|^{2}=1.
\end{align}

\bibliography{main}

\begin{thebibliography}{95}%
\makeatletter
\providecommand \@ifxundefined [1]{%
 \@ifx{#1\undefined}
}%
\providecommand \@ifnum [1]{%
 \ifnum #1\expandafter \@firstoftwo
 \else \expandafter \@secondoftwo
 \fi
}%
\providecommand \@ifx [1]{%
 \ifx #1\expandafter \@firstoftwo
 \else \expandafter \@secondoftwo
 \fi
}%
\providecommand \natexlab [1]{#1}%
\providecommand \enquote  [1]{``#1''}%
\providecommand \bibnamefont  [1]{#1}%
\providecommand \bibfnamefont [1]{#1}%
\providecommand \citenamefont [1]{#1}%
\providecommand \href@noop [0]{\@secondoftwo}%
\providecommand \href [0]{\begingroup \@sanitize@url \@href}%
\providecommand \@href[1]{\@@startlink{#1}\@@href}%
\providecommand \@@href[1]{\endgroup#1\@@endlink}%
\providecommand \@sanitize@url [0]{\catcode `\\12\catcode `\$12\catcode
  `\&12\catcode `\#12\catcode `\^12\catcode `\_12\catcode `\%12\relax}%
\providecommand \@@startlink[1]{}%
\providecommand \@@endlink[0]{}%
\providecommand \url  [0]{\begingroup\@sanitize@url \@url }%
\providecommand \@url [1]{\endgroup\@href {#1}{\urlprefix }}%
\providecommand \urlprefix  [0]{URL }%
\providecommand \Eprint [0]{\href }%
\providecommand \doibase [0]{https://doi.org/}%
\providecommand \selectlanguage [0]{\@gobble}%
\providecommand \bibinfo  [0]{\@secondoftwo}%
\providecommand \bibfield  [0]{\@secondoftwo}%
\providecommand \translation [1]{[#1]}%
\providecommand \BibitemOpen [0]{}%
\providecommand \bibitemStop [0]{}%
\providecommand \bibitemNoStop [0]{.\EOS\space}%
\providecommand \EOS [0]{\spacefactor3000\relax}%
\providecommand \BibitemShut  [1]{\csname bibitem#1\endcsname}%
\let\auto@bib@innerbib\@empty
\bibitem [{\citenamefont {Allen}\ \emph {et~al.}(1992)\citenamefont {Allen},
  \citenamefont {Beijersbergen}, \citenamefont {Spreeuw},\ and\ \citenamefont
  {Woerdman}}]{allen1992orbital}%
  \BibitemOpen
  \bibfield  {author} {\bibinfo {author} {\bibfnamefont {L.}~\bibnamefont
  {Allen}}, \bibinfo {author} {\bibfnamefont {M.~W.}\ \bibnamefont
  {Beijersbergen}}, \bibinfo {author} {\bibfnamefont {R.~J.~C.}\ \bibnamefont
  {Spreeuw}},\ and\ \bibinfo {author} {\bibfnamefont {J.~P.}\ \bibnamefont
  {Woerdman}},\ }\bibfield  {title} {\bibinfo {title} {Orbital angular momentum
  of light and the transformation of laguerre-gaussian laser modes},\ }\href
  {https://doi.org/10.1103/PhysRevA.45.8185} {\bibfield  {journal} {\bibinfo
  {journal} {Phys. Rev. A}\ }\textbf {\bibinfo {volume} {45}},\ \bibinfo
  {pages} {8185} (\bibinfo {year} {1992})}\BibitemShut {NoStop}%
\bibitem [{\citenamefont {Allen}\ \emph {et~al.}(1999)\citenamefont {Allen},
  \citenamefont {Padgett},\ and\ \citenamefont {Babiker}}]{allen1999iv}%
  \BibitemOpen
  \bibfield  {author} {\bibinfo {author} {\bibfnamefont {L.}~\bibnamefont
  {Allen}}, \bibinfo {author} {\bibfnamefont {M.}~\bibnamefont {Padgett}},\
  and\ \bibinfo {author} {\bibfnamefont {M.}~\bibnamefont {Babiker}},\
  }\bibfield  {title} {\bibinfo {title} {Iv the orbital angular momentum of
  light}\ }(\bibinfo  {publisher} {Elsevier},\ \bibinfo {year} {1999})\ pp.\
  \bibinfo {pages} {291--372}\BibitemShut {NoStop}%
\bibitem [{\citenamefont {Bliokh}\ and\ \citenamefont
  {Nori}(2015)}]{bliokh2015transverse}%
  \BibitemOpen
  \bibfield  {author} {\bibinfo {author} {\bibfnamefont {K.~Y.}\ \bibnamefont
  {Bliokh}}\ and\ \bibinfo {author} {\bibfnamefont {F.}~\bibnamefont {Nori}},\
  }\bibfield  {title} {\bibinfo {title} {Transverse and longitudinal angular
  momenta of light},\ }\href
  {https://www.sciencedirect.com/science/article/pii/S0370157315003336}
  {\bibfield  {journal} {\bibinfo  {journal} {Physics Reports}\ }\textbf
  {\bibinfo {volume} {592}},\ \bibinfo {pages} {1} (\bibinfo {year}
  {2015})}\BibitemShut {NoStop}%
\bibitem [{\citenamefont {Molina-Terriza}\ \emph {et~al.}(2007)\citenamefont
  {Molina-Terriza}, \citenamefont {Torres},\ and\ \citenamefont
  {Torner}}]{molina2007twisted}%
  \BibitemOpen
  \bibfield  {author} {\bibinfo {author} {\bibfnamefont {G.}~\bibnamefont
  {Molina-Terriza}}, \bibinfo {author} {\bibfnamefont {J.~P.}\ \bibnamefont
  {Torres}},\ and\ \bibinfo {author} {\bibfnamefont {L.}~\bibnamefont
  {Torner}},\ }\bibfield  {title} {\bibinfo {title} {Twisted photons},\ }\href
  {https://doi.org/10.1038/nphys607} {\bibfield  {journal} {\bibinfo  {journal}
  {Nature physics}\ }\textbf {\bibinfo {volume} {3}},\ \bibinfo {pages} {305}
  (\bibinfo {year} {2007})}\BibitemShut {NoStop}%
\bibitem [{\citenamefont {Götte}\ \emph {et~al.}(2008)\citenamefont {Götte},
  \citenamefont {O'Holleran}, \citenamefont {Preece}, \citenamefont
  {Flossmann},\ and\ \citenamefont {Padgett}}]{Gotte2008Light}%
  \BibitemOpen
  \bibfield  {author} {\bibinfo {author} {\bibfnamefont {J.~B.}\ \bibnamefont
  {Götte}}, \bibinfo {author} {\bibfnamefont {K.}~\bibnamefont {O'Holleran}},
  \bibinfo {author} {\bibfnamefont {D.}~\bibnamefont {Preece}}, \bibinfo
  {author} {\bibfnamefont {F.}~\bibnamefont {Flossmann}},\ and\ \bibinfo
  {author} {\bibfnamefont {M.~J.}\ \bibnamefont {Padgett}},\ }\bibfield
  {title} {\bibinfo {title} {Light beams with fractional orbital angular
  momentum and their vortex structure},\ }\href
  {https://doi.org/10.1364/JOSAB.14.003054} {\bibfield  {journal} {\bibinfo
  {journal} {Optics Express}\ }\textbf {\bibinfo {volume} {16}},\ \bibinfo
  {pages} {993} (\bibinfo {year} {2008})}\BibitemShut {NoStop}%
\bibitem [{\citenamefont {Yu}\ \emph {et~al.}(2011)\citenamefont {Yu},
  \citenamefont {Genevet}, \citenamefont {Kats}, \citenamefont {Aieta},
  \citenamefont {Tetienne}, \citenamefont {Capasso},\ and\ \citenamefont
  {Gaburro}}]{yu2011light}%
  \BibitemOpen
  \bibfield  {author} {\bibinfo {author} {\bibfnamefont {N.}~\bibnamefont
  {Yu}}, \bibinfo {author} {\bibfnamefont {P.}~\bibnamefont {Genevet}},
  \bibinfo {author} {\bibfnamefont {M.~A.}\ \bibnamefont {Kats}}, \bibinfo
  {author} {\bibfnamefont {F.}~\bibnamefont {Aieta}}, \bibinfo {author}
  {\bibfnamefont {J.-P.}\ \bibnamefont {Tetienne}}, \bibinfo {author}
  {\bibfnamefont {F.}~\bibnamefont {Capasso}},\ and\ \bibinfo {author}
  {\bibfnamefont {Z.}~\bibnamefont {Gaburro}},\ }\bibfield  {title} {\bibinfo
  {title} {Light propagation with phase discontinuities: Generalized laws of
  reflection and refraction},\ }\href
  {https://science.sciencemag.org/content/sci/334/6054/333.full.pdf} {\bibfield
   {journal} {\bibinfo  {journal} {Science}\ }\textbf {\bibinfo {volume}
  {334}},\ \bibinfo {pages} {333} (\bibinfo {year} {2011})}\BibitemShut
  {NoStop}%
\bibitem [{\citenamefont {Naidoo}\ \emph {et~al.}(2016)\citenamefont {Naidoo},
  \citenamefont {Roux}, \citenamefont {Dudley}, \citenamefont {Litvin},
  \citenamefont {Piccirillo}, \citenamefont {Marrucci},\ and\ \citenamefont
  {Forbes}}]{Naidoo2016controlled}%
  \BibitemOpen
  \bibfield  {author} {\bibinfo {author} {\bibfnamefont {D.}~\bibnamefont
  {Naidoo}}, \bibinfo {author} {\bibfnamefont {F.~S.}\ \bibnamefont {Roux}},
  \bibinfo {author} {\bibfnamefont {A.}~\bibnamefont {Dudley}}, \bibinfo
  {author} {\bibfnamefont {I.}~\bibnamefont {Litvin}}, \bibinfo {author}
  {\bibfnamefont {B.}~\bibnamefont {Piccirillo}}, \bibinfo {author}
  {\bibfnamefont {L.}~\bibnamefont {Marrucci}},\ and\ \bibinfo {author}
  {\bibfnamefont {A.}~\bibnamefont {Forbes}},\ }\bibfield  {title} {\bibinfo
  {title} {Controlled generation of higher-order poincar{\'e} sphere beams from
  a laser},\ }\href {https://doi.org/10.1038/nphoton.2016.37} {\bibfield
  {journal} {\bibinfo  {journal} {Nature Photonics}\ }\textbf {\bibinfo
  {volume} {10}},\ \bibinfo {pages} {327} (\bibinfo {year} {2016})}\BibitemShut
  {NoStop}%
\bibitem [{\citenamefont {{Devlin}}\ \emph {et~al.}(2017)\citenamefont
  {{Devlin}}, \citenamefont {{Ambrosio}}, \citenamefont {{Rubin}},
  \citenamefont {{Mueller}},\ and\ \citenamefont
  {{Capasso}}}]{devlin2017arbitrary}%
  \BibitemOpen
  \bibfield  {author} {\bibinfo {author} {\bibfnamefont {R.~C.}\ \bibnamefont
  {{Devlin}}}, \bibinfo {author} {\bibfnamefont {A.}~\bibnamefont
  {{Ambrosio}}}, \bibinfo {author} {\bibfnamefont {N.~A.}\ \bibnamefont
  {{Rubin}}}, \bibinfo {author} {\bibfnamefont {J.~P.~B.}\ \bibnamefont
  {{Mueller}}},\ and\ \bibinfo {author} {\bibfnamefont {F.}~\bibnamefont
  {{Capasso}}},\ }\bibfield  {title} {\bibinfo {title} {Arbitrary
  spin-to–orbital angular momentum conversion of light},\ }\href
  {https://science.sciencemag.org/content/358/6365/896} {\bibfield  {journal}
  {\bibinfo  {journal} {Science}\ }\textbf {\bibinfo {volume} {358}},\ \bibinfo
  {pages} {896} (\bibinfo {year} {2017})}\BibitemShut {NoStop}%
\bibitem [{\citenamefont {Forbes}\ \emph {et~al.}(2021)\citenamefont {Forbes},
  \citenamefont {de~Oliveira},\ and\ \citenamefont
  {Dennis}}]{Forbes2021Structured}%
  \BibitemOpen
  \bibfield  {author} {\bibinfo {author} {\bibfnamefont {A.}~\bibnamefont
  {Forbes}}, \bibinfo {author} {\bibfnamefont {M.}~\bibnamefont
  {de~Oliveira}},\ and\ \bibinfo {author} {\bibfnamefont {M.~R.}\ \bibnamefont
  {Dennis}},\ }\bibfield  {title} {\bibinfo {title} {Structured light},\ }\href
  {https://doi.org/10.1038/s41566-021-00780-4} {\bibfield  {journal} {\bibinfo
  {journal} {Nature Photonics}\ }\textbf {\bibinfo {volume} {15}},\ \bibinfo
  {pages} {253} (\bibinfo {year} {2021})}\BibitemShut {NoStop}%
\bibitem [{\citenamefont {Hayenga}\ \emph {et~al.}(2019)\citenamefont
  {Hayenga}, \citenamefont {Parto}, \citenamefont {Ren}, \citenamefont {Wu},
  \citenamefont {Hokmabadi}, \citenamefont {Wolff}, \citenamefont {El-Ganainy},
  \citenamefont {Mortensen}, \citenamefont {Christodoulides},\ and\
  \citenamefont {Khajavikhan}}]{hayenga2019direct}%
  \BibitemOpen
  \bibfield  {author} {\bibinfo {author} {\bibfnamefont {W.~E.}\ \bibnamefont
  {Hayenga}}, \bibinfo {author} {\bibfnamefont {M.}~\bibnamefont {Parto}},
  \bibinfo {author} {\bibfnamefont {J.}~\bibnamefont {Ren}}, \bibinfo {author}
  {\bibfnamefont {F.~O.}\ \bibnamefont {Wu}}, \bibinfo {author} {\bibfnamefont
  {M.~P.}\ \bibnamefont {Hokmabadi}}, \bibinfo {author} {\bibfnamefont
  {C.}~\bibnamefont {Wolff}}, \bibinfo {author} {\bibfnamefont
  {R.}~\bibnamefont {El-Ganainy}}, \bibinfo {author} {\bibfnamefont {N.~A.}\
  \bibnamefont {Mortensen}}, \bibinfo {author} {\bibfnamefont {D.~N.}\
  \bibnamefont {Christodoulides}},\ and\ \bibinfo {author} {\bibfnamefont
  {M.}~\bibnamefont {Khajavikhan}},\ }\bibfield  {title} {\bibinfo {title}
  {Direct generation of tunable orbital angular momentum beams in microring
  lasers with broadband exceptional points},\ }\href
  {https://doi.org/10.1021/acsphotonics.9b00779} {\bibfield  {journal}
  {\bibinfo  {journal} {ACS Photonics}\ }\textbf {\bibinfo {volume} {6}},\
  \bibinfo {pages} {1895} (\bibinfo {year} {2019})}\BibitemShut {NoStop}%
\bibitem [{\citenamefont {Zhang}\ \emph {et~al.}(2020)\citenamefont {Zhang},
  \citenamefont {Qiao}, \citenamefont {Midya}, \citenamefont {Liu},
  \citenamefont {Sun}, \citenamefont {Wu}, \citenamefont {Liu}, \citenamefont
  {Agarwal}, \citenamefont {Jornet}, \citenamefont {Longhi} \emph
  {et~al.}}]{zhang2020tunable}%
  \BibitemOpen
  \bibfield  {author} {\bibinfo {author} {\bibfnamefont {Z.}~\bibnamefont
  {Zhang}}, \bibinfo {author} {\bibfnamefont {X.}~\bibnamefont {Qiao}},
  \bibinfo {author} {\bibfnamefont {B.}~\bibnamefont {Midya}}, \bibinfo
  {author} {\bibfnamefont {K.}~\bibnamefont {Liu}}, \bibinfo {author}
  {\bibfnamefont {J.}~\bibnamefont {Sun}}, \bibinfo {author} {\bibfnamefont
  {T.}~\bibnamefont {Wu}}, \bibinfo {author} {\bibfnamefont {W.}~\bibnamefont
  {Liu}}, \bibinfo {author} {\bibfnamefont {R.}~\bibnamefont {Agarwal}},
  \bibinfo {author} {\bibfnamefont {J.~M.}\ \bibnamefont {Jornet}}, \bibinfo
  {author} {\bibfnamefont {S.}~\bibnamefont {Longhi}}, \emph {et~al.},\
  }\bibfield  {title} {\bibinfo {title} {Tunable topological charge vortex
  microlaser},\ }\href {https://doi.org/10.1126/science.aba8996} {\bibfield
  {journal} {\bibinfo  {journal} {Science}\ }\textbf {\bibinfo {volume}
  {368}},\ \bibinfo {pages} {760} (\bibinfo {year} {2020})}\BibitemShut
  {NoStop}%
\bibitem [{\citenamefont {Sroor}\ \emph {et~al.}(2020)\citenamefont {Sroor},
  \citenamefont {Huang}, \citenamefont {Sephton}, \citenamefont {Naidoo},
  \citenamefont {Vall{\'e}s}, \citenamefont {Ginis}, \citenamefont {Qiu},
  \citenamefont {Ambrosio}, \citenamefont {Capasso},\ and\ \citenamefont
  {Forbes}}]{Sroor2020high}%
  \BibitemOpen
  \bibfield  {author} {\bibinfo {author} {\bibfnamefont {H.}~\bibnamefont
  {Sroor}}, \bibinfo {author} {\bibfnamefont {Y.-W.}\ \bibnamefont {Huang}},
  \bibinfo {author} {\bibfnamefont {B.}~\bibnamefont {Sephton}}, \bibinfo
  {author} {\bibfnamefont {D.}~\bibnamefont {Naidoo}}, \bibinfo {author}
  {\bibfnamefont {A.}~\bibnamefont {Vall{\'e}s}}, \bibinfo {author}
  {\bibfnamefont {V.}~\bibnamefont {Ginis}}, \bibinfo {author} {\bibfnamefont
  {C.-W.}\ \bibnamefont {Qiu}}, \bibinfo {author} {\bibfnamefont
  {A.}~\bibnamefont {Ambrosio}}, \bibinfo {author} {\bibfnamefont
  {F.}~\bibnamefont {Capasso}},\ and\ \bibinfo {author} {\bibfnamefont
  {A.}~\bibnamefont {Forbes}},\ }\bibfield  {title} {\bibinfo {title}
  {High-purity orbital angular momentum states from a visible metasurface
  laser},\ }\href {https://doi.org/10.1038/s41566-020-0623-z} {\bibfield
  {journal} {\bibinfo  {journal} {Nature Photonics}\ }\textbf {\bibinfo
  {volume} {14}},\ \bibinfo {pages} {498} (\bibinfo {year} {2020})}\BibitemShut
  {NoStop}%
\bibitem [{\citenamefont {Cai}\ \emph {et~al.}(2012)\citenamefont {Cai},
  \citenamefont {Wang}, \citenamefont {Strain}, \citenamefont {Johnson-Morris},
  \citenamefont {Zhu}, \citenamefont {Sorel}, \citenamefont
  {O{\textquoteright}Brien}, \citenamefont {Thompson},\ and\ \citenamefont
  {Yu}}]{Cai2012Integrated}%
  \BibitemOpen
  \bibfield  {author} {\bibinfo {author} {\bibfnamefont {X.}~\bibnamefont
  {Cai}}, \bibinfo {author} {\bibfnamefont {J.}~\bibnamefont {Wang}}, \bibinfo
  {author} {\bibfnamefont {M.~J.}\ \bibnamefont {Strain}}, \bibinfo {author}
  {\bibfnamefont {B.}~\bibnamefont {Johnson-Morris}}, \bibinfo {author}
  {\bibfnamefont {J.}~\bibnamefont {Zhu}}, \bibinfo {author} {\bibfnamefont
  {M.}~\bibnamefont {Sorel}}, \bibinfo {author} {\bibfnamefont {J.~L.}\
  \bibnamefont {O{\textquoteright}Brien}}, \bibinfo {author} {\bibfnamefont
  {M.~G.}\ \bibnamefont {Thompson}},\ and\ \bibinfo {author} {\bibfnamefont
  {S.}~\bibnamefont {Yu}},\ }\bibfield  {title} {\bibinfo {title} {Integrated
  compact optical vortex beam emitters},\ }\href
  {https://doi.org/10.1126/science.1226528} {\bibfield  {journal} {\bibinfo
  {journal} {Science}\ }\textbf {\bibinfo {volume} {338}},\ \bibinfo {pages}
  {363} (\bibinfo {year} {2012})}\BibitemShut {NoStop}%
\bibitem [{\citenamefont {Wang}\ \emph {et~al.}(2018)\citenamefont {Wang},
  \citenamefont {Nie}, \citenamefont {Liang}, \citenamefont {Wang},
  \citenamefont {Li},\ and\ \citenamefont {Jia}}]{Wang2018Recent}%
  \BibitemOpen
  \bibfield  {author} {\bibinfo {author} {\bibfnamefont {X.}~\bibnamefont
  {Wang}}, \bibinfo {author} {\bibfnamefont {Z.}~\bibnamefont {Nie}}, \bibinfo
  {author} {\bibfnamefont {Y.}~\bibnamefont {Liang}}, \bibinfo {author}
  {\bibfnamefont {J.}~\bibnamefont {Wang}}, \bibinfo {author} {\bibfnamefont
  {T.}~\bibnamefont {Li}},\ and\ \bibinfo {author} {\bibfnamefont
  {B.}~\bibnamefont {Jia}},\ }\bibfield  {title} {\bibinfo {title} {Recent
  advances on optical vortex generation},\ }\href
  {https://doi.org/doi:10.1515/nanoph-2018-0072} {\bibfield  {journal}
  {\bibinfo  {journal} {Nanophotonics}\ }\textbf {\bibinfo {volume} {7}},\
  \bibinfo {pages} {1533} (\bibinfo {year} {2018})}\BibitemShut {NoStop}%
\bibitem [{\citenamefont {Chen}\ \emph {et~al.}(2021)\citenamefont {Chen},
  \citenamefont {Wei}, \citenamefont {Zhao}, \citenamefont {Liu}, \citenamefont
  {Su}, \citenamefont {Yao}, \citenamefont {Yu}, \citenamefont {Liu},\ and\
  \citenamefont {Wang}}]{Chen2021Bright}%
  \BibitemOpen
  \bibfield  {author} {\bibinfo {author} {\bibfnamefont {B.}~\bibnamefont
  {Chen}}, \bibinfo {author} {\bibfnamefont {Y.}~\bibnamefont {Wei}}, \bibinfo
  {author} {\bibfnamefont {T.}~\bibnamefont {Zhao}}, \bibinfo {author}
  {\bibfnamefont {S.}~\bibnamefont {Liu}}, \bibinfo {author} {\bibfnamefont
  {R.}~\bibnamefont {Su}}, \bibinfo {author} {\bibfnamefont {B.}~\bibnamefont
  {Yao}}, \bibinfo {author} {\bibfnamefont {Y.}~\bibnamefont {Yu}}, \bibinfo
  {author} {\bibfnamefont {J.}~\bibnamefont {Liu}},\ and\ \bibinfo {author}
  {\bibfnamefont {X.}~\bibnamefont {Wang}},\ }\bibfield  {title} {\bibinfo
  {title} {Bright solid-state sources for single photons with orbital angular
  momentum},\ }\href {https://doi.org/10.1038/s41565-020-00827-7} {\bibfield
  {journal} {\bibinfo  {journal} {Nature Nanotechnology}\ }\textbf {\bibinfo
  {volume} {16}},\ \bibinfo {pages} {302} (\bibinfo {year} {2021})}\BibitemShut
  {NoStop}%
\bibitem [{\citenamefont {Zhu}\ \emph {et~al.}(2021)\citenamefont {Zhu},
  \citenamefont {Tang}, \citenamefont {Li}, \citenamefont {Tai},\ and\
  \citenamefont {Li}}]{Zhu2021optical}%
  \BibitemOpen
  \bibfield  {author} {\bibinfo {author} {\bibfnamefont {L.}~\bibnamefont
  {Zhu}}, \bibinfo {author} {\bibfnamefont {M.}~\bibnamefont {Tang}}, \bibinfo
  {author} {\bibfnamefont {H.}~\bibnamefont {Li}}, \bibinfo {author}
  {\bibfnamefont {Y.}~\bibnamefont {Tai}},\ and\ \bibinfo {author}
  {\bibfnamefont {X.}~\bibnamefont {Li}},\ }\bibfield  {title} {\bibinfo
  {title} {Optical vortex lattice: an exploitation of orbital angular
  momentum},\ }\href {https://doi.org/doi:10.1515/nanoph-2021-0139} {\bibfield
  {journal} {\bibinfo  {journal} {Nanophotonics}\ }\textbf {\bibinfo {volume}
  {10}},\ \bibinfo {pages} {2487} (\bibinfo {year} {2021})}\BibitemShut
  {NoStop}%
\bibitem [{\citenamefont {Du}\ and\ \citenamefont {Wang}(2017)}]{Du2017Chip}%
  \BibitemOpen
  \bibfield  {author} {\bibinfo {author} {\bibfnamefont {J.}~\bibnamefont
  {Du}}\ and\ \bibinfo {author} {\bibfnamefont {J.}~\bibnamefont {Wang}},\
  }\bibfield  {title} {\bibinfo {title} {Chip-scale optical vortex lattice
  generator on a silicon platform},\ }\href
  {https://doi.org/10.1364/OL.42.005054} {\bibfield  {journal} {\bibinfo
  {journal} {Opt. Lett.}\ }\textbf {\bibinfo {volume} {42}},\ \bibinfo {pages}
  {5054} (\bibinfo {year} {2017})}\BibitemShut {NoStop}%
\bibitem [{\citenamefont {Jhajj}\ \emph {et~al.}(2016)\citenamefont {Jhajj},
  \citenamefont {Larkin}, \citenamefont {Rosenthal}, \citenamefont {Zahedpour},
  \citenamefont {Wahlstrand},\ and\ \citenamefont
  {Milchberg}}]{Jhajj2016spatiotemporal}%
  \BibitemOpen
  \bibfield  {author} {\bibinfo {author} {\bibfnamefont {N.}~\bibnamefont
  {Jhajj}}, \bibinfo {author} {\bibfnamefont {I.}~\bibnamefont {Larkin}},
  \bibinfo {author} {\bibfnamefont {E.~W.}\ \bibnamefont {Rosenthal}}, \bibinfo
  {author} {\bibfnamefont {S.}~\bibnamefont {Zahedpour}}, \bibinfo {author}
  {\bibfnamefont {J.~K.}\ \bibnamefont {Wahlstrand}},\ and\ \bibinfo {author}
  {\bibfnamefont {H.~M.}\ \bibnamefont {Milchberg}},\ }\bibfield  {title}
  {\bibinfo {title} {Spatiotemporal optical vortices},\ }\href
  {https://doi.org/10.1103/PhysRevX.6.031037} {\bibfield  {journal} {\bibinfo
  {journal} {Phys. Rev. X}\ }\textbf {\bibinfo {volume} {6}},\ \bibinfo {pages}
  {031037} (\bibinfo {year} {2016})}\BibitemShut {NoStop}%
\bibitem [{\citenamefont {Hancock}\ \emph {et~al.}(2019)\citenamefont
  {Hancock}, \citenamefont {Zahedpour}, \citenamefont {Goffin},\ and\
  \citenamefont {Milchberg}}]{Hancock2019free}%
  \BibitemOpen
  \bibfield  {author} {\bibinfo {author} {\bibfnamefont {S.~W.}\ \bibnamefont
  {Hancock}}, \bibinfo {author} {\bibfnamefont {S.}~\bibnamefont {Zahedpour}},
  \bibinfo {author} {\bibfnamefont {A.}~\bibnamefont {Goffin}},\ and\ \bibinfo
  {author} {\bibfnamefont {H.~M.}\ \bibnamefont {Milchberg}},\ }\bibfield
  {title} {\bibinfo {title} {Free-space propagation of spatiotemporal optical
  vortices},\ }\href {https://doi.org/10.1364/OPTICA.6.001547} {\bibfield
  {journal} {\bibinfo  {journal} {Optica}\ }\textbf {\bibinfo {volume} {6}},\
  \bibinfo {pages} {1547} (\bibinfo {year} {2019})}\BibitemShut {NoStop}%
\bibitem [{\citenamefont {Chong}\ \emph {et~al.}(2020)\citenamefont {Chong},
  \citenamefont {Wan}, \citenamefont {Chen},\ and\ \citenamefont
  {Zhan}}]{Chong2020Generation}%
  \BibitemOpen
  \bibfield  {author} {\bibinfo {author} {\bibfnamefont {A.}~\bibnamefont
  {Chong}}, \bibinfo {author} {\bibfnamefont {C.}~\bibnamefont {Wan}}, \bibinfo
  {author} {\bibfnamefont {J.}~\bibnamefont {Chen}},\ and\ \bibinfo {author}
  {\bibfnamefont {Q.}~\bibnamefont {Zhan}},\ }\bibfield  {title} {\bibinfo
  {title} {Generation of spatiotemporal optical vortices with controllable
  transverse orbital angular momentum},\ }\href
  {https://doi.org/10.1038/s41566-020-0587-z} {\bibfield  {journal} {\bibinfo
  {journal} {Nature Photonics}\ }\textbf {\bibinfo {volume} {14}},\ \bibinfo
  {pages} {350} (\bibinfo {year} {2020})}\BibitemShut {NoStop}%
\bibitem [{\citenamefont {Gahagan}\ and\ \citenamefont
  {Swartzlander}(1996)}]{Gahagan1996Optical}%
  \BibitemOpen
  \bibfield  {author} {\bibinfo {author} {\bibfnamefont {K.~T.}\ \bibnamefont
  {Gahagan}}\ and\ \bibinfo {author} {\bibfnamefont {G.~A.}\ \bibnamefont
  {Swartzlander}},\ }\bibfield  {title} {\bibinfo {title} {Optical vortex
  trapping of particles},\ }\href {https://doi.org/10.1364/OL.21.000827}
  {\bibfield  {journal} {\bibinfo  {journal} {Opt. Lett.}\ }\textbf {\bibinfo
  {volume} {21}},\ \bibinfo {pages} {827} (\bibinfo {year} {1996})}\BibitemShut
  {NoStop}%
\bibitem [{\citenamefont {Malik}\ \emph {et~al.}(2016)\citenamefont {Malik},
  \citenamefont {Erhard}, \citenamefont {Huber}, \citenamefont {Krenn},
  \citenamefont {Fickler},\ and\ \citenamefont
  {Zeilinger}}]{Malik2016multiphoton}%
  \BibitemOpen
  \bibfield  {author} {\bibinfo {author} {\bibfnamefont {M.}~\bibnamefont
  {Malik}}, \bibinfo {author} {\bibfnamefont {M.}~\bibnamefont {Erhard}},
  \bibinfo {author} {\bibfnamefont {M.}~\bibnamefont {Huber}}, \bibinfo
  {author} {\bibfnamefont {M.}~\bibnamefont {Krenn}}, \bibinfo {author}
  {\bibfnamefont {R.}~\bibnamefont {Fickler}},\ and\ \bibinfo {author}
  {\bibfnamefont {A.}~\bibnamefont {Zeilinger}},\ }\bibfield  {title} {\bibinfo
  {title} {Multi-photon entanglement in high dimensions},\ }\href
  {https://doi.org/10.1038/nphoton.2016.12} {\bibfield  {journal} {\bibinfo
  {journal} {Nature Photonics}\ }\textbf {\bibinfo {volume} {10}},\ \bibinfo
  {pages} {248} (\bibinfo {year} {2016})}\BibitemShut {NoStop}%
\bibitem [{\citenamefont {Ding}\ \emph {et~al.}(2015)\citenamefont {Ding},
  \citenamefont {Zhang}, \citenamefont {Zhou}, \citenamefont {Shi},
  \citenamefont {Xiang}, \citenamefont {Wang}, \citenamefont {Jiang},
  \citenamefont {Shi},\ and\ \citenamefont {Guo}}]{Ding2015Quantum}%
  \BibitemOpen
  \bibfield  {author} {\bibinfo {author} {\bibfnamefont {D.-S.}\ \bibnamefont
  {Ding}}, \bibinfo {author} {\bibfnamefont {W.}~\bibnamefont {Zhang}},
  \bibinfo {author} {\bibfnamefont {Z.-Y.}\ \bibnamefont {Zhou}}, \bibinfo
  {author} {\bibfnamefont {S.}~\bibnamefont {Shi}}, \bibinfo {author}
  {\bibfnamefont {G.-Y.}\ \bibnamefont {Xiang}}, \bibinfo {author}
  {\bibfnamefont {X.-S.}\ \bibnamefont {Wang}}, \bibinfo {author}
  {\bibfnamefont {Y.-K.}\ \bibnamefont {Jiang}}, \bibinfo {author}
  {\bibfnamefont {B.-S.}\ \bibnamefont {Shi}},\ and\ \bibinfo {author}
  {\bibfnamefont {G.-C.}\ \bibnamefont {Guo}},\ }\bibfield  {title} {\bibinfo
  {title} {Quantum storage of orbital angular momentum entanglement in an
  atomic ensemble},\ }\href {https://doi.org/10.1103/PhysRevLett.114.050502}
  {\bibfield  {journal} {\bibinfo  {journal} {Phys. Rev. Lett.}\ }\textbf
  {\bibinfo {volume} {114}},\ \bibinfo {pages} {050502} (\bibinfo {year}
  {2015})}\BibitemShut {NoStop}%
\bibitem [{\citenamefont {Malik}\ \emph {et~al.}(2014)\citenamefont {Malik},
  \citenamefont {Mirhosseini}, \citenamefont {Lavery}, \citenamefont {Leach},
  \citenamefont {Padgett},\ and\ \citenamefont {Boyd}}]{Malik2014Direct}%
  \BibitemOpen
  \bibfield  {author} {\bibinfo {author} {\bibfnamefont {M.}~\bibnamefont
  {Malik}}, \bibinfo {author} {\bibfnamefont {M.}~\bibnamefont {Mirhosseini}},
  \bibinfo {author} {\bibfnamefont {M.~P.~J.}\ \bibnamefont {Lavery}}, \bibinfo
  {author} {\bibfnamefont {J.}~\bibnamefont {Leach}}, \bibinfo {author}
  {\bibfnamefont {M.~J.}\ \bibnamefont {Padgett}},\ and\ \bibinfo {author}
  {\bibfnamefont {R.~W.}\ \bibnamefont {Boyd}},\ }\bibfield  {title} {\bibinfo
  {title} {Direct measurement of a 27-dimensional orbital-angular-momentum
  state vector},\ }\href {https://doi.org/10.1038/ncomms4115} {\bibfield
  {journal} {\bibinfo  {journal} {Nature Communications}\ }\textbf {\bibinfo
  {volume} {5}},\ \bibinfo {pages} {3115} (\bibinfo {year} {2014})}\BibitemShut
  {NoStop}%
\bibitem [{\citenamefont {Babazadeh}\ \emph {et~al.}(2017)\citenamefont
  {Babazadeh}, \citenamefont {Erhard}, \citenamefont {Wang}, \citenamefont
  {Malik}, \citenamefont {Nouroozi}, \citenamefont {Krenn},\ and\ \citenamefont
  {Zeilinger}}]{Babazadeh2017high}%
  \BibitemOpen
  \bibfield  {author} {\bibinfo {author} {\bibfnamefont {A.}~\bibnamefont
  {Babazadeh}}, \bibinfo {author} {\bibfnamefont {M.}~\bibnamefont {Erhard}},
  \bibinfo {author} {\bibfnamefont {F.}~\bibnamefont {Wang}}, \bibinfo {author}
  {\bibfnamefont {M.}~\bibnamefont {Malik}}, \bibinfo {author} {\bibfnamefont
  {R.}~\bibnamefont {Nouroozi}}, \bibinfo {author} {\bibfnamefont
  {M.}~\bibnamefont {Krenn}},\ and\ \bibinfo {author} {\bibfnamefont
  {A.}~\bibnamefont {Zeilinger}},\ }\bibfield  {title} {\bibinfo {title}
  {High-dimensional single-photon quantum gates: Concepts and experiments},\
  }\href {https://doi.org/10.1103/PhysRevLett.119.180510} {\bibfield  {journal}
  {\bibinfo  {journal} {Phys. Rev. Lett.}\ }\textbf {\bibinfo {volume} {119}},\
  \bibinfo {pages} {180510} (\bibinfo {year} {2017})}\BibitemShut {NoStop}%
\bibitem [{\citenamefont {Zhuang}(2004)}]{Zhuang2004Unraveling}%
  \BibitemOpen
  \bibfield  {author} {\bibinfo {author} {\bibfnamefont {X.}~\bibnamefont
  {Zhuang}},\ }\bibfield  {title} {\bibinfo {title} {Unraveling dna
  condensation with optical tweezers},\ }\href
  {https://doi.org/10.1126/science.1100603} {\bibfield  {journal} {\bibinfo
  {journal} {Science}\ }\textbf {\bibinfo {volume} {305}},\ \bibinfo {pages}
  {188} (\bibinfo {year} {2004})}\BibitemShut {NoStop}%
\bibitem [{\citenamefont {Fang}\ \emph {et~al.}(2021)\citenamefont {Fang},
  \citenamefont {Han}, \citenamefont {Ge}, \citenamefont {Guo}, \citenamefont
  {Yu}, \citenamefont {Deng}, \citenamefont {Wu}, \citenamefont {Gong},\ and\
  \citenamefont {Liu}}]{Fang2021photo}%
  \BibitemOpen
  \bibfield  {author} {\bibinfo {author} {\bibfnamefont {Y.}~\bibnamefont
  {Fang}}, \bibinfo {author} {\bibfnamefont {M.}~\bibnamefont {Han}}, \bibinfo
  {author} {\bibfnamefont {P.}~\bibnamefont {Ge}}, \bibinfo {author}
  {\bibfnamefont {Z.}~\bibnamefont {Guo}}, \bibinfo {author} {\bibfnamefont
  {X.}~\bibnamefont {Yu}}, \bibinfo {author} {\bibfnamefont {Y.}~\bibnamefont
  {Deng}}, \bibinfo {author} {\bibfnamefont {C.}~\bibnamefont {Wu}}, \bibinfo
  {author} {\bibfnamefont {Q.}~\bibnamefont {Gong}},\ and\ \bibinfo {author}
  {\bibfnamefont {Y.}~\bibnamefont {Liu}},\ }\bibfield  {title} {\bibinfo
  {title} {Photoelectronic mapping of the spin--orbit interaction of intense
  light fields},\ }\href {https://doi.org/10.1038/s41566-020-00709-3}
  {\bibfield  {journal} {\bibinfo  {journal} {Nature Photonics}\ }\textbf
  {\bibinfo {volume} {15}},\ \bibinfo {pages} {115} (\bibinfo {year}
  {2021})}\BibitemShut {NoStop}%
\bibitem [{\citenamefont {Coullet}\ \emph {et~al.}(1989)\citenamefont
  {Coullet}, \citenamefont {Gil},\ and\ \citenamefont
  {Rocca}}]{COULLET1989optical}%
  \BibitemOpen
  \bibfield  {author} {\bibinfo {author} {\bibfnamefont {P.}~\bibnamefont
  {Coullet}}, \bibinfo {author} {\bibfnamefont {L.}~\bibnamefont {Gil}},\ and\
  \bibinfo {author} {\bibfnamefont {F.}~\bibnamefont {Rocca}},\ }\bibfield
  {title} {\bibinfo {title} {Optical vortices},\ }\href
  {https://doi.org/https://doi.org/10.1016/0030-4018(89)90180-6} {\bibfield
  {journal} {\bibinfo  {journal} {Optics Communications}\ }\textbf {\bibinfo
  {volume} {73}},\ \bibinfo {pages} {403} (\bibinfo {year} {1989})}\BibitemShut
  {NoStop}%
\bibitem [{\citenamefont {Soskin}\ \emph {et~al.}(1997)\citenamefont {Soskin},
  \citenamefont {Gorshkov}, \citenamefont {Vasnetsov}, \citenamefont {Malos},\
  and\ \citenamefont {Heckenberg}}]{Soskin1997topological}%
  \BibitemOpen
  \bibfield  {author} {\bibinfo {author} {\bibfnamefont {M.~S.}\ \bibnamefont
  {Soskin}}, \bibinfo {author} {\bibfnamefont {V.~N.}\ \bibnamefont
  {Gorshkov}}, \bibinfo {author} {\bibfnamefont {M.~V.}\ \bibnamefont
  {Vasnetsov}}, \bibinfo {author} {\bibfnamefont {J.~T.}\ \bibnamefont
  {Malos}},\ and\ \bibinfo {author} {\bibfnamefont {N.~R.}\ \bibnamefont
  {Heckenberg}},\ }\bibfield  {title} {\bibinfo {title} {Topological charge and
  angular momentum of light beams carrying optical vortices},\ }\href
  {https://doi.org/10.1103/PhysRevA.56.4064} {\bibfield  {journal} {\bibinfo
  {journal} {Phys. Rev. A}\ }\textbf {\bibinfo {volume} {56}},\ \bibinfo
  {pages} {4064} (\bibinfo {year} {1997})}\BibitemShut {NoStop}%
\bibitem [{\citenamefont {Berry}\ and\ \citenamefont
  {Dennis}(2001)}]{Berry2001Polarization}%
  \BibitemOpen
  \bibfield  {author} {\bibinfo {author} {\bibfnamefont {M.~V.}\ \bibnamefont
  {Berry}}\ and\ \bibinfo {author} {\bibfnamefont {M.~R.}\ \bibnamefont
  {Dennis}},\ }\bibfield  {title} {\bibinfo {title} {Polarization singularities
  in isotropic random vector waves},\ }\href
  {https://doi.org/10.1098/rspa.2000.0660} {\bibfield  {journal} {\bibinfo
  {journal} {Proceedings of the Royal Society A: Mathematical, Physical and
  Engineering Science}\ }\textbf {\bibinfo {volume} {457}},\ \bibinfo {pages}
  {141} (\bibinfo {year} {2001})}\BibitemShut {NoStop}%
\bibitem [{\citenamefont {Berry}(2004)}]{Berry2004optical}%
  \BibitemOpen
  \bibfield  {author} {\bibinfo {author} {\bibfnamefont {M.~V.}\ \bibnamefont
  {Berry}},\ }\bibfield  {title} {\bibinfo {title} {Optical vortices evolving
  from helicoidal integer and fractional phase steps},\ }\href
  {https://doi.org/10.1088/1464-4258/6/2/018} {\bibfield  {journal} {\bibinfo
  {journal} {Journal of Optics A: Pure and Applied Optics}\ }\textbf {\bibinfo
  {volume} {6}},\ \bibinfo {pages} {259} (\bibinfo {year} {2004})}\BibitemShut
  {NoStop}%
\bibitem [{\citenamefont {Dennis}(2002)}]{Dennis2002polarization}%
  \BibitemOpen
  \bibfield  {author} {\bibinfo {author} {\bibfnamefont {M.}~\bibnamefont
  {Dennis}},\ }\bibfield  {title} {\bibinfo {title} {Polarization singularities
  in paraxial vector fields: morphology and statistics},\ }\href
  {https://doi.org/https://doi.org/10.1016/S0030-4018(02)02088-6} {\bibfield
  {journal} {\bibinfo  {journal} {Optics Communications}\ }\textbf {\bibinfo
  {volume} {213}},\ \bibinfo {pages} {201} (\bibinfo {year}
  {2002})}\BibitemShut {NoStop}%
\bibitem [{\citenamefont {Freund}(2002)}]{freund2002polarization}%
  \BibitemOpen
  \bibfield  {author} {\bibinfo {author} {\bibfnamefont {I.}~\bibnamefont
  {Freund}},\ }\bibfield  {title} {\bibinfo {title} {Polarization singularity
  indices in gaussian laser beams},\ }\href
  {https://doi.org/https://doi.org/10.1016/S0030-4018(01)01725-4} {\bibfield
  {journal} {\bibinfo  {journal} {Optics Communications}\ }\textbf {\bibinfo
  {volume} {201}},\ \bibinfo {pages} {251} (\bibinfo {year}
  {2002})}\BibitemShut {NoStop}%
\bibitem [{\citenamefont {Kotlyar}\ \emph {et~al.}(2021)\citenamefont
  {Kotlyar}, \citenamefont {Kovalev},\ and\ \citenamefont
  {Nalimov}}]{Kotlyar2020Topological}%
  \BibitemOpen
  \bibfield  {author} {\bibinfo {author} {\bibfnamefont {V.~V.}\ \bibnamefont
  {Kotlyar}}, \bibinfo {author} {\bibfnamefont {A.~A.}\ \bibnamefont
  {Kovalev}},\ and\ \bibinfo {author} {\bibfnamefont {A.~G.}\ \bibnamefont
  {Nalimov}},\ }\bibfield  {title} {\bibinfo {title} {Conservation of the
  half-integer topological charge on propagation of a superposition of two
  bessel-gaussian beams},\ }\href {https://doi.org/10.1103/PhysRevA.104.033507}
  {\bibfield  {journal} {\bibinfo  {journal} {Phys. Rev. A}\ }\textbf {\bibinfo
  {volume} {104}},\ \bibinfo {pages} {033507} (\bibinfo {year}
  {2021})}\BibitemShut {NoStop}%
\bibitem [{\citenamefont {Wen}\ \emph {et~al.}(2019)\citenamefont {Wen},
  \citenamefont {Wang}, \citenamefont {Yang}, \citenamefont {Zhang},\ and\
  \citenamefont {Zhu}}]{Wen2019vortex}%
  \BibitemOpen
  \bibfield  {author} {\bibinfo {author} {\bibfnamefont {J.}~\bibnamefont
  {Wen}}, \bibinfo {author} {\bibfnamefont {L.-G.}\ \bibnamefont {Wang}},
  \bibinfo {author} {\bibfnamefont {X.}~\bibnamefont {Yang}}, \bibinfo {author}
  {\bibfnamefont {J.}~\bibnamefont {Zhang}},\ and\ \bibinfo {author}
  {\bibfnamefont {S.-Y.}\ \bibnamefont {Zhu}},\ }\bibfield  {title} {\bibinfo
  {title} {Vortex strength and beam propagation factor of fractional vortex
  beams},\ }\href {https://doi.org/10.1364/OE.27.005893} {\bibfield  {journal}
  {\bibinfo  {journal} {Opt. Express}\ }\textbf {\bibinfo {volume} {27}},\
  \bibinfo {pages} {5893} (\bibinfo {year} {2019})}\BibitemShut {NoStop}%
\bibitem [{\citenamefont {Leach}\ \emph {et~al.}(2004)\citenamefont {Leach},
  \citenamefont {Yao},\ and\ \citenamefont {Padgett}}]{Leach2004observation}%
  \BibitemOpen
  \bibfield  {author} {\bibinfo {author} {\bibfnamefont {J.}~\bibnamefont
  {Leach}}, \bibinfo {author} {\bibfnamefont {E.}~\bibnamefont {Yao}},\ and\
  \bibinfo {author} {\bibfnamefont {M.~J.}\ \bibnamefont {Padgett}},\
  }\bibfield  {title} {\bibinfo {title} {Observation of the vortex structure of
  a non-integer vortex beam},\ }\href
  {https://doi.org/10.1088/1367-2630/6/1/071} {\bibfield  {journal} {\bibinfo
  {journal} {New Journal of Physics}\ }\textbf {\bibinfo {volume} {6}},\
  \bibinfo {pages} {71} (\bibinfo {year} {2004})}\BibitemShut {NoStop}%
\bibitem [{\citenamefont {Tao}\ \emph {et~al.}(2005)\citenamefont {Tao},
  \citenamefont {Yuan}, \citenamefont {Lin}, \citenamefont {Peng},\ and\
  \citenamefont {Niu}}]{Tao2005Fractional}%
  \BibitemOpen
  \bibfield  {author} {\bibinfo {author} {\bibfnamefont {S.~H.}\ \bibnamefont
  {Tao}}, \bibinfo {author} {\bibfnamefont {X.-C.}\ \bibnamefont {Yuan}},
  \bibinfo {author} {\bibfnamefont {J.}~\bibnamefont {Lin}}, \bibinfo {author}
  {\bibfnamefont {X.}~\bibnamefont {Peng}},\ and\ \bibinfo {author}
  {\bibfnamefont {H.~B.}\ \bibnamefont {Niu}},\ }\bibfield  {title} {\bibinfo
  {title} {Fractional optical vortex beam induced rotation of particles},\
  }\href {https://doi.org/10.1364/OPEX.13.007726} {\bibfield  {journal}
  {\bibinfo  {journal} {Opt. Express}\ }\textbf {\bibinfo {volume} {13}},\
  \bibinfo {pages} {7726} (\bibinfo {year} {2005})}\BibitemShut {NoStop}%
\bibitem [{\citenamefont {Guo}\ \emph {et~al.}(2010)\citenamefont {Guo},
  \citenamefont {Yu},\ and\ \citenamefont {Hong}}]{Guo2010Optical}%
  \BibitemOpen
  \bibfield  {author} {\bibinfo {author} {\bibfnamefont {C.-S.}\ \bibnamefont
  {Guo}}, \bibinfo {author} {\bibfnamefont {Y.-N.}\ \bibnamefont {Yu}},\ and\
  \bibinfo {author} {\bibfnamefont {Z.}~\bibnamefont {Hong}},\ }\bibfield
  {title} {\bibinfo {title} {Optical sorting using an array of optical vortices
  with fractional topological charge},\ }\href
  {https://doi.org/https://doi.org/10.1016/j.optcom.2009.12.063} {\bibfield
  {journal} {\bibinfo  {journal} {Optics Communications}\ }\textbf {\bibinfo
  {volume} {283}},\ \bibinfo {pages} {1889} (\bibinfo {year}
  {2010})}\BibitemShut {NoStop}%
\bibitem [{\citenamefont {Fang}\ \emph {et~al.}(2017)\citenamefont {Fang},
  \citenamefont {Lu}, \citenamefont {Wang}, \citenamefont {Zhang},\ and\
  \citenamefont {Chen}}]{Fang2017Fractional}%
  \BibitemOpen
  \bibfield  {author} {\bibinfo {author} {\bibfnamefont {Y.}~\bibnamefont
  {Fang}}, \bibinfo {author} {\bibfnamefont {Q.}~\bibnamefont {Lu}}, \bibinfo
  {author} {\bibfnamefont {X.}~\bibnamefont {Wang}}, \bibinfo {author}
  {\bibfnamefont {W.}~\bibnamefont {Zhang}},\ and\ \bibinfo {author}
  {\bibfnamefont {L.}~\bibnamefont {Chen}},\ }\bibfield  {title} {\bibinfo
  {title} {Fractional-topological-charge-induced vortex birth and splitting of
  light fields on the submicron scale},\ }\href
  {https://doi.org/10.1103/PhysRevA.95.023821} {\bibfield  {journal} {\bibinfo
  {journal} {Phys. Rev. A}\ }\textbf {\bibinfo {volume} {95}},\ \bibinfo
  {pages} {023821} (\bibinfo {year} {2017})}\BibitemShut {NoStop}%
\bibitem [{\citenamefont {Dennis}\ \emph {et~al.}(2009)\citenamefont {Dennis},
  \citenamefont {O'Holleran},\ and\ \citenamefont
  {Padgett}}]{Dennis2009Chapter}%
  \BibitemOpen
  \bibfield  {author} {\bibinfo {author} {\bibfnamefont {M.~R.}\ \bibnamefont
  {Dennis}}, \bibinfo {author} {\bibfnamefont {K.}~\bibnamefont {O'Holleran}},\
  and\ \bibinfo {author} {\bibfnamefont {M.~J.}\ \bibnamefont {Padgett}},\
  }\bibfield  {title} {\bibinfo {title} {Chapter 5 singular optics: Optical
  vortices and polarization singularities},\ }\href
  {https://doi.org/10.1016/S0079-6638(08)00205-9} {\bibfield  {journal}
  {\bibinfo  {journal} {Progress in Optics}\ }\textbf {\bibinfo {volume}
  {53}},\ \bibinfo {pages} {293} (\bibinfo {year} {2009})}\BibitemShut
  {NoStop}%
\bibitem [{\citenamefont {Siviloglou}\ \emph {et~al.}(2007)\citenamefont
  {Siviloglou}, \citenamefont {Broky}, \citenamefont {Dogariu},\ and\
  \citenamefont {Christodoulides}}]{Siviloglou2007observation}%
  \BibitemOpen
  \bibfield  {author} {\bibinfo {author} {\bibfnamefont {G.~A.}\ \bibnamefont
  {Siviloglou}}, \bibinfo {author} {\bibfnamefont {J.}~\bibnamefont {Broky}},
  \bibinfo {author} {\bibfnamefont {A.}~\bibnamefont {Dogariu}},\ and\ \bibinfo
  {author} {\bibfnamefont {D.~N.}\ \bibnamefont {Christodoulides}},\ }\bibfield
   {title} {\bibinfo {title} {Observation of accelerating airy beams},\ }\href
  {https://doi.org/10.1103/PhysRevLett.99.213901} {\bibfield  {journal}
  {\bibinfo  {journal} {Phys. Rev. Lett.}\ }\textbf {\bibinfo {volume} {99}},\
  \bibinfo {pages} {213901} (\bibinfo {year} {2007})}\BibitemShut {NoStop}%
\bibitem [{\citenamefont {Wan}\ \emph {et~al.}(2007)\citenamefont {Wan},
  \citenamefont {Jia},\ and\ \citenamefont {Fleischer}}]{Wan2007dispersive}%
  \BibitemOpen
  \bibfield  {author} {\bibinfo {author} {\bibfnamefont {W.}~\bibnamefont
  {Wan}}, \bibinfo {author} {\bibfnamefont {S.}~\bibnamefont {Jia}},\ and\
  \bibinfo {author} {\bibfnamefont {J.~W.}\ \bibnamefont {Fleischer}},\
  }\bibfield  {title} {\bibinfo {title} {Dispersive superfluid-like shock waves
  in nonlinear optics},\ }\href {https://doi.org/10.1038/nphys486} {\bibfield
  {journal} {\bibinfo  {journal} {Nature Physics}\ }\textbf {\bibinfo {volume}
  {3}},\ \bibinfo {pages} {46} (\bibinfo {year} {2007})}\BibitemShut {NoStop}%
\bibitem [{\citenamefont {Zhang}\ \emph {et~al.}(2019)\citenamefont {Zhang},
  \citenamefont {Li}, \citenamefont {Malpuech}, \citenamefont {Zhang},
  \citenamefont {Bleu}, \citenamefont {Koniakhin}, \citenamefont {Li},
  \citenamefont {Zhang}, \citenamefont {Xiao},\ and\ \citenamefont
  {Solnyshkov}}]{zhang2019particle}%
  \BibitemOpen
  \bibfield  {author} {\bibinfo {author} {\bibfnamefont {Z.}~\bibnamefont
  {Zhang}}, \bibinfo {author} {\bibfnamefont {F.}~\bibnamefont {Li}}, \bibinfo
  {author} {\bibfnamefont {G.}~\bibnamefont {Malpuech}}, \bibinfo {author}
  {\bibfnamefont {Y.}~\bibnamefont {Zhang}}, \bibinfo {author} {\bibfnamefont
  {O.}~\bibnamefont {Bleu}}, \bibinfo {author} {\bibfnamefont {S.}~\bibnamefont
  {Koniakhin}}, \bibinfo {author} {\bibfnamefont {C.}~\bibnamefont {Li}},
  \bibinfo {author} {\bibfnamefont {Y.}~\bibnamefont {Zhang}}, \bibinfo
  {author} {\bibfnamefont {M.}~\bibnamefont {Xiao}},\ and\ \bibinfo {author}
  {\bibfnamefont {D.~D.}\ \bibnamefont {Solnyshkov}},\ }\bibfield  {title}
  {\bibinfo {title} {Particlelike behavior of topological defects in linear
  wave packets in photonic graphene},\ }\href
  {https://doi.org/10.1103/PhysRevLett.122.233905} {\bibfield  {journal}
  {\bibinfo  {journal} {Phys. Rev. Lett.}\ }\textbf {\bibinfo {volume} {122}},\
  \bibinfo {pages} {233905} (\bibinfo {year} {2019})}\BibitemShut {NoStop}%
\bibitem [{\citenamefont {Glauber}(1963)}]{Glauber1963coherence}%
  \BibitemOpen
  \bibfield  {author} {\bibinfo {author} {\bibfnamefont {R.~J.}\ \bibnamefont
  {Glauber}},\ }\bibfield  {title} {\bibinfo {title} {The quantum theory of
  optical coherence},\ }\href {https://doi.org/10.1103/PhysRev.130.2529}
  {\bibfield  {journal} {\bibinfo  {journal} {Phys. Rev.}\ }\textbf {\bibinfo
  {volume} {130}},\ \bibinfo {pages} {2529} (\bibinfo {year}
  {1963})}\BibitemShut {NoStop}%
\bibitem [{\citenamefont {Soskin}\ and\ \citenamefont
  {Vasnetsov}(2001)}]{Soskin2001Singular}%
  \BibitemOpen
  \bibfield  {author} {\bibinfo {author} {\bibfnamefont {M.~S.}\ \bibnamefont
  {Soskin}}\ and\ \bibinfo {author} {\bibfnamefont {M.~V.}\ \bibnamefont
  {Vasnetsov}},\ }\bibfield  {title} {\bibinfo {title} {Singular optics},\
  }\href {https://doi.org/10.1016/S0079-6638(01)80018-4} {\bibfield  {journal}
  {\bibinfo  {journal} {Progress in Optics}\ }\textbf {\bibinfo {volume}
  {42}},\ \bibinfo {pages} {219} (\bibinfo {year} {2001})}\BibitemShut
  {NoStop}%
\bibitem [{\citenamefont {Yang}\ and\ \citenamefont
  {Jacob}(2021)}]{yang2021quantum}%
  \BibitemOpen
  \bibfield  {author} {\bibinfo {author} {\bibfnamefont {L.-P.}\ \bibnamefont
  {Yang}}\ and\ \bibinfo {author} {\bibfnamefont {Z.}~\bibnamefont {Jacob}},\
  }\bibfield  {title} {\bibinfo {title} {Non-classical photonic spin texture of
  quantum structured light},\ }\href
  {https://doi.org/10.1038/s42005-021-00726-w} {\bibfield  {journal} {\bibinfo
  {journal} {Communications Physics}\ }\textbf {\bibinfo {volume} {4}},\
  \bibinfo {pages} {221} (\bibinfo {year} {2021})}\BibitemShut {NoStop}%
\bibitem [{\citenamefont {Berestetskii}\ \emph {et~al.}(1982)\citenamefont
  {Berestetskii}, \citenamefont {Lifshitz},\ and\ \citenamefont
  {Pitaevski}}]{QED1982Landau}%
  \BibitemOpen
  \bibfield  {author} {\bibinfo {author} {\bibfnamefont {V.~B.}\ \bibnamefont
  {Berestetskii}}, \bibinfo {author} {\bibfnamefont {E.~M.}\ \bibnamefont
  {Lifshitz}},\ and\ \bibinfo {author} {\bibfnamefont {L.~P.}\ \bibnamefont
  {Pitaevski}},\ }\href@noop {} {\emph {\bibinfo {title} {Quantum
  Electrodynamics}}},\ \bibinfo {edition} {second edition}\ ed.\ (\bibinfo
  {publisher} {Pergamon Press},\ \bibinfo {address} {Oxford},\ \bibinfo {year}
  {1982})\ Chap.~\bibinfo {chapter} {4}\BibitemShut {NoStop}%
\bibitem [{\citenamefont {Basistiy}\ \emph {et~al.}(2004)\citenamefont
  {Basistiy}, \citenamefont {ko}, \citenamefont {Slyusar}, \citenamefont
  {Soskin},\ and\ \citenamefont {Vasnetsov}}]{Basistiy2004Synthesis}%
  \BibitemOpen
  \bibfield  {author} {\bibinfo {author} {\bibfnamefont {I.~V.}\ \bibnamefont
  {Basistiy}}, \bibinfo {author} {\bibfnamefont {V.~A.~P.}\ \bibnamefont {ko}},
  \bibinfo {author} {\bibfnamefont {V.~V.}\ \bibnamefont {Slyusar}}, \bibinfo
  {author} {\bibfnamefont {M.~S.}\ \bibnamefont {Soskin}},\ and\ \bibinfo
  {author} {\bibfnamefont {M.~V.}\ \bibnamefont {Vasnetsov}},\ }\bibfield
  {title} {\bibinfo {title} {Synthesis and analysis of optical vortices with
  fractional topological charges},\ }\href
  {https://doi.org/10.1088/1464-4258/6/5/003} {\bibfield  {journal} {\bibinfo
  {journal} {Journal of Optics A: Pure and Applied Optics}\ }\textbf {\bibinfo
  {volume} {6}},\ \bibinfo {pages} {S166} (\bibinfo {year} {2004})}\BibitemShut
  {NoStop}%
\bibitem [{\citenamefont {Annett}(2005)}]{Annett2005super}%
  \BibitemOpen
  \bibfield  {author} {\bibinfo {author} {\bibfnamefont {J.~F.}\ \bibnamefont
  {Annett}},\ }\href@noop {} {\emph {\bibinfo {title} {Superconductivity,
  superfluids and condensates}}}\ (\bibinfo  {publisher} {Oxford University
  Press},\ \bibinfo {year} {2005})\ \bibinfo {note} {chap. 2}\BibitemShut
  {NoStop}%
\bibitem [{\citenamefont {Pethick}\ and\ \citenamefont
  {Smith}(2008)}]{pethick2008bose}%
  \BibitemOpen
  \bibfield  {author} {\bibinfo {author} {\bibfnamefont {C.~J.}\ \bibnamefont
  {Pethick}}\ and\ \bibinfo {author} {\bibfnamefont {H.}~\bibnamefont
  {Smith}},\ }\href@noop {} {\emph {\bibinfo {title} {Bose--Einstein
  condensation in dilute gases}}}\ (\bibinfo  {publisher} {Cambridge university
  press},\ \bibinfo {year} {2008})\ \bibinfo {note} {chap. 7}\BibitemShut
  {NoStop}%
\bibitem [{\citenamefont {Siegman}(1971)}]{Siegman1971An}%
  \BibitemOpen
  \bibfield  {author} {\bibinfo {author} {\bibfnamefont {A.~E.}\ \bibnamefont
  {Siegman}},\ }\href@noop {} {\emph {\bibinfo {title} {An introduction to
  lasers and masers}}}\ (\bibinfo  {publisher} {McGraw-Hill},\ \bibinfo {year}
  {1971})\ \bibinfo {note} {chap. 16}\BibitemShut {NoStop}%
\bibitem [{\citenamefont {Zhan}(2009)}]{Zhan2009cylindrical}%
  \BibitemOpen
  \bibfield  {author} {\bibinfo {author} {\bibfnamefont {Q.}~\bibnamefont
  {Zhan}},\ }\bibfield  {title} {\bibinfo {title} {Cylindrical vector beams:
  from mathematical concepts to applications},\ }\href
  {https://doi.org/10.1364/AOP.1.000001} {\bibfield  {journal} {\bibinfo
  {journal} {Adv. Opt. Photon.}\ }\textbf {\bibinfo {volume} {1}},\ \bibinfo
  {pages} {1} (\bibinfo {year} {2009})}\BibitemShut {NoStop}%
\bibitem [{\citenamefont {Rozas}\ \emph
  {et~al.}(1997{\natexlab{a}})\citenamefont {Rozas}, \citenamefont {Sacks},\
  and\ \citenamefont {Swartzlander}}]{Rozas1997experimental}%
  \BibitemOpen
  \bibfield  {author} {\bibinfo {author} {\bibfnamefont {D.}~\bibnamefont
  {Rozas}}, \bibinfo {author} {\bibfnamefont {Z.~S.}\ \bibnamefont {Sacks}},\
  and\ \bibinfo {author} {\bibfnamefont {G.~A.}\ \bibnamefont {Swartzlander}},\
  }\bibfield  {title} {\bibinfo {title} {Experimental observation of fluidlike
  motion of optical vortices},\ }\href
  {https://doi.org/10.1103/PhysRevLett.79.3399} {\bibfield  {journal} {\bibinfo
   {journal} {Phys. Rev. Lett.}\ }\textbf {\bibinfo {volume} {79}},\ \bibinfo
  {pages} {3399} (\bibinfo {year} {1997}{\natexlab{a}})}\BibitemShut {NoStop}%
\bibitem [{\citenamefont {An}\ \emph {et~al.}(2012)\citenamefont {An},
  \citenamefont {Liu}, \citenamefont {Lin},\ and\ \citenamefont
  {Liu}}]{An2012universal}%
  \BibitemOpen
  \bibfield  {author} {\bibinfo {author} {\bibfnamefont {Z.}~\bibnamefont
  {An}}, \bibinfo {author} {\bibfnamefont {F.~Q.}\ \bibnamefont {Liu}},
  \bibinfo {author} {\bibfnamefont {Y.}~\bibnamefont {Lin}},\ and\ \bibinfo
  {author} {\bibfnamefont {C.}~\bibnamefont {Liu}},\ }\bibfield  {title}
  {\bibinfo {title} {The universal definition of spin current},\ }\href
  {https://doi.org/10.1038/srep00388} {\bibfield  {journal} {\bibinfo
  {journal} {Scientific Reports}\ }\textbf {\bibinfo {volume} {2}},\ \bibinfo
  {pages} {388} (\bibinfo {year} {2012})}\BibitemShut {NoStop}%
\bibitem [{\citenamefont {Sun}\ and\ \citenamefont
  {Xie}(2005)}]{Sun2005definition}%
  \BibitemOpen
  \bibfield  {author} {\bibinfo {author} {\bibfnamefont {Q.-f.}\ \bibnamefont
  {Sun}}\ and\ \bibinfo {author} {\bibfnamefont {X.~C.}\ \bibnamefont {Xie}},\
  }\bibfield  {title} {\bibinfo {title} {Definition of the spin current: The
  angular spin current and its physical consequences},\ }\href
  {https://doi.org/10.1103/PhysRevB.72.245305} {\bibfield  {journal} {\bibinfo
  {journal} {Phys. Rev. B}\ }\textbf {\bibinfo {volume} {72}},\ \bibinfo
  {pages} {245305} (\bibinfo {year} {2005})}\BibitemShut {NoStop}%
\bibitem [{\citenamefont {Landau}\ and\ \citenamefont
  {Lifshitz}(1987)}]{Landau1987fluid}%
  \BibitemOpen
  \bibfield  {author} {\bibinfo {author} {\bibfnamefont {L.~D.}\ \bibnamefont
  {Landau}}\ and\ \bibinfo {author} {\bibfnamefont {E.~M.}\ \bibnamefont
  {Lifshitz}},\ }\href@noop {} {\emph {\bibinfo {title} {Fluid Mechanics}}}\
  (\bibinfo  {publisher} {Pergamon Press},\ \bibinfo {year} {1987})\ \bibinfo
  {note} {chap. 8}\BibitemShut {NoStop}%
\bibitem [{\citenamefont {Vyas}\ \emph {et~al.}(2013)\citenamefont {Vyas},
  \citenamefont {Kozawa},\ and\ \citenamefont {Sato}}]{Vyas2013polarization}%
  \BibitemOpen
  \bibfield  {author} {\bibinfo {author} {\bibfnamefont {S.}~\bibnamefont
  {Vyas}}, \bibinfo {author} {\bibfnamefont {Y.}~\bibnamefont {Kozawa}},\ and\
  \bibinfo {author} {\bibfnamefont {S.}~\bibnamefont {Sato}},\ }\bibfield
  {title} {\bibinfo {title} {Polarization singularities in superposition of
  vector beams},\ }\href {https://doi.org/10.1364/OE.21.008972} {\bibfield
  {journal} {\bibinfo  {journal} {Opt. Express}\ }\textbf {\bibinfo {volume}
  {21}},\ \bibinfo {pages} {8972} (\bibinfo {year} {2013})}\BibitemShut
  {NoStop}%
\bibitem [{\citenamefont {Otte}\ \emph {et~al.}(2016)\citenamefont {Otte},
  \citenamefont {Alpmann},\ and\ \citenamefont {Denz}}]{Otte2016higher}%
  \BibitemOpen
  \bibfield  {author} {\bibinfo {author} {\bibfnamefont {E.}~\bibnamefont
  {Otte}}, \bibinfo {author} {\bibfnamefont {C.}~\bibnamefont {Alpmann}},\ and\
  \bibinfo {author} {\bibfnamefont {C.}~\bibnamefont {Denz}},\ }\bibfield
  {title} {\bibinfo {title} {Higher-order polarization singularitites in
  tailored vector beams},\ }\href
  {https://doi.org/10.1088/2040-8978/18/7/074012} {\bibfield  {journal}
  {\bibinfo  {journal} {Journal of Optics}\ }\textbf {\bibinfo {volume} {18}},\
  \bibinfo {pages} {074012} (\bibinfo {year} {2016})}\BibitemShut {NoStop}%
\bibitem [{\citenamefont {Freund}\ \emph {et~al.}(2002)\citenamefont {Freund},
  \citenamefont {Mokhun}, \citenamefont {Soskin}, \citenamefont {Angelsky},\
  and\ \citenamefont {Mokhun}}]{Freund2002stokes}%
  \BibitemOpen
  \bibfield  {author} {\bibinfo {author} {\bibfnamefont {I.}~\bibnamefont
  {Freund}}, \bibinfo {author} {\bibfnamefont {A.~I.}\ \bibnamefont {Mokhun}},
  \bibinfo {author} {\bibfnamefont {M.~S.}\ \bibnamefont {Soskin}}, \bibinfo
  {author} {\bibfnamefont {O.~V.}\ \bibnamefont {Angelsky}},\ and\ \bibinfo
  {author} {\bibfnamefont {I.~I.}\ \bibnamefont {Mokhun}},\ }\bibfield  {title}
  {\bibinfo {title} {Stokes singularity relations},\ }\href
  {https://doi.org/10.1364/OL.27.000545} {\bibfield  {journal} {\bibinfo
  {journal} {Opt. Lett.}\ }\textbf {\bibinfo {volume} {27}},\ \bibinfo {pages}
  {545} (\bibinfo {year} {2002})}\BibitemShut {NoStop}%
\bibitem [{\citenamefont {Kulkarni}\ and\ \citenamefont
  {Gruev}(2012)}]{Kulkarni2012polarization}%
  \BibitemOpen
  \bibfield  {author} {\bibinfo {author} {\bibfnamefont {M.}~\bibnamefont
  {Kulkarni}}\ and\ \bibinfo {author} {\bibfnamefont {V.}~\bibnamefont
  {Gruev}},\ }\bibfield  {title} {\bibinfo {title} {Integrated
  spectral-polarization imaging sensor with aluminum nanowire polarization
  filters},\ }\href {https://doi.org/10.1364/OE.20.022997} {\bibfield
  {journal} {\bibinfo  {journal} {Opt. Express}\ }\textbf {\bibinfo {volume}
  {20}},\ \bibinfo {pages} {22997} (\bibinfo {year} {2012})}\BibitemShut
  {NoStop}%
\bibitem [{\citenamefont {Garcia}\ \emph {et~al.}(2017)\citenamefont {Garcia},
  \citenamefont {Edmiston}, \citenamefont {Marinov}, \citenamefont {Vail},\
  and\ \citenamefont {Gruev}}]{Garcia2017polarization}%
  \BibitemOpen
  \bibfield  {author} {\bibinfo {author} {\bibfnamefont {M.}~\bibnamefont
  {Garcia}}, \bibinfo {author} {\bibfnamefont {C.}~\bibnamefont {Edmiston}},
  \bibinfo {author} {\bibfnamefont {R.}~\bibnamefont {Marinov}}, \bibinfo
  {author} {\bibfnamefont {A.}~\bibnamefont {Vail}},\ and\ \bibinfo {author}
  {\bibfnamefont {V.}~\bibnamefont {Gruev}},\ }\bibfield  {title} {\bibinfo
  {title} {Bio-inspired color-polarization imager for real-time in situ
  imaging},\ }\href {https://doi.org/10.1364/OPTICA.4.001263} {\bibfield
  {journal} {\bibinfo  {journal} {Optica}\ }\textbf {\bibinfo {volume} {4}},\
  \bibinfo {pages} {1263} (\bibinfo {year} {2017})}\BibitemShut {NoStop}%
\bibitem [{\citenamefont {Andersen}\ \emph {et~al.}(2006)\citenamefont
  {Andersen}, \citenamefont {Ryu}, \citenamefont {Clad\'e}, \citenamefont
  {Natarajan}, \citenamefont {Vaziri}, \citenamefont {Helmerson},\ and\
  \citenamefont {Phillips}}]{anderson2006quantized}%
  \BibitemOpen
  \bibfield  {author} {\bibinfo {author} {\bibfnamefont {M.~F.}\ \bibnamefont
  {Andersen}}, \bibinfo {author} {\bibfnamefont {C.}~\bibnamefont {Ryu}},
  \bibinfo {author} {\bibfnamefont {P.}~\bibnamefont {Clad\'e}}, \bibinfo
  {author} {\bibfnamefont {V.}~\bibnamefont {Natarajan}}, \bibinfo {author}
  {\bibfnamefont {A.}~\bibnamefont {Vaziri}}, \bibinfo {author} {\bibfnamefont
  {K.}~\bibnamefont {Helmerson}},\ and\ \bibinfo {author} {\bibfnamefont
  {W.~D.}\ \bibnamefont {Phillips}},\ }\bibfield  {title} {\bibinfo {title}
  {Quantized rotation of atoms from photons with orbital angular momentum},\
  }\href {https://doi.org/10.1103/PhysRevLett.97.170406} {\bibfield  {journal}
  {\bibinfo  {journal} {Phys. Rev. Lett.}\ }\textbf {\bibinfo {volume} {97}},\
  \bibinfo {pages} {170406} (\bibinfo {year} {2006})}\BibitemShut {NoStop}%
\bibitem [{\citenamefont {Rosales-Guzm{\'{a}}n}\ \emph
  {et~al.}(2018)\citenamefont {Rosales-Guzm{\'{a}}n}, \citenamefont {Ndagano},\
  and\ \citenamefont {Forbes}}]{Rosales2018review}%
  \BibitemOpen
  \bibfield  {author} {\bibinfo {author} {\bibfnamefont {C.}~\bibnamefont
  {Rosales-Guzm{\'{a}}n}}, \bibinfo {author} {\bibfnamefont {B.}~\bibnamefont
  {Ndagano}},\ and\ \bibinfo {author} {\bibfnamefont {A.}~\bibnamefont
  {Forbes}},\ }\bibfield  {title} {\bibinfo {title} {A review of complex vector
  light fields and their applications},\ }\href
  {https://doi.org/10.1088/2040-8986/aaeb7d} {\bibfield  {journal} {\bibinfo
  {journal} {Journal of Optics}\ }\textbf {\bibinfo {volume} {20}},\ \bibinfo
  {pages} {123001} (\bibinfo {year} {2018})}\BibitemShut {NoStop}%
\bibitem [{\citenamefont {Shen}\ \emph {et~al.}(2019)\citenamefont {Shen},
  \citenamefont {Wang}, \citenamefont {Xie}, \citenamefont {Min}, \citenamefont
  {Fu}, \citenamefont {Liu}, \citenamefont {Gong},\ and\ \citenamefont
  {Yuan}}]{Shen2019optical}%
  \BibitemOpen
  \bibfield  {author} {\bibinfo {author} {\bibfnamefont {Y.}~\bibnamefont
  {Shen}}, \bibinfo {author} {\bibfnamefont {X.}~\bibnamefont {Wang}}, \bibinfo
  {author} {\bibfnamefont {Z.}~\bibnamefont {Xie}}, \bibinfo {author}
  {\bibfnamefont {C.}~\bibnamefont {Min}}, \bibinfo {author} {\bibfnamefont
  {X.}~\bibnamefont {Fu}}, \bibinfo {author} {\bibfnamefont {Q.}~\bibnamefont
  {Liu}}, \bibinfo {author} {\bibfnamefont {M.}~\bibnamefont {Gong}},\ and\
  \bibinfo {author} {\bibfnamefont {X.}~\bibnamefont {Yuan}},\ }\bibfield
  {title} {\bibinfo {title} {Optical vortices 30 years on: Oam manipulation
  from topological charge to multiple singularities},\ }\href
  {https://doi.org/10.1038/s41377-019-0194-2} {\bibfield  {journal} {\bibinfo
  {journal} {Light: Science {\&} Applications}\ }\textbf {\bibinfo {volume}
  {8}},\ \bibinfo {pages} {90} (\bibinfo {year} {2019})}\BibitemShut {NoStop}%
\bibitem [{\citenamefont {Enderlein}\ and\ \citenamefont
  {Pampaloni}(2004)}]{Enderlein2004unified}%
  \BibitemOpen
  \bibfield  {author} {\bibinfo {author} {\bibfnamefont {J.}~\bibnamefont
  {Enderlein}}\ and\ \bibinfo {author} {\bibfnamefont {F.}~\bibnamefont
  {Pampaloni}},\ }\bibfield  {title} {\bibinfo {title} {Unified operator
  approach for deriving hermite--gaussian and laguerre--gaussian laser modes},\
  }\href {https://doi.org/10.1364/JOSAA.21.001553} {\bibfield  {journal}
  {\bibinfo  {journal} {J. Opt. Soc. Am. A}\ }\textbf {\bibinfo {volume}
  {21}},\ \bibinfo {pages} {1553} (\bibinfo {year} {2004})}\BibitemShut
  {NoStop}%
\bibitem [{\citenamefont {Gatteschi}(2002)}]{GATTESCHI2002Asymptotics}%
  \BibitemOpen
  \bibfield  {author} {\bibinfo {author} {\bibfnamefont {L.}~\bibnamefont
  {Gatteschi}},\ }\bibfield  {title} {\bibinfo {title} {Asymptotics and bounds
  for the zeros of laguerre polynomials: a survey},\ }\href
  {https://doi.org/https://doi.org/10.1016/S0377-0427(01)00549-0} {\bibfield
  {journal} {\bibinfo  {journal} {Journal of Computational and Applied
  Mathematics}\ }\textbf {\bibinfo {volume} {144}},\ \bibinfo {pages} {7}
  (\bibinfo {year} {2002})},\ \bibinfo {note} {selected papers of the Int.
  Symp. on Applied Mathematics, August 2000, Dalian, China}\BibitemShut
  {NoStop}%
\bibitem [{\citenamefont {Gori}\ \emph {et~al.}(1987)\citenamefont {Gori},
  \citenamefont {Guattari},\ and\ \citenamefont {Padovani}}]{gori1987bessel}%
  \BibitemOpen
  \bibfield  {author} {\bibinfo {author} {\bibfnamefont {F.}~\bibnamefont
  {Gori}}, \bibinfo {author} {\bibfnamefont {G.}~\bibnamefont {Guattari}},\
  and\ \bibinfo {author} {\bibfnamefont {C.}~\bibnamefont {Padovani}},\
  }\bibfield  {title} {\bibinfo {title} {Bessel-gauss beams},\ }\href
  {https://doi.org/10.1016/0030-4018(87)90276-8} {\bibfield  {journal}
  {\bibinfo  {journal} {Optics communications}\ }\textbf {\bibinfo {volume}
  {64}},\ \bibinfo {pages} {491} (\bibinfo {year} {1987})}\BibitemShut
  {NoStop}%
\bibitem [{\citenamefont {Kovalev}\ and\ \citenamefont
  {Kotlyar}(2021)}]{kovalev2021propagation}%
  \BibitemOpen
  \bibfield  {author} {\bibinfo {author} {\bibfnamefont {A.~A.}\ \bibnamefont
  {Kovalev}}\ and\ \bibinfo {author} {\bibfnamefont {V.~V.}\ \bibnamefont
  {Kotlyar}},\ }\bibfield  {title} {\bibinfo {title} {Propagation-invariant
  laser beams with an array of phase singularities},\ }\href
  {https://doi.org/10.1103/PhysRevA.103.063502} {\bibfield  {journal} {\bibinfo
   {journal} {Phys. Rev. A}\ }\textbf {\bibinfo {volume} {103}},\ \bibinfo
  {pages} {063502} (\bibinfo {year} {2021})}\BibitemShut {NoStop}%
\bibitem [{\citenamefont {Pleinert}\ \emph {et~al.}(2021)\citenamefont
  {Pleinert}, \citenamefont {Rueda}, \citenamefont {Lutz},\ and\ \citenamefont
  {von Zanthier}}]{Zanthier2021highorder}%
  \BibitemOpen
  \bibfield  {author} {\bibinfo {author} {\bibfnamefont {M.-O.}\ \bibnamefont
  {Pleinert}}, \bibinfo {author} {\bibfnamefont {A.}~\bibnamefont {Rueda}},
  \bibinfo {author} {\bibfnamefont {E.}~\bibnamefont {Lutz}},\ and\ \bibinfo
  {author} {\bibfnamefont {J.}~\bibnamefont {von Zanthier}},\ }\bibfield
  {title} {\bibinfo {title} {Testing higher-order quantum interference with
  many-particle states},\ }\href
  {https://doi.org/10.1103/PhysRevLett.126.190401} {\bibfield  {journal}
  {\bibinfo  {journal} {Phys. Rev. Lett.}\ }\textbf {\bibinfo {volume} {126}},\
  \bibinfo {pages} {190401} (\bibinfo {year} {2021})}\BibitemShut {NoStop}%
\bibitem [{\citenamefont {Torres}\ \emph {et~al.}(2003)\citenamefont {Torres},
  \citenamefont {Alexandrescu},\ and\ \citenamefont
  {Torner}}]{Torres2003quantum}%
  \BibitemOpen
  \bibfield  {author} {\bibinfo {author} {\bibfnamefont {J.~P.}\ \bibnamefont
  {Torres}}, \bibinfo {author} {\bibfnamefont {A.}~\bibnamefont
  {Alexandrescu}},\ and\ \bibinfo {author} {\bibfnamefont {L.}~\bibnamefont
  {Torner}},\ }\bibfield  {title} {\bibinfo {title} {Quantum spiral bandwidth
  of entangled two-photon states},\ }\href
  {https://doi.org/10.1103/PhysRevA.68.050301} {\bibfield  {journal} {\bibinfo
  {journal} {Phys. Rev. A}\ }\textbf {\bibinfo {volume} {68}},\ \bibinfo
  {pages} {050301} (\bibinfo {year} {2003})}\BibitemShut {NoStop}%
\bibitem [{\citenamefont {Saleh}\ \emph {et~al.}(2000)\citenamefont {Saleh},
  \citenamefont {Abouraddy}, \citenamefont {Sergienko},\ and\ \citenamefont
  {Teich}}]{Saleh2000photonpair}%
  \BibitemOpen
  \bibfield  {author} {\bibinfo {author} {\bibfnamefont {B.~E.~A.}\
  \bibnamefont {Saleh}}, \bibinfo {author} {\bibfnamefont {A.~F.}\ \bibnamefont
  {Abouraddy}}, \bibinfo {author} {\bibfnamefont {A.~V.}\ \bibnamefont
  {Sergienko}},\ and\ \bibinfo {author} {\bibfnamefont {M.~C.}\ \bibnamefont
  {Teich}},\ }\bibfield  {title} {\bibinfo {title} {Duality between partial
  coherence and partial entanglement},\ }\href
  {https://doi.org/10.1103/PhysRevA.62.043816} {\bibfield  {journal} {\bibinfo
  {journal} {Phys. Rev. A}\ }\textbf {\bibinfo {volume} {62}},\ \bibinfo
  {pages} {043816} (\bibinfo {year} {2000})}\BibitemShut {NoStop}%
\bibitem [{\citenamefont {Liu}\ \emph {et~al.}(2019)\citenamefont {Liu},
  \citenamefont {Su}, \citenamefont {Wei}, \citenamefont {Yao}, \citenamefont
  {Silva}, \citenamefont {Yu}, \citenamefont {Iles-Smith}, \citenamefont
  {Srinivasan}, \citenamefont {Rastelli}, \citenamefont {Li},\ and\
  \citenamefont {Wang}}]{Liu2019photonpair}%
  \BibitemOpen
  \bibfield  {author} {\bibinfo {author} {\bibfnamefont {J.}~\bibnamefont
  {Liu}}, \bibinfo {author} {\bibfnamefont {R.}~\bibnamefont {Su}}, \bibinfo
  {author} {\bibfnamefont {Y.}~\bibnamefont {Wei}}, \bibinfo {author}
  {\bibfnamefont {B.}~\bibnamefont {Yao}}, \bibinfo {author} {\bibfnamefont
  {S.~F. C.~d.}\ \bibnamefont {Silva}}, \bibinfo {author} {\bibfnamefont
  {Y.}~\bibnamefont {Yu}}, \bibinfo {author} {\bibfnamefont {J.}~\bibnamefont
  {Iles-Smith}}, \bibinfo {author} {\bibfnamefont {K.}~\bibnamefont
  {Srinivasan}}, \bibinfo {author} {\bibfnamefont {A.}~\bibnamefont
  {Rastelli}}, \bibinfo {author} {\bibfnamefont {J.}~\bibnamefont {Li}},\ and\
  \bibinfo {author} {\bibfnamefont {X.}~\bibnamefont {Wang}},\ }\bibfield
  {title} {\bibinfo {title} {A solid-state source of strongly entangled photon
  pairs with high brightness and indistinguishability},\ }\href
  {https://doi.org/10.1038/s41565-019-0435-9} {\bibfield  {journal} {\bibinfo
  {journal} {Nature Nanotechnology}\ }\textbf {\bibinfo {volume} {14}},\
  \bibinfo {pages} {586} (\bibinfo {year} {2019})}\BibitemShut {NoStop}%
\bibitem [{\citenamefont {Shalm}\ \emph {et~al.}(2015)\citenamefont {Shalm},
  \citenamefont {Meyer-Scott}, \citenamefont {Christensen}, \citenamefont
  {Bierhorst}, \citenamefont {Wayne}, \citenamefont {Stevens}, \citenamefont
  {Gerrits}, \citenamefont {Glancy}, \citenamefont {Hamel}, \citenamefont
  {Allman}, \citenamefont {Coakley}, \citenamefont {Dyer}, \citenamefont
  {Hodge}, \citenamefont {Lita}, \citenamefont {Verma}, \citenamefont
  {Lambrocco}, \citenamefont {Tortorici}, \citenamefont {Migdall},
  \citenamefont {Zhang}, \citenamefont {Kumor}, \citenamefont {Farr},
  \citenamefont {Marsili}, \citenamefont {Shaw}, \citenamefont {Stern},
  \citenamefont {Abell\'an}, \citenamefont {Amaya}, \citenamefont {Pruneri},
  \citenamefont {Jennewein}, \citenamefont {Mitchell}, \citenamefont {Kwiat},
  \citenamefont {Bienfang}, \citenamefont {Mirin}, \citenamefont {Knill},\ and\
  \citenamefont {Nam}}]{Shalm2015strong}%
  \BibitemOpen
  \bibfield  {author} {\bibinfo {author} {\bibfnamefont {L.~K.}\ \bibnamefont
  {Shalm}}, \bibinfo {author} {\bibfnamefont {E.}~\bibnamefont {Meyer-Scott}},
  \bibinfo {author} {\bibfnamefont {B.~G.}\ \bibnamefont {Christensen}},
  \bibinfo {author} {\bibfnamefont {P.}~\bibnamefont {Bierhorst}}, \bibinfo
  {author} {\bibfnamefont {M.~A.}\ \bibnamefont {Wayne}}, \bibinfo {author}
  {\bibfnamefont {M.~J.}\ \bibnamefont {Stevens}}, \bibinfo {author}
  {\bibfnamefont {T.}~\bibnamefont {Gerrits}}, \bibinfo {author} {\bibfnamefont
  {S.}~\bibnamefont {Glancy}}, \bibinfo {author} {\bibfnamefont {D.~R.}\
  \bibnamefont {Hamel}}, \bibinfo {author} {\bibfnamefont {M.~S.}\ \bibnamefont
  {Allman}}, \bibinfo {author} {\bibfnamefont {K.~J.}\ \bibnamefont {Coakley}},
  \bibinfo {author} {\bibfnamefont {S.~D.}\ \bibnamefont {Dyer}}, \bibinfo
  {author} {\bibfnamefont {C.}~\bibnamefont {Hodge}}, \bibinfo {author}
  {\bibfnamefont {A.~E.}\ \bibnamefont {Lita}}, \bibinfo {author}
  {\bibfnamefont {V.~B.}\ \bibnamefont {Verma}}, \bibinfo {author}
  {\bibfnamefont {C.}~\bibnamefont {Lambrocco}}, \bibinfo {author}
  {\bibfnamefont {E.}~\bibnamefont {Tortorici}}, \bibinfo {author}
  {\bibfnamefont {A.~L.}\ \bibnamefont {Migdall}}, \bibinfo {author}
  {\bibfnamefont {Y.}~\bibnamefont {Zhang}}, \bibinfo {author} {\bibfnamefont
  {D.~R.}\ \bibnamefont {Kumor}}, \bibinfo {author} {\bibfnamefont {W.~H.}\
  \bibnamefont {Farr}}, \bibinfo {author} {\bibfnamefont {F.}~\bibnamefont
  {Marsili}}, \bibinfo {author} {\bibfnamefont {M.~D.}\ \bibnamefont {Shaw}},
  \bibinfo {author} {\bibfnamefont {J.~A.}\ \bibnamefont {Stern}}, \bibinfo
  {author} {\bibfnamefont {C.}~\bibnamefont {Abell\'an}}, \bibinfo {author}
  {\bibfnamefont {W.}~\bibnamefont {Amaya}}, \bibinfo {author} {\bibfnamefont
  {V.}~\bibnamefont {Pruneri}}, \bibinfo {author} {\bibfnamefont
  {T.}~\bibnamefont {Jennewein}}, \bibinfo {author} {\bibfnamefont {M.~W.}\
  \bibnamefont {Mitchell}}, \bibinfo {author} {\bibfnamefont {P.~G.}\
  \bibnamefont {Kwiat}}, \bibinfo {author} {\bibfnamefont {J.~C.}\ \bibnamefont
  {Bienfang}}, \bibinfo {author} {\bibfnamefont {R.~P.}\ \bibnamefont {Mirin}},
  \bibinfo {author} {\bibfnamefont {E.}~\bibnamefont {Knill}},\ and\ \bibinfo
  {author} {\bibfnamefont {S.~W.}\ \bibnamefont {Nam}},\ }\bibfield  {title}
  {\bibinfo {title} {Strong loophole-free test of local realism},\ }\href
  {https://doi.org/10.1103/PhysRevLett.115.250402} {\bibfield  {journal}
  {\bibinfo  {journal} {Phys. Rev. Lett.}\ }\textbf {\bibinfo {volume} {115}},\
  \bibinfo {pages} {250402} (\bibinfo {year} {2015})}\BibitemShut {NoStop}%
\bibitem [{\citenamefont {Ac\'{\i}n}\ \emph {et~al.}(2007)\citenamefont
  {Ac\'{\i}n}, \citenamefont {Brunner}, \citenamefont {Gisin}, \citenamefont
  {Massar}, \citenamefont {Pironio},\ and\ \citenamefont
  {Scarani}}]{Antonio2007keydistribution}%
  \BibitemOpen
  \bibfield  {author} {\bibinfo {author} {\bibfnamefont {A.}~\bibnamefont
  {Ac\'{\i}n}}, \bibinfo {author} {\bibfnamefont {N.}~\bibnamefont {Brunner}},
  \bibinfo {author} {\bibfnamefont {N.}~\bibnamefont {Gisin}}, \bibinfo
  {author} {\bibfnamefont {S.}~\bibnamefont {Massar}}, \bibinfo {author}
  {\bibfnamefont {S.}~\bibnamefont {Pironio}},\ and\ \bibinfo {author}
  {\bibfnamefont {V.}~\bibnamefont {Scarani}},\ }\bibfield  {title} {\bibinfo
  {title} {Device-independent security of quantum cryptography against
  collective attacks},\ }\href {https://doi.org/10.1103/PhysRevLett.98.230501}
  {\bibfield  {journal} {\bibinfo  {journal} {Phys. Rev. Lett.}\ }\textbf
  {\bibinfo {volume} {98}},\ \bibinfo {pages} {230501} (\bibinfo {year}
  {2007})}\BibitemShut {NoStop}%
\bibitem [{\citenamefont {Cai}\ and\ \citenamefont {Zhu}(2005)}]{cai2005ghost}%
  \BibitemOpen
  \bibfield  {author} {\bibinfo {author} {\bibfnamefont {Y.}~\bibnamefont
  {Cai}}\ and\ \bibinfo {author} {\bibfnamefont {S.-Y.}\ \bibnamefont {Zhu}},\
  }\bibfield  {title} {\bibinfo {title} {Ghost imaging with incoherent and
  partially coherent light radiation},\ }\href
  {https://doi.org/10.1103/PhysRevE.71.056607} {\bibfield  {journal} {\bibinfo
  {journal} {Phys. Rev. E}\ }\textbf {\bibinfo {volume} {71}},\ \bibinfo
  {pages} {056607} (\bibinfo {year} {2005})}\BibitemShut {NoStop}%
\bibitem [{\citenamefont {Chan}\ \emph {et~al.}(2009)\citenamefont {Chan},
  \citenamefont {O'Sullivan},\ and\ \citenamefont {Boyd}}]{Chan2009twocolor}%
  \BibitemOpen
  \bibfield  {author} {\bibinfo {author} {\bibfnamefont {K.~W.~C.}\
  \bibnamefont {Chan}}, \bibinfo {author} {\bibfnamefont {M.~N.}\ \bibnamefont
  {O'Sullivan}},\ and\ \bibinfo {author} {\bibfnamefont {R.~W.}\ \bibnamefont
  {Boyd}},\ }\bibfield  {title} {\bibinfo {title} {Two-color ghost imaging},\
  }\href {https://doi.org/10.1103/PhysRevA.79.033808} {\bibfield  {journal}
  {\bibinfo  {journal} {Phys. Rev. A}\ }\textbf {\bibinfo {volume} {79}},\
  \bibinfo {pages} {033808} (\bibinfo {year} {2009})}\BibitemShut {NoStop}%
\bibitem [{\citenamefont {Loudon}(2000)}]{loudon2000quantum}%
  \BibitemOpen
  \bibfield  {author} {\bibinfo {author} {\bibfnamefont {R.}~\bibnamefont
  {Loudon}},\ }\href@noop {} {\emph {\bibinfo {title} {The quantum theory of
  light}}}\ (\bibinfo  {publisher} {OUP Oxford},\ \bibinfo {year}
  {2000})\BibitemShut {NoStop}%
\bibitem [{\citenamefont {Fearn}\ and\ \citenamefont
  {Loudon}(1989)}]{Fearn1989twophoton}%
  \BibitemOpen
  \bibfield  {author} {\bibinfo {author} {\bibfnamefont {H.}~\bibnamefont
  {Fearn}}\ and\ \bibinfo {author} {\bibfnamefont {R.}~\bibnamefont {Loudon}},\
  }\bibfield  {title} {\bibinfo {title} {Theory of two-photon interference},\
  }\href {https://doi.org/10.1364/JOSAB.6.000917} {\bibfield  {journal}
  {\bibinfo  {journal} {J. Opt. Soc. Am. B}\ }\textbf {\bibinfo {volume} {6}},\
  \bibinfo {pages} {917} (\bibinfo {year} {1989})}\BibitemShut {NoStop}%
\bibitem [{\citenamefont {Chang}\ \emph {et~al.}(2016)\citenamefont {Chang},
  \citenamefont {Gonz\'alez-Tudela}, \citenamefont {S\'anchez Mu\~noz},
  \citenamefont {Navarrete-Benlloch},\ and\ \citenamefont
  {Shi}}]{chang2016determinstic}%
  \BibitemOpen
  \bibfield  {author} {\bibinfo {author} {\bibfnamefont {Y.}~\bibnamefont
  {Chang}}, \bibinfo {author} {\bibfnamefont {A.}~\bibnamefont
  {Gonz\'alez-Tudela}}, \bibinfo {author} {\bibfnamefont {C.}~\bibnamefont
  {S\'anchez Mu\~noz}}, \bibinfo {author} {\bibfnamefont {C.}~\bibnamefont
  {Navarrete-Benlloch}},\ and\ \bibinfo {author} {\bibfnamefont
  {T.}~\bibnamefont {Shi}},\ }\bibfield  {title} {\bibinfo {title}
  {Deterministic down-converter and continuous photon-pair source within the
  bad-cavity limit},\ }\href {https://doi.org/10.1103/PhysRevLett.117.203602}
  {\bibfield  {journal} {\bibinfo  {journal} {Phys. Rev. Lett.}\ }\textbf
  {\bibinfo {volume} {117}},\ \bibinfo {pages} {203602} (\bibinfo {year}
  {2016})}\BibitemShut {NoStop}%
\bibitem [{\citenamefont {Bin}\ \emph {et~al.}(2020)\citenamefont {Bin},
  \citenamefont {L\"u}, \citenamefont {Laussy}, \citenamefont {Nori},\ and\
  \citenamefont {Wu}}]{Bin2020Nphoton}%
  \BibitemOpen
  \bibfield  {author} {\bibinfo {author} {\bibfnamefont {Q.}~\bibnamefont
  {Bin}}, \bibinfo {author} {\bibfnamefont {X.-Y.}\ \bibnamefont {L\"u}},
  \bibinfo {author} {\bibfnamefont {F.~P.}\ \bibnamefont {Laussy}}, \bibinfo
  {author} {\bibfnamefont {F.}~\bibnamefont {Nori}},\ and\ \bibinfo {author}
  {\bibfnamefont {Y.}~\bibnamefont {Wu}},\ }\bibfield  {title} {\bibinfo
  {title} {$n$-phonon bundle emission via the stokes process},\ }\href
  {https://doi.org/10.1103/PhysRevLett.124.053601} {\bibfield  {journal}
  {\bibinfo  {journal} {Phys. Rev. Lett.}\ }\textbf {\bibinfo {volume} {124}},\
  \bibinfo {pages} {053601} (\bibinfo {year} {2020})}\BibitemShut {NoStop}%
\bibitem [{\citenamefont {Moreno-Cardoner}\ \emph {et~al.}(2021)\citenamefont
  {Moreno-Cardoner}, \citenamefont {Goncalves},\ and\ \citenamefont
  {Chang}}]{chang2021quantum}%
  \BibitemOpen
  \bibfield  {author} {\bibinfo {author} {\bibfnamefont {M.}~\bibnamefont
  {Moreno-Cardoner}}, \bibinfo {author} {\bibfnamefont {D.}~\bibnamefont
  {Goncalves}},\ and\ \bibinfo {author} {\bibfnamefont {D.~E.}\ \bibnamefont
  {Chang}},\ }\bibfield  {title} {\bibinfo {title} {Quantum nonlinear optics
  based on two-dimensional rydberg atom arrays},\ }\href
  {https://arxiv.org/abs/2101.01936} {\bibfield  {journal} {\bibinfo  {journal}
  {arXiv preprint:2101.01936}\ } (\bibinfo {year} {2021})}\BibitemShut
  {NoStop}%
\bibitem [{\citenamefont {Zhang}\ \emph {et~al.}(2021)\citenamefont {Zhang},
  \citenamefont {Walther}, \citenamefont {Mølmer},\ and\ \citenamefont
  {Pohl}}]{zhang2021photon}%
  \BibitemOpen
  \bibfield  {author} {\bibinfo {author} {\bibfnamefont {L.}~\bibnamefont
  {Zhang}}, \bibinfo {author} {\bibfnamefont {V.}~\bibnamefont {Walther}},
  \bibinfo {author} {\bibfnamefont {K.}~\bibnamefont {Mølmer}},\ and\ \bibinfo
  {author} {\bibfnamefont {T.}~\bibnamefont {Pohl}},\ }\bibfield  {title}
  {\bibinfo {title} {Photon-photon interactions in rydberg-atom arrays},\
  }\href {https://arxiv.org/abs/2101.11375} {\bibfield  {journal} {\bibinfo
  {journal} {arXiv preprint:2101.11375}\ } (\bibinfo {year}
  {2021})}\BibitemShut {NoStop}%
\bibitem [{\citenamefont {Kalhor}\ \emph {et~al.}(2021)\citenamefont {Kalhor},
  \citenamefont {Yang}, \citenamefont {Bauer},\ and\ \citenamefont
  {Jacob}}]{Kalhor2020probe}%
  \BibitemOpen
  \bibfield  {author} {\bibinfo {author} {\bibfnamefont {F.}~\bibnamefont
  {Kalhor}}, \bibinfo {author} {\bibfnamefont {L.-P.}\ \bibnamefont {Yang}},
  \bibinfo {author} {\bibfnamefont {L.}~\bibnamefont {Bauer}},\ and\ \bibinfo
  {author} {\bibfnamefont {Z.}~\bibnamefont {Jacob}},\ }\bibfield  {title}
  {\bibinfo {title} {Quantum sensing of photonic spin density using a single
  spin qubit},\ }\href {https://doi.org/10.1103/PhysRevResearch.3.043007}
  {\bibfield  {journal} {\bibinfo  {journal} {Phys. Rev. Research}\ }\textbf
  {\bibinfo {volume} {3}},\ \bibinfo {pages} {043007} (\bibinfo {year}
  {2021})}\BibitemShut {NoStop}%
\bibitem [{\citenamefont {Arnaut}\ and\ \citenamefont
  {Barbosa}(2000)}]{Arnaut2000photonpair}%
  \BibitemOpen
  \bibfield  {author} {\bibinfo {author} {\bibfnamefont {H.~H.}\ \bibnamefont
  {Arnaut}}\ and\ \bibinfo {author} {\bibfnamefont {G.~A.}\ \bibnamefont
  {Barbosa}},\ }\bibfield  {title} {\bibinfo {title} {Orbital and intrinsic
  angular momentum of single photons and entangled pairs of photons generated
  by parametric down-conversion},\ }\href
  {https://doi.org/10.1103/PhysRevLett.85.286} {\bibfield  {journal} {\bibinfo
  {journal} {Phys. Rev. Lett.}\ }\textbf {\bibinfo {volume} {85}},\ \bibinfo
  {pages} {286} (\bibinfo {year} {2000})}\BibitemShut {NoStop}%
\bibitem [{\citenamefont {Keller}\ and\ \citenamefont
  {Rubin}(1997)}]{Keller1997theory}%
  \BibitemOpen
  \bibfield  {author} {\bibinfo {author} {\bibfnamefont {T.~E.}\ \bibnamefont
  {Keller}}\ and\ \bibinfo {author} {\bibfnamefont {M.~H.}\ \bibnamefont
  {Rubin}},\ }\bibfield  {title} {\bibinfo {title} {Theory of two-photon
  entanglement for spontaneous parametric down-conversion driven by a narrow
  pump pulse},\ }\href {https://doi.org/10.1103/PhysRevA.56.1534} {\bibfield
  {journal} {\bibinfo  {journal} {Phys. Rev. A}\ }\textbf {\bibinfo {volume}
  {56}},\ \bibinfo {pages} {1534} (\bibinfo {year} {1997})}\BibitemShut
  {NoStop}%
\bibitem [{\citenamefont {Maurer}\ \emph {et~al.}(2011)\citenamefont {Maurer},
  \citenamefont {Jesacher}, \citenamefont {Bernet},\ and\ \citenamefont
  {Ritsch-Marte}}]{maurer2011spatial}%
  \BibitemOpen
  \bibfield  {author} {\bibinfo {author} {\bibfnamefont {C.}~\bibnamefont
  {Maurer}}, \bibinfo {author} {\bibfnamefont {A.}~\bibnamefont {Jesacher}},
  \bibinfo {author} {\bibfnamefont {S.}~\bibnamefont {Bernet}},\ and\ \bibinfo
  {author} {\bibfnamefont {M.}~\bibnamefont {Ritsch-Marte}},\ }\bibfield
  {title} {\bibinfo {title} {What spatial light modulators can do for optical
  microscopy},\ }\href {https://doi.org/10.1002/lpor.200900047} {\bibfield
  {journal} {\bibinfo  {journal} {Laser \& Photonics Reviews}\ }\textbf
  {\bibinfo {volume} {5}},\ \bibinfo {pages} {81} (\bibinfo {year}
  {2011})}\BibitemShut {NoStop}%
\bibitem [{\citenamefont {Lloyd}\ and\ \citenamefont
  {S.}(2008)}]{Lloyd2008Enhanced}%
  \BibitemOpen
  \bibfield  {author} {\bibinfo {author} {\bibnamefont {Lloyd}}\ and\ \bibinfo
  {author} {\bibnamefont {S.}},\ }\bibfield  {title} {\bibinfo {title}
  {Enhanced sensitivity of photodetection via quantum illumination},\ }\href
  {https://doi.org/10.1126/science.1160627} {\bibfield  {journal} {\bibinfo
  {journal} {Science}\ }\textbf {\bibinfo {volume} {321}},\ \bibinfo {pages}
  {1463} (\bibinfo {year} {2008})}\BibitemShut {NoStop}%
\bibitem [{\citenamefont {Tan}\ \emph {et~al.}(2008)\citenamefont {Tan},
  \citenamefont {Erkmen}, \citenamefont {Giovannetti}, \citenamefont {Guha},
  \citenamefont {Lloyd}, \citenamefont {Maccone}, \citenamefont {Pirandola},\
  and\ \citenamefont {Shapiro}}]{Tan2008quantum}%
  \BibitemOpen
  \bibfield  {author} {\bibinfo {author} {\bibfnamefont {S.-H.}\ \bibnamefont
  {Tan}}, \bibinfo {author} {\bibfnamefont {B.~I.}\ \bibnamefont {Erkmen}},
  \bibinfo {author} {\bibfnamefont {V.}~\bibnamefont {Giovannetti}}, \bibinfo
  {author} {\bibfnamefont {S.}~\bibnamefont {Guha}}, \bibinfo {author}
  {\bibfnamefont {S.}~\bibnamefont {Lloyd}}, \bibinfo {author} {\bibfnamefont
  {L.}~\bibnamefont {Maccone}}, \bibinfo {author} {\bibfnamefont
  {S.}~\bibnamefont {Pirandola}},\ and\ \bibinfo {author} {\bibfnamefont
  {J.~H.}\ \bibnamefont {Shapiro}},\ }\bibfield  {title} {\bibinfo {title}
  {Quantum illumination with gaussian states},\ }\href
  {https://doi.org/10.1103/PhysRevLett.101.253601} {\bibfield  {journal}
  {\bibinfo  {journal} {Phys. Rev. Lett.}\ }\textbf {\bibinfo {volume} {101}},\
  \bibinfo {pages} {253601} (\bibinfo {year} {2008})}\BibitemShut {NoStop}%
\bibitem [{\citenamefont {Yang}\ \emph {et~al.}(2020)\citenamefont {Yang},
  \citenamefont {Khosravi},\ and\ \citenamefont {Jacob}}]{yang2020quantum}%
  \BibitemOpen
  \bibfield  {author} {\bibinfo {author} {\bibfnamefont {L.-P.}\ \bibnamefont
  {Yang}}, \bibinfo {author} {\bibfnamefont {F.}~\bibnamefont {Khosravi}},\
  and\ \bibinfo {author} {\bibfnamefont {Z.}~\bibnamefont {Jacob}},\ }\bibfield
   {title} {\bibinfo {title} {Quantum spin operator of the photon},\ }\href
  {https://arxiv.org/abs/2004.03771} {\bibfield  {journal} {\bibinfo  {journal}
  {arXiv preprint:2004.03771}\ } (\bibinfo {year} {2020})}\BibitemShut
  {NoStop}%
\bibitem [{\citenamefont {Barnett}\ and\ \citenamefont
  {Allen}(1994)}]{barnett1994orbital}%
  \BibitemOpen
  \bibfield  {author} {\bibinfo {author} {\bibfnamefont {S.~M.}\ \bibnamefont
  {Barnett}}\ and\ \bibinfo {author} {\bibfnamefont {L.}~\bibnamefont
  {Allen}},\ }\bibfield  {title} {\bibinfo {title} {Orbital angular momentum
  and nonparaxial light beams},\ }\href
  {https://www.sciencedirect.com/science/article/abs/pii/0030401894902690}
  {\bibfield  {journal} {\bibinfo  {journal} {Optics communications}\ }\textbf
  {\bibinfo {volume} {110}},\ \bibinfo {pages} {670} (\bibinfo {year}
  {1994})}\BibitemShut {NoStop}%
\bibitem [{\citenamefont {Berry}(1998)}]{berry1998paraxial}%
  \BibitemOpen
  \bibfield  {author} {\bibinfo {author} {\bibfnamefont {M.~V.}\ \bibnamefont
  {Berry}},\ }\bibfield  {title} {\bibinfo {title} {Paraxial beams of spinning
  light},\ }in\ \href
  {https://www.spiedigitallibrary.org/conference-proceedings-of-spie/3487/0000/Paraxial-beams-of-spinning-light/10.1117/12.317704.full}
  {\emph {\bibinfo {booktitle} {International conference on singular
  optics}}},\ Vol.\ \bibinfo {volume} {3487}\ (\bibinfo {organization}
  {International Society for Optics and Photonics},\ \bibinfo {year} {1998})\
  pp.\ \bibinfo {pages} {6--11}\BibitemShut {NoStop}%
\bibitem [{\citenamefont {Li}\ \emph {et~al.}(2020)\citenamefont {Li},
  \citenamefont {Rodriguez-Fajardo}, \citenamefont {Chen},\ and\ \citenamefont
  {Forbes}}]{li2020spin}%
  \BibitemOpen
  \bibfield  {author} {\bibinfo {author} {\bibfnamefont {H.}~\bibnamefont
  {Li}}, \bibinfo {author} {\bibfnamefont {V.}~\bibnamefont
  {Rodriguez-Fajardo}}, \bibinfo {author} {\bibfnamefont {P.}~\bibnamefont
  {Chen}},\ and\ \bibinfo {author} {\bibfnamefont {A.}~\bibnamefont {Forbes}},\
  }\bibfield  {title} {\bibinfo {title} {Spin and orbital angular momentum
  dynamics in counterpropagating vectorially structured light},\ }\href
  {https://doi.org/10.1103/PhysRevA.102.063533} {\bibfield  {journal} {\bibinfo
   {journal} {Phys. Rev. A}\ }\textbf {\bibinfo {volume} {102}},\ \bibinfo
  {pages} {063533} (\bibinfo {year} {2020})}\BibitemShut {NoStop}%
\bibitem [{\citenamefont {Bialynicki-Birula}(1994)}]{Bia1994On}%
  \BibitemOpen
  \bibfield  {author} {\bibinfo {author} {\bibfnamefont {I.}~\bibnamefont
  {Bialynicki-Birula}},\ }\bibfield  {title} {\bibinfo {title} {On the wave
  function of the photon},\ }\href {https://doi.org/10.12693/aphyspola.86.97}
  {\bibfield  {journal} {\bibinfo  {journal} {Acta Phys. Pol. A}\ }\textbf
  {\bibinfo {volume} {86}},\ \bibinfo {pages} {97} (\bibinfo {year}
  {1994})}\BibitemShut {NoStop}%
\bibitem [{\citenamefont {Sipe}(1995)}]{Sipe1995photon}%
  \BibitemOpen
  \bibfield  {author} {\bibinfo {author} {\bibfnamefont {J.~E.}\ \bibnamefont
  {Sipe}},\ }\bibfield  {title} {\bibinfo {title} {Photon wave functions},\
  }\href {https://doi.org/10.1103/PhysRevA.52.1875} {\bibfield  {journal}
  {\bibinfo  {journal} {Phys. Rev. A}\ }\textbf {\bibinfo {volume} {52}},\
  \bibinfo {pages} {1875} (\bibinfo {year} {1995})}\BibitemShut {NoStop}%
\bibitem [{\citenamefont {Rozas}\ \emph
  {et~al.}(1997{\natexlab{b}})\citenamefont {Rozas}, \citenamefont {Law},\ and\
  \citenamefont {Swartzlander}}]{Rozas1997propagation}%
  \BibitemOpen
  \bibfield  {author} {\bibinfo {author} {\bibfnamefont {D.}~\bibnamefont
  {Rozas}}, \bibinfo {author} {\bibfnamefont {C.~T.}\ \bibnamefont {Law}},\
  and\ \bibinfo {author} {\bibfnamefont {G.~A.}\ \bibnamefont {Swartzlander}},\
  }\bibfield  {title} {\bibinfo {title} {Propagation dynamics of optical
  vortices},\ }\href {https://doi.org/10.1364/JOSAB.14.003054} {\bibfield
  {journal} {\bibinfo  {journal} {J. Opt. Soc. Am. B}\ }\textbf {\bibinfo
  {volume} {14}},\ \bibinfo {pages} {3054} (\bibinfo {year}
  {1997}{\natexlab{b}})}\BibitemShut {NoStop}%
\end{thebibliography}%
\end{document}